\documentclass[12pt]{article}
\usepackage{epsfig}   

\def\sec{\ifmmode {}^{\prime\prime}\else ${}^{\prime\prime}$\fi~}

\def\magdot{\ifmmode {}^{\rm m}\!\!\!.\, \else ${}^{\rm m}\!\!\!.\,$\fi}
\def\daydot{\ifmmode {}^{\rm d}\!\!\!.\, \else ${}^{\rm d}\!\!\!.\,$\fi}
\def\degdot{\hbox{$.\!\!^{\circ}$}}
\def\asec{\ifmmode ^{\prime\prime}\else$^{\prime\prime}$\fi}

\def\sun{\hbox{$\odot$}}

\def\la{\mathrel{\mathchoice {\vcenter{\offinterlineskip\halign{\hfil
$\displaystyle##$\hfil\cr<\cr\sim\cr}}}
{\vcenter{\offinterlineskip\halign{\hfil$\textstyle##$\hfil\cr
<\cr\sim\cr}}}
{\vcenter{\offinterlineskip\halign{\hfil$\scriptstyle##$\hfil\cr
<\cr\sim\cr}}}
{\vcenter{\offinterlineskip\halign{\hfil$\scriptscriptstyle##$\hfil\cr
<\cr\sim\cr}}}}}

\def\ga{\mathrel{\mathchoice {\vcenter{\offinterlineskip\halign{\hfil
$\displaystyle##$\hfil\cr>\cr\sim\cr}}}
{\vcenter{\offinterlineskip\halign{\hfil$\textstyle##$\hfil\cr
>\cr\sim\cr}}}
{\vcenter{\offinterlineskip\halign{\hfil$\scriptstyle##$\hfil\cr
>\cr\sim\cr}}}
{\vcenter{\offinterlineskip\halign{\hfil$\scriptscriptstyle##$\hfil\cr
>\cr\sim\cr}}}}}

\newcommand{\bc}{\begin{center}}
\newcommand{\ec}{\end{center}}

\begin{document}

\thispagestyle{empty}
 
\setcounter{page}{1}
\begin{verse}
2004, {\em Astrophysics \& Space Physics Reviews,} vol. 12, p. 1 \\
\end{verse}
\vspace*{4em}

\bc
{\bf \large THE JETS AND SUPERCRITICAL ACCRETION DISK IN SS433}\\

\vspace*{1em}
{\large Sergei Fabrika}  \\

\vspace*{1em}
{Special Astrophysical Observatory, Russia \\ fabrika@sao.ru }

\ec

\abstract{
The review describes observations and investigations of the unique object
SS433 obtained after 23 years of studying this massive binary system. The
main difference between SS433 and other known X-ray binaries is the action
of a constant supercritical regime for the accretion of gas onto the
relativistic star (most likely a black hole), which has lead to the formation
of a supercritical accretion disk and collimated relativistic jets. The
properties of the jets are, to a large extent, determined by their
interaction with the disk wind. The precession of the disk and jets, as
well as the eclipsing in the binary system, make SS433 a unique laboratory
for studies of mechanisms for the microquasar phenomenon. The review describes
the main ideas and results emerging from studies of the formation of the jets
and supercritical accretion disk in SS433. Essentially all photometric and
spectroscopic properties of SS433 are determined by the accretion disk and
its orientation, but the disk itself is not observed, being located beneath
the photosphere of the dense wind. Observational manifestations of the
wind and of gas flows in the system are described, as well as possible
properties of the material lost by the system in its equatorial plane.
Little is known about the structure of the central regions where the hot
bases of the jets are located immediately above the plane of the disk.
The available X-ray, ultraviolet, and optical observations paint a picture
in which the bases of the jets are surrounded by cocoons of hot gas
reradiating emission from the inner regions of the jet channel. Direct
investigations of this channel in the supercritical accretion disk of
SS433 are not possible; however, a similar object oriented face-on would
likely be an extremely bright X-ray source, such as those observed in
other galaxies.}


\clearpage

\thispagestyle{empty}
 
\tableofcontents
    
\thispagestyle{empty}

\newpage

\setcounter{page}{1}

\section{Introduction}

The well known and unique object SS433 was first identified in the survey
of stars exhibiting H$\alpha$ emission of Stephenson and Sanduleak (1977),
which included 455 objects in the Galactic plane. SS433 was a variable
non-thermal radio source (Feldman {\it et al.\/} 1978; Seaquest {\it et al.\/} 1978) and
variable X-ray so urce (Marshall {\it et al.\/} 1978). The first slit spectra of
this object (Ciatti {\it et al.\/} 1978; Mammano and Vittone 1978; Clark and Murdin
1978) revealed bright and variable lines whose origin was unclear. Bruce
Margon and his colleagues (Margon 1979; Margon {\it et al.\/} 1979ab) identified
these emission lines with lines of hydrogen and neutral helium shifted
by tens of thousand   of km/s to the red and the blue, with a pair of lines
for each transition. The huge observed shifts of these lines could not arise
due to Zeeman splitting (Liebert {\it et al.\/} 1979), and it was evident that the
line shifts were due to the Doppler effect associated with rapidly moving
gas. It was then elucidated that the shifted H and He\,I lines arise in two
oppositely directed jets of gas (Fabian and Rees 1979; Milgrom 1979a;
Margon {\it et al.\/} 1979c) that change their position in space in a periodic
fashion (they ``precess''), leading to ``motion'' of the lines in the
spectrum. This was the beginning of intensive studies of SS433, as a binary
system with unique properties.

The main characteristic property of SS433 that distinguishes it from other
binary systems containing relativistic objects is that,   in SS433, a
continuous (non-transient) regime of supercritical accretion of gas onto
the relativistic star is realized. In this case, a supercritical accretion
disk forms, together with narrow jets of gas that propagate with the
relativistic speed of 79\,000~km/s from the inner regions of the disk,
perpendicular to the disk plane. The second component of the system --
the donor star -- obviously fills its critical Roche lobe, providing
a powerful and approximately continuous flow of gas into the region of the
relativistic star at a rate   of $\sim 10^{-4}\,M_{\sun}$/yr. Essentially,
the reason SS433 is unique among other massive X-ray binary stars (with
black holes or neutron stars) can be revealed by determining the origin
of the very high rate of mass transfer in this system (van den Heuvel
1981; Shklovskii 1981).

It is interesting that no direct observational evidence has been found
for either an accretion disk or a ``normal'' or ``optical'' star in the
SS433 system. Nevertheless, there is no doubt that these two components
are present in SS433. This opinion is not only due to the accumulated
experience with studies of dozens of close X-ray binary systems having
neutron stars or black holes as their relativistic stellar components.
There are many indirect pieces of evidence and observational manifestations
of these two components, and all the main properties of SS433 can be well
described using modern concepts about supercritical disk accretion, first
discussed by Shakura and Sunyaev (1973).

SS433 is a close binary and a massive eclipsing system with an orbital
period of 13.1~days (Crampton {\it et al.\/} 1980; Cherepashchuk 1981). Eclipses
of both bodies are clearly observable in the optical and near infrared,
as well as eclipses of the base of the relativistic jets in the X-ray.
The source of the jets (accretion disk or object at the disk center) is
appreciably brighter than the secondary (donor) star. The accretion disk of
SS433 precesses, changing its orientation in space with a period of $P_{pr}=
162$~days, with the jets repeating this precessional motion. Essentially,
we observe in SS433 only a dense wind outflowing from an accretion disk,
and two bright regions in the central part of the disk, at the places of
exit of the relativistic jets. Observationally, the
star in SS433 is manifest only as an object that periodically eclipses
the accretion disk and gaseous flows forming the disk, reflects the
radiation of bright central regions, and perturbs the disk wind. The
precession of the accretion disk cardinally changes the photometric
properties (orbital light curve) and appreciably affects the spectral
properties of the system. Below, we will use the term ``accretion disk''
to mean not only the disk itself, which must be present, but also the
disk wind, and will use the term ``optical'' or ``normal'' star for
the secondary, in spite of the fact that very little is known about this
star.

In this review, we will describe the main properties of SS433's relativistic
jets and accretion disk~-- the machine generating the jets~-- as they
are currently (2002) understood, focusing primarily
on the available observations and their interpretation. Spectral and
photometric studies of SS433 as a binary system will also be described,
since these results are required to understand the nature of the disk and
jets. The bulk of observational data on SS433 were obtained in the first
years of investigations of this object, during the ``SS433
boom''. The main ideas and models attempting to explain the
behavior of SS433 were also proposed in these early years. To a
large extent, these ideas obtained confirmation in subsequent
observations. Therefore, previously published reviews on SS433
remain valuable. We refer the reader to these reviews, not only
for information on the history of studies of SS433~-- an object
that has played and continues to play a fundamental role in modern
astrophysics -- but also because of the importance of these reviews.
Margon (1984) summarized the results of five years of investigation of
this object, while Cherepashchuk (1989) reviewed the results of photometric
studies. Milgrom (1981), Petterson (1981), and Katz (1986) addressed
models and theoretical concepts for SS433. Reviews have also been published
by Clark (1985), Zwitter {\it et al.\/} (1989), and Vermeulen (1996). Of course,
the results of new observations, especially X-ray and radio interferometric
observations, as well as numerical simulations, have also made a fundamental
contribution to our understanding of SS433.

Here in the Introduction, we present a very brief list of the main parameters
of SS433 (many of which will be discussed in detail below), to provide a
basis so that the subsequent chapters of the review can be read independently.

\bigskip

\subsection{Parameters of SS433}

SS433 is the variable star V1343 Aquilae, located at a distance   of 5.0~kpc
roughly in the Galactic plane ($l=39.7^{\circ}, b=-2.2^{\circ}$). This is
a relatively bright red star: $V=14.0,$ $(U-B)=0.8$, $(B-V)=2.1$,
$(V-R)=2.2$ (Goranskii {\it et al.\/} 1998a). A finding chart of SS433 and photometric
data for the surrounding stars can be found in Leibowitz and
Mendelson (1982). SS433 is strongly absorbed, $A_V \approx 8^m$, and the
intrinsic luminosity of the object assuming isotropic radiation of its
emission is $L_{bol} \sim 10^{40}$~erg/s (Cherepashchuk {\it et al.\/} 1982;
Dolan {\it et al.\/} 1997). It is one of the brightest stars in the Galaxy, and has
its spectral maximum in the ultraviolet. There is an infrared excess in the
L and K bands, in which the mean brightnesses of the object   are 7\magdot0 and
8\magdot0, respectively (Giles {\it et al.\/} 1980; Kodaira {\it et al.\/} 1985). This
infrared excess is associated with free--free radiation by gas in the
immediate vicinity of the system. The X-ray luminosity of SS433 is about
$\sim 10^{36}$~erg/s (Brinkmann {\it et al.\/} 1991; Kotani {\it et al.\/} 1996; Marshall
{\it et al.\/} 2002). The 1--10~keV X-ray emission is primarily due to hot
($\sim 10^8$~K) gas of the jets located above the photosphere of the
accretion disk.

In addition to emission lines associated with both jets that shift in
accordance with the precessional and nutational periods, the optical
spectrum of SS433 shows very bright and variable ``stationary'' lines of
hydrogen, He\,I, He\,II, C\,III, and N\,III, as well as
other weaker Fe\,II emission
lines (Murdin {\it et al.\/} 1980; Crampton and Hutchings 1981a). Together with the
H\,I and He\,I emission lines, these last lines show clear P\,Cyg profiles at
certain precessional phases. All these lines are formed both in the wind
flowing from the accretion disk and in gaseous flows in the system. No lines
from the normal star have been detected (Gies {\it et al.\/} 2002a), despite numerous
attempts to do this. However the latest data show that the donor star in SS433
is an evolved A supergiant (Gies {\it et al.\/} 2002b).

The radiation of SS433 is very variable in all accessible wavelength
ranges. In addition to sporadic variability (flares), active and quiescent
states are observed. In quiescent states, optical, IR, and X-ray variability
with the orbital and precessional periods is observed. In active states,
which last from 30 to 90 days, the mean brightness of the object increases
by approximately a factor of 1.5, and powerful flares with characteristic
time scales of hours to days (Irsmambetova 1997) are observed against this
enhanced background level. In addition, in active states SS433 ``reddens'',
i.e., the gas
exchange and flow of gas from the system are increased. The active periods are
especially clearly visible in the radio, where long series of observations
are available (Bonsignori-Facondi {\it et al.\/} 1986; Fiedler {\it et al.\/} 1987).

\bigskip

\subsection{The Jets of SS433}

The most striking phenomenon associated with SS433 is its jets. Depending
on the distance from the source, or alternatively on the temperature of
the jets, radiation mechanism and, accordingly, observational methods used,
we can distinguish the X-ray jets ($\sim 10^{10-13}$~cm), optical
($\sim 10^{14-15}$~cm), and radio ($\ga 10^{15}$~cm); the extended
  X-ray jets
are also observed ($> 10^{17}$~cm). However, this division is somewhat
arbitrary; for example, radio emission is observed virtually along the
entire extent of the optical jets.

The jets are manifest in the optical spectra as ``moving'' emission lines of
hydrogen and He\,I.  The lines are shifted in the spectrum due to variations
in the inclination of the jets to the line of sight during their precession.
The jets are surprisingly narrow, and their opening angle in the region
in which the hydrogen lines are emitted (whose distance from the central
object corresponds to 1--3 days of flight)   is $1\degdot0-1\degdot5$
(Borisov and Fabrika 1987). Clouds of gas with a normal ``astrophysical''
temperature of $\sim 10\,000$~K move in the optical jets (Davidson and McCray
1980). A continuous source of heating is required to maintain the emission
of the gas in the optical jets. The X-ray jets (Marshall {\it et al.\/} 2002) are
short (only several hundreds of seconds of flight), and give rise to lines
of highly ionized heavy elements. The jet X-rays are radiated by hot gas
($T\sim 10^8$\,K), which cools as the jets propagate due to expansion and
radiative cooling. The opening angle of the X-ray jets is $\approx 1\degdot2$.
The gas in the SS433 jets moves along strictly ballistic trajectories.
The flux of kinetic energy, or kinetic luminosity, of the jets is enormous:
$L_k \sim 10^{39}$~erg/s (Panferov and Fabrika 1997; Marshall {\it et al.\/} 2002).

The well-known precessing jet pattern is observed in the radio on scales
of several arcseconds (Hjellming and Johnston 1981). The radio flux from
the synchrotron-radiating jets is about 1~Jy, and their luminosity is
$10^{30-31}$~erg/s. The jets are also clearly visible on VLBI scales
(Vermeulen {\it et al.\/} 1987), right down to the smallest available resolution
of about 2~mas (Paragi {\it et al.\/} 1999, 2000), where the effect of self-absorption
is already strong in inner regions $\sim 20$~AU in size. The jets of
SS433 excite the radio nebula W50, which closely resembles a supernova
remnant. W50 is extended in the direction of the jet precessional axis
(PA$\approx 100^{\circ}$) with SS433 at its centre, and the nebula
stretches on both sides in this orientation to 50--70~pc. The large-scale X-ray jets
extend in this same direction (Brinkmann {\it et al.\/} 1996), and end in optical
filaments (Zealey {\it et al.\/} 1980).

A ``kinematic model'' of SS433 (Abell and Margon 1979) predicts the
position of the jets in space and the positions of lines in the spectrum
very well. This is a geometric model for the jet precession, and will be
considered in more detail below. In spite of some instability in the
precessional period, this model has been fully confirmed over long time
intervals (Eikenberry {\it et al.\/} 2001). In addition to their precessional
motion, the jets undergo small amplitude so-called nutational oscillations
with a period of 6.28~days, which is half the synodic orbital period.
The nutational nodding of the jets (accretion disk) is due to periodic
tidal perturbations of the disk by the gravitational field of the
donor star (Katz {\it et al.\/} 1982) or to perturbations of the accretion
flow. The most successful precession scenario for SS433 is forced precession
of the donor star, whose rotational axis does not coincide with the orbital axis,
and the drifting or ``slaved'' accretion disk (Shakura 1972; van den Heuvel
{\it et al.\/} 1980; Whitmire and Matese 1980; Katz 1980).

Further in the review, we will describe the results of observations of the
SS433 jets, observational manifestations of the accretion disk, and SS433 as a
binary system, as well as our current understanding of the physical processes
associated with the jets and their formation and collimation. It is still
early to state that the ``mystery of SS433'' has been solved; the most
interesting works in connection with many questions, especially the formation
of the jets and the inner structure of the central object, probably still
lie ahead. However, the progress in our understanding of SS433 that has
already been attained is no less surprising than SS433 itself. This object
has exerted a huge influence on modern astrophysics, in our understanding
of critical stages in the evolution of close binary systems and of the
jets ejected from young stars, active galactic nuclei and microquasars
(Mirabel and Rodrigues 1999). Microquasars appear to be the closest
relatives   of SS433. The main, but far from only, characteristic
distinguishing SS433 from microquasars and some X-ray novae, in which
episodes of supercritical accretion are possible during flares, is the
presence of a constant, appreciably supercritical accretion regime onto
the relativistic star. SS433 remains the only stellar-mass object in which
we can directly observe an operating supercritical accretion disk, together
with the ejection and propagation of the jets. Moreover, this disk (and
also all gaseous flows in the system and wind from the disk and jets) is
constantly turning with the precessional period and is eclipsed with the
orbital period, making the system a true gift for researchers and a unique
astrophysical laboratory. The two jets of SS433 should be intrinsically
identical, but often appear very different. The changing orientation of the
jets also presents splendid opportunities for studies of the behavior of
gas moving at relativistic speeds and of the associated relativistic effects
themselves.

In all, 614 papers on SS433 were listed in the ADS server  
(http://adsabs.harvard.edu) in January 2002; references to SS433 were made
in the abstracts of an additional 199 articles (i.e., these latter papers
were concerned with objects or phenomena directly related to SS433). In
the year of its discovery, 1978, five articles and other communications
were published; the maximum publication rate for papers on SS433 was in
the following three years from 1979 to 1981, with on average 75 papers per
year. Further, the number of SS433 publications steadily decreased,
as was quite natural. The average number of papers for 1982--1989 was
27 per year, while the average for 1990--2001 was 14 per year. The intensity
of studies of SS433 again began to rise in 1996--2000, due in part to
the ability to obtain new observations in the X-ray (ASCA, ROSAT, CHANDRA)
and radio (Very Long Baseline Interferometry), as well as to the discovery
of a new related class of object~-- microquasars. The importance of
this last factor is confirmed by the number of publications in which
the name SS433 appears only in the abstract: the number of such
articles had a broad maximum in 1981--1986 (while the frequency of direct
investigations steadily decreased), but an even higher maximum   in 1996--1999.

Naturally, it is not possible to reflect the results of more than 800
publications on SS433 in this review, although a substantial majority of
them made appreciable contributions. Many gave new life to old topics
of study or stimulated subsequent investigations, and the history of some
important turning points in studies of SS433 reads almost like a detective
story. The main goal of this review is to describe our current understanding
of SS433.

\bigskip

\section{The Optical Jets}

\subsection{The Jet Spectral Lines}

The brightest optical lines radiated in the SS433 jets are
hydrogen lines, namely H$\alpha^{\pm}$, where the ``$+$'' line forms in
the receding jet and the ``$-$'' in the approaching jet. The mean
equivalent width of   the H$\alpha$ lines is several tens of \AA, and the
jet lines are strongly variable. As a rule, higher-order Balmer lines
emitted by the jets have not been studied in detail, since SS433 is
relatively weak in the blue, and it is often difficult to distinguish
the properties of individual lines in this region of the spectrum (which
has an appreciably richer line structure than the red) due to blending
of lines from the jets and numerous ``stationary'' lines. Figure~1 shows
spectra of SS433 in the blue obtained on the 6-m telescope of the Special
Astrophysical Observatory by Goranskii {\it et al.\/} (1987)  on June 1 and 2, 1986
as part of a program of coordinated observations of SS433. Only the strongest
stationary and moving lines are marked; weaker lines belong primarily to
He\,I. Among the stationary lines, only the hydrogen, He\,I and Fe\,II lines
show P\,Cyg profiles. As a rule, spectroscopy in the blue part of the
spectrum has been used for investigations of SS433 as a binary system.


\begin{figure}[t]
\centerline{\psfig{figure=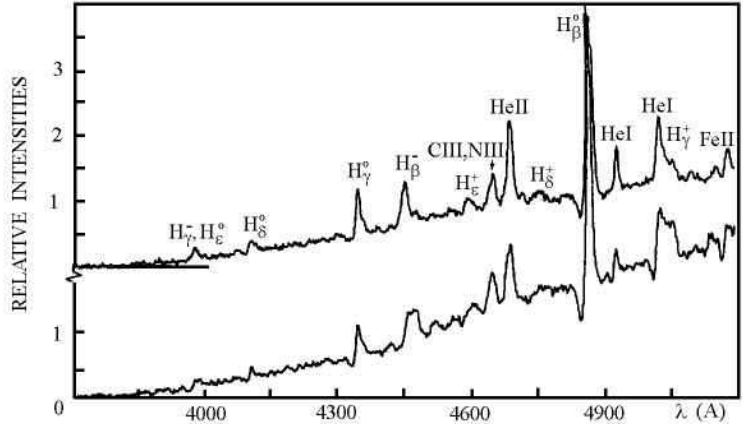,width=100mm}}
\caption{Blue spectra of SS433 obtained on June~1 (upper) and June~2
(lower), 1986. A small shift of the moving lines by 1~day can be seen,
as well as an enhancement of the absorption components of the stationary
lines.
}
\end{figure}

\begin{figure}[t]
\centerline{\psfig{figure=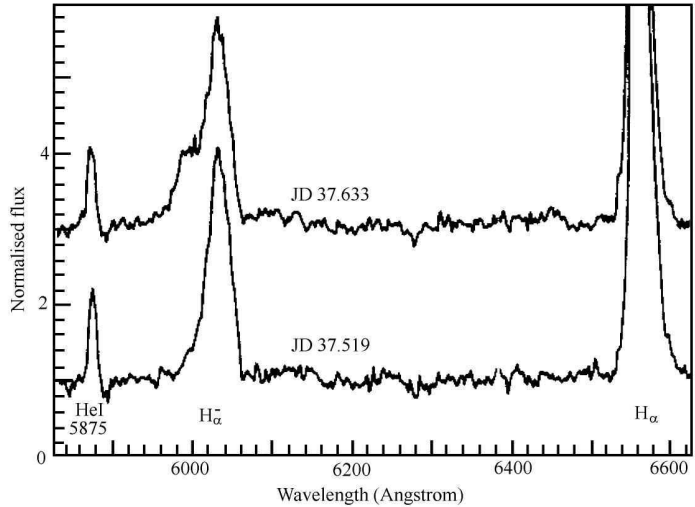,width=93mm}}
\caption{Fragments of two spectra of SS433 (Vermeulen {\it et al.\/} 1993a) 
that show rapid variations in the H$\alpha$ profile of the approaching jet.
}
\end{figure}

\begin{figure}[t]
\centerline{\psfig{figure=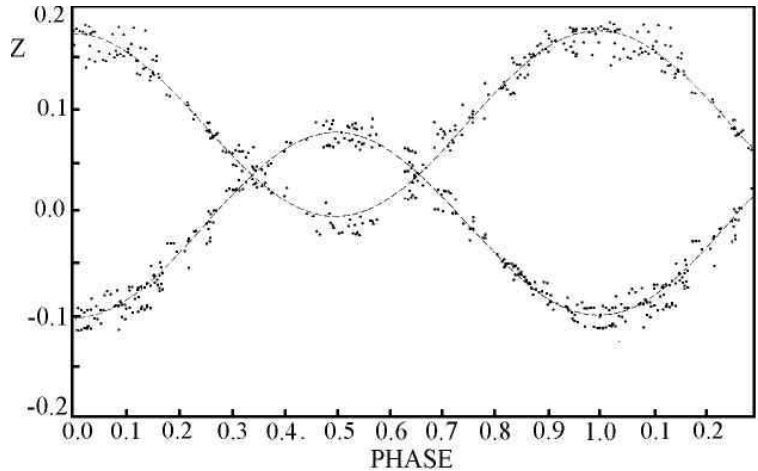,width=100mm}}
\caption{Precessional curves of the radial velocities of the shifted
lines from the  approaching (lower curve) and receding (upper curve) jets,
derived from spectroscopic data obtained during the first two years in
which SS433 was studied (Ciatti {\it et al.\/} 1981). The scatter of the data
about the curves is due primarily to the nutational variability.
}
\end{figure}

The moving H$\alpha^{\pm}$ lines are about an order of magnitude less
intense than the stationary H$\alpha$ line. The prominent moving lines
include He\,I lines from the strongest transitions; the He\,I$^\pm$ lines are
about an order of magnitude weaker than the H$\alpha^\pm$ lines, suggesting
an absence of strong chemical anomalies in the SS433 gas.   No
He\,II\,$\lambda4686$ emission has been detected from the jets, although this
is most likely due to the limited signal/noise ratio of the spectra
(Vermeulen {\it et al.\/} 1993a). According to our estimates based on spectroscopic
observations of SS433 conducted on the 6-m telescope of the Special
Astrophysical Observatory, the intensity of this line in the jets does
not exceed 1~\% of the continuum intensity.

Virtually all data on
variability of the optical jets and on the geometric and kinematic
structure of the jets have been derived from observations of the
H$\alpha^{\pm}$ lines. Figure~2 presents fragments of two spectra of
SS433 containing the H$\alpha^-$ line (Vermeulen {\it et al.\/} 1993a) and the
stationary H$\alpha$ and He\,I\,$\lambda5876$ lines. The spectra obtained on the
1.2-m Calar Alto telescope on May 21, 1987, also during an observing
campaign targetting SS433, show rapid variability of the jet lines. New
volumes of gas emitting in the H$\alpha$ line appeared in the jet over
a time interval of three hours.

\bigskip

\subsection{Kinematic Model and Precession of the Jets}

Variations in the radial velocities of the jets with the precessional phase
derived from the H$\alpha^{\pm}$ lines (Ciatti {\it et al.\/} 1981) using data for the
first two years of SS433 studies are shown in Fig.~3. The mean radial-velocity
curves are shown, with the scatter of the data about these curves being due
to the nutational variability. During the precessional period, the jets
lie in the plane of the sky twice (the radial velocities of the two jets
coincide), two crossovers of the moving lines are observed, and, accordingly,
the jet lines also move away from each other twice. The time of maximum
separation of the lines to the blue and red corresponds to the minimum
inclination of the jets and the axis of the accretion disk to the line of
sight, which occurs at precessional phase $\psi=0$, also called the $T_3$
moment. The two crossovers are usually denoted times $T_1$ and
$T_2$, and their precessional phases are 0.34 and 0.66. It is obvious
that the phases for the extrema and crossovers of the radial-velocity
curves (Fig.~3) are determined only by the orientation of SS433
relative to the observer and not by any physical processes occurring in the
system. This, generally speaking, trivial fact nevertheless is sometimes
forgotten in interpretations of the complex phenomena observed
  in SS433.

During the precessional cycle, the lines of the two jets change places,
so that the jet that recedes from us during most of the precessional period
is denoted with a ``$+$'' and the opposite jet is denoted with a ``$-$''.
At the times $T_{1,2}$, the radial velocities of the jet lines coincide,
but are not equal to zero. This is due to the well known transverse Doppler
effect, or time dilation, which is clearly observed in this way only in
SS433 (among macroscopic objects). The Doppler shifts of the spectral
lines are described by the well known formula $$\lambda = \lambda_0
\gamma (1-\beta \cos \eta),$$ where $\lambda$ and $\lambda_0$ are the
shifted and laboratory wavelengths, $\eta$ is the angle between the jet
and the line of sight,
$$\gamma = \frac1{\sqrt{1-\beta^2}}$$
is the Lorentz factor
and the jet velocity $V_j$ is expressed in units of the speed of light
$\beta=V_j/c$. At the times of the crossovers, the jet radial velocities
are $V_r^{\pm}/c=\gamma-1$. Thus, the velocity of propagation of the
jets can be directly measured in SS433, and, consequently, the geometric
parameters of the jets, inclination of the system, and distance to the
object (from radio images of the precessing jets) can also be determined.

\renewcommand{\thefigure}{4}
\begin{figure}[t]
\centerline{\psfig{figure=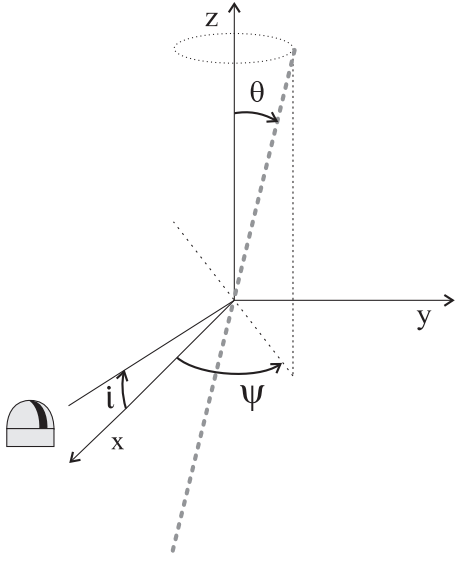,width=60mm}}
\caption{Geometrical schematic of the jet precession. The precessional
axis is $Z$ and the line of sight is in the $XOZ$ plane.
}
\end{figure}

The behavior of the moving lines of SS433 is described well by the kinematic
model for the precessing jets (Abell and Margon 1979). Figure~4 shows a
geometric schematic of the jet precession. We adopt the following notation.
The angle between the jets and the precessional axis (the precession angle)
is $\theta$, the angle between the precessional axis (orbital axis)
and the line of sight $i$, the precessional period $P_{pr}$, and the
precessional phase $\psi$. The angle between the approaching jet and the
line of sight is $\eta$, $\cos \eta = \sin i \cos \theta +
\sin \theta \cos i \cos \psi$, and this angle is minimum when $\psi = 0$.
The radial velocities of the two  jets $V_r^{\pm}/c=z^{\pm}$, or equivalently
the positions of the lines in the spectrum, can be computed using the formula
$$ 1+z^{\pm} = \gamma(1 \pm \beta \sin\theta \sin i \cos\psi
                      \pm \beta \cos\theta \cos i),   $$
where the $+$ and $-$ signs correspond to the receding and approaching jets.
This kinematic model was tested and refined after four years (Anderson
{\it et al.\/} 1983), ten years (Margon and Anderson 1989) and twenty years
(Eikenberry {\it et al.\/} 2001) of spectroscopic studies of the SS433 jets.
The analysis of Eikenberry {\it et al.\/} (2001) is based on 433 values for $z^+$
and 482 values for $z^-$. To avoid uncertainty associated with the 6.3-day
period of the jet nutation, the data were smoothed with a broader time
filter. Thus, the kinematic model can be used to study only the precessional
motion and possible long-term deviations and secular variations in the
precessional clock of SS433. The mean values of the precessional parameters
of SS433 were determined to high accuracy by Eikenberry {\it et al.\/} (2001):
$\beta=0.2647 \pm 0.0008$, $\theta=20\degdot92 \pm 0\degdot08$,
$i=78\degdot05 \pm 0\degdot05$, $P_{pr}=162\daydot375 \pm 0\daydot011$;
the date corresponding to the $T_1$ moment is $JD=2443563\daydot23 \pm
0\daydot011$. The epoch $T_3$ of the maximum separation of the lines in the
spectrum, or the precessional phase $\psi=0$, occurs at $JD=2443507\daydot47$.
These values are the result of finding the optimal five-parameter
precessional model. The actual kinematic parameters could differ slightly
from these values; for example, a simple average of the jet velocity
yields $\beta=0.254 \pm 0.0011$, which is 3\,200~km/s lower than the velocity
given by the kinematic model.

Note that we should not smooth the nutational variability to derive the
real precessional trajectory. The nutational deviations are due to tidal
perturbations of the accretion disk by the gravitational field of the donor
star (Katz {\it et al.\/} 1982; Collins and Newsom 1986). These perturbations lead
to a periodic decrease in the angle between the plane of the disk and the
plane of the orbit. Therefore, the real surface of the
precession cone passes closer to the outer extrema of the nutational
trajectory of the jets. Taking this effect into account leads to a slight
increase in the precession angle $\theta$ by an amount equal to the nutational
angle, $\approx 3^{\circ}$.

The precessional period has been stable over a long time interval of about
20 years (Eikenberry {\it et al.\/} 2001), $\dot P_{pr} < 5 \cdot 10^{-5}$, in
spite of numerous reports of variations of this period in the first few years
of studies of the object (Anderson {\it et al.\/} 1983). This behavior is associated
with real instability of the precessional cycle on time scales of weeks to
months, which cannot be distinguished at any precessional phases. This
instability leads to the appearance of real (but random) variational
trends of the period over times of several hundred days. Instability in
the precession has also been detected in optical photometric measurements
(Goranskii {\it et al.\/} 1998b). The mean brightness of the system varies with the
precessional phase by roughly 0\magdot5 (Kemp {\it et al.\/} 1986; Gladyshev
{\it et al.\/} 1987); SS433 becomes brighter at the time ${T_3}$, when the disk is
maximally turned toward the observer (the angle between the disk axis and
the line of sight is 57$^\circ$). Optical photometry is used as an
independent method for studying the precessional clock.

In a model of driven precession (for example, of the precession of the
rotational axis of the normal star), the precessional and orbital periods
are directly related. Variations in the orbital period   of SS433 based
on the times of eclipses (Fabrika {\it et al.\/} 1990; Goranskii {\it et al.\/} 1998b)
in O--C diagrams also show instability of about the same amplitude and
at about the same times as the instability in the precessional period.
Small-scale instability in the precessional and orbital clocks could
be associated with variations in the rate of mass transfer between the components
in the active and passive states   of SS433. Like the precessional period, the
orbital period is also stable over long time scales (Goranskii {\it et al.\/} 1998b).

The instability of the precessional cycle (Margon and Anderson 1989;
Baykal {\it et al.\/} 1993; Eikenberry {\it et al.\/} 2001) resembles random deviations
of the precessional phase from the computed ephemerides with amplitudes
to $\Delta \psi \approx 0.1$ (7--15 days), and is probably associated with
both real variations in the phase and variations in the inclination
and speed of the jets. None of the parameters $\theta, \beta$ or
$P_{pr}$ by itself can explain the observed precessional
``noise''. The statistical behavior of the deviations is well
described as white noise in frequency, or as a random walk of
the phase of the precessional period (Baykal {\it et al.\/}
1993). The precessional noise of SS433 is quantitatively similar
to the noise in the 35-day (precessional) period of the X-ray
source Her\,X-1.

Almost no observations of the moving lines have been made over
the past ten years (at least the results have not been published); however,
investigating the origin of these instabilities requires series of spectral
monitoring observations. For example, it would be useful to compare the
times when the instabilities appear with periods of activity of SS433.
In their analysis of the periodicity in the anticorrelated shifts of the
lines of both jets in the spectrum, Frasca   {\it et al.\/}
(1984) detected about ten harmonics, including harmonics
corresponding to periods of 80, 155 and 1500~days. No
periodicity in the absolute velocity of motion of the jets was
found. Such analyses should be continued and refined using
additional data.

\bigskip

\subsection{How the Moving Lines Move}

The SS433 jets are strictly antisymmetrical, and, as a rule, the profiles
of lines radiating in the opposite jets show mirror symmetry. The arrival
times for the signals from the two jets should differ only slightly.
The region of maximum brightness in the hydrogen lines is separated
from the central source by approximately one day of flight of the
gas in the jets. Even when the inclination of the jets to the line of sight is
maximum (at the time $T_3$), the time delay for the radiation of the
receding jet is about 0.2--0.25~day. Therefore, rapid reconstruction
of the jet structure such as that shown in Fig.~2, or even more rapid
line profile variations (on time scales of less than an hour; Kopylov {\it et al.\/}
1986), are often observed only in one (as a rule, the closer) jet. However,
overall, the differing arrival times of the signals in no way disrupts
the observed symmetry of the jets.

\renewcommand{\thefigure}{5}
\begin{figure}[t]
\centerline{\psfig{figure=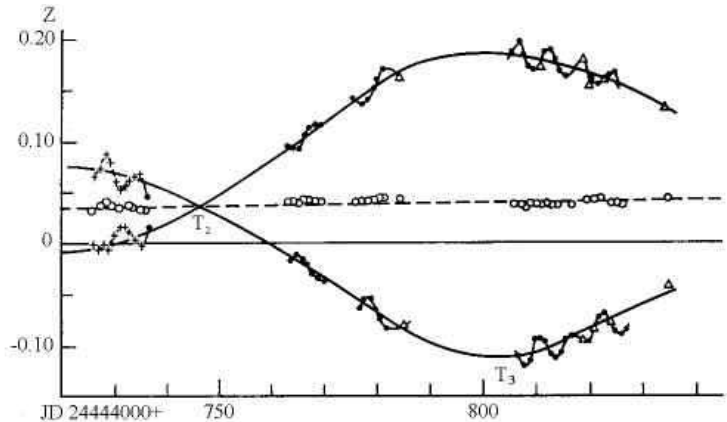,width=100mm}}
\caption{Fragments of the radial-velocity curves for moving lines according
to the data of Kopylov {\it et al.\/} (1987), constructed using the strongest
components of the H$\alpha^{\pm}$ lines. The nutational behaviour of the
jets is completely antisymmetric.
}
\end{figure}

In addition to the regular nutational motion of the two jets   (Fig.~5),
short-term (time scales of several days) disruption or jitter of the jets
is sometimes observed. The amplitude of the jitter reaches 3\,000--5\,000~km/s,
which is equivalent to variations in the jet inclination by
$2\degdot5-4^{\circ}$. The jitter amplitude is comparable to (or slightly
higher than) the amplitude of the nutational motion. The origin of the
jitter is probably closely related to the nature of the nutational shifts
of the flows of accretion gas, as well as to the conditions for the development
of a certain disk inclination (or, more precisely, to the disruption of
these conditions), and the time for the passage of matter through the disk.
For example, in a
slaved-disk model, the instantaneous inclination of the disk depends on
the inclination of the star, the orbital phase (periodic gravitational
moment perturbing the disk), specific geometry for heating of the star
by the bright source (shadowing of part of the stellar surface by the
edge of the disk or by clouds of gas), and the accretion rate (state of
activity).

The jet lines sometimes ``disappear'' for up to several days,
after which they appear in the expected position, in accordance with
the ephemerides (Kopylov {\it et al.\/} 1985; Vermeulen {\it et al.\/} 1993a). It is not
ruled out that such ``switching off'' of the engine is somehow related
to active periods; in both papers cited above, the switching off of the
jet emission coincided with powerful photometric flares of the object.
It remains completely unclear whether these disappearances are associated
with the cessation of jet activity, or possibly with the disruption of
the jet-collimation mechanism and the onset of thermal instabilities
that result in the formation of clouds of cool gas in the jets. A detailed
analysis of the times of disappearance of the jets could shed light on
the mechanism for collimation and acceleration of the jets.

The answer to the question of ``how the moving lines move'' (Grandi and
Stone 1982) is now well understood. The jet gas travels along strictly
ballistic trajectories (straight lines), along which it was ejected from
the source. The source -- central region of the accretion disk --
participates in the continuous precessional and nutational motion of
the system. The ejection of the gas in the jets is modulated, and occurs
in clumps, with on average one to three clumps ejected per day. These
clumps are sometimes called ``bullets'', and appear in the spectra
(the ``young'' jet) as fragments of line profiles or individual lines
that have the invariable position in the spectrum. The radiation of these
bullets is already appreciably weaker after a day, and they remain as
weakening ``remnants'' (the ``old'' jet) in the spectrum for up to four,
or sometimes even six, days after their appearance. The more rapidly the
lines are shifted through the spectrum (i.e., the more rapidly the angle
between the jets and the line of sight changes), the less energy is
accumulated at the given wavelength. Therefore, the moving lines and
their numerous components are most clearly visible at the phases of
extrema of the precessional and nutational periods, when the inclination
of the jets to the line of sight varies slowly. In terms of the
precessional motion, these are phases 0.0 (the time $T_3$) and 0.5.

The angular velocity of the nutational motion is fairly high, so that a line
will be spread over virtually the entire spectrum between the extrema of
the nutational curve, and, on the contrary, will be appreciably strengthened
at the extrema. This is a geometrical projection effect (Borisov and
Fabrika 1987). Thus, the profiles of the moving lines depend on the
nutational and precessional phases. As a rule, there is one bright
component (FWHM\,$=1\,000-1\,500$~km/s) formed by the effect of projection,
as well as several weaker secondary components within an interval of
several thousands of km/s. The secondary components are remnants
(``tracks'') of either the previous bright component or the largest
bullets.

Many interesting data on the SS433 jets were obtained in coordinated
observations in May/June 1987 (Vermeulen {\it et al.\/} 1993a), during which
about 200 spectra were obtained over 20 days at various observatories
around the world. In addition, the campaign included radio interferometric
observations of the jets (Vermeulen {\it et al.\/} 1993b), radio monitoring
(Vermeulen {\it et al.\/}~1993c), optical photometry (Aslanov {\it et al.\/}
1993) and X-ray observations (Kawai {\it et al.\/} 1989).
Figure~6 shows the result of spectral observations of the moving
H$\alpha^{\pm}$ lines. The non-stationary injection of new
``bullets'' of material into the jets against the background of
the regular precessional and nutational motions is clearly
visible.

\renewcommand{\thefigure}{6}
\begin{figure}[p]
\vspace*{5mm}
\centerline{\psfig{figure=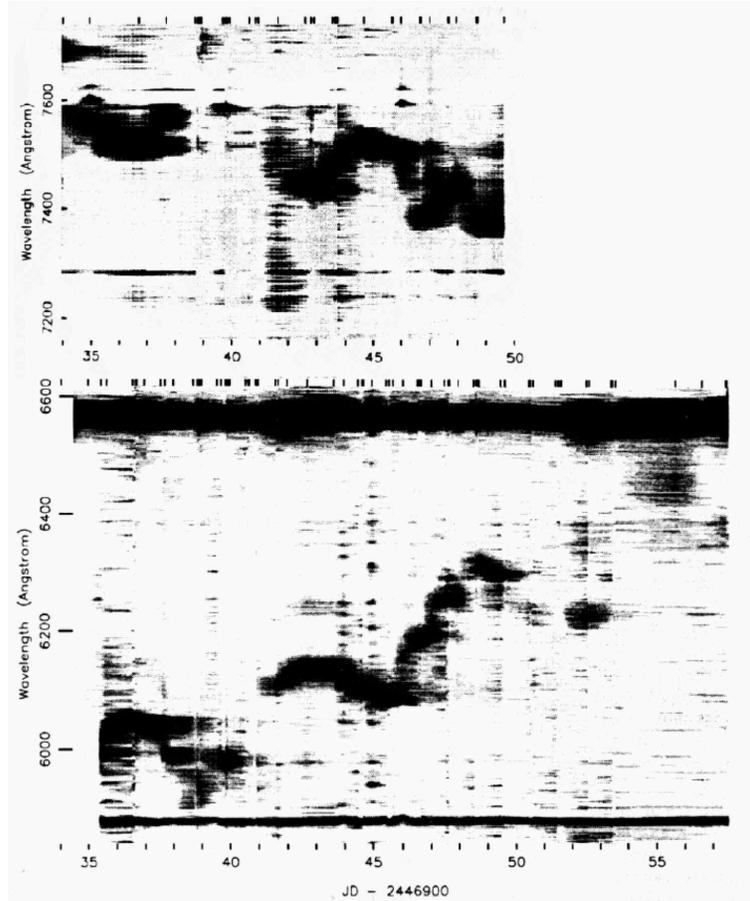,width=100mm}}
\caption{Grey-scale representation of the behaviour of the moving
H$\alpha^+$ (upper) and H$\alpha^-$ (lower) lines based on the coordinated
observations reported by Vermeulen {\it et al.\/} (1993a). The vertical axes
plot the wavelength and the horizontal axes the observation epochs.
The darkest places correspond to the brightest parts of the line profiles.
The stationary H$\alpha$ and He\,I\,$\lambda 5876$ lines (lower) and
He\,I\,$\lambda 7281$ line (upper) are visible as horizontal bands. The
vertical bars in the upper parts of both panels mark the positions of
individual observations.
}
\end{figure}

\bigskip

\subsection{Geometric and Kinematic Parameters of the Jets}

Although the bullets appear in the jets relatively suddenly, over a few
hours, the weakening of their emission lasts over several
days,   and can be studied in detail. Kopylov {\it et al.\/} (1987) and
Vermeulen {\it et al.\/} (1993a)
present light curves for individual clumps of material. Borisov and Fabrika
(1987) used the data of Kopylov {\it et al.\/} (1987) to derive the brightness
profile along the jets in H$\alpha$ emission at the phase of weakening of
the emission ($R \ge R_m$):
$$F(H\alpha) \propto \exp(-(R-R_m)/R_f),$$
where the maximum radiation occurs in the jet at a distance of   $R_m
\approx 4 \cdot 10^{14}$~cm from the source (0.6 days of flight), and
the characteristic scale for the decay of the radiation is $ R_f =
(6.7 \pm 0.5)\cdot 10^{14}$~cm ($1.0 \pm 0.07$ days of flight). This
behavior is obeyed for variations of $F(H\alpha)$ by more than 1.5 orders
of magnitude, and the radiation of the remnants of the moving lines can
be reliably traced for four days.

Modeling of the profiles of the moving H$\alpha^-$ line (see also
Panferov and Fabrika 1993) showed that, at the phase of ignition of
the emission, i.e., when $R < R_m$, the brightness profile of the jet
can be described by the relation $F(H\alpha) \propto \exp((R-R_m)/R_{in})$,
where $ R_{in} \le 1.7 \cdot 10^{14}$~cm (0.25 days of flight). New clumps
of gas at the base of the jets cool and begin to intensely radiate
hydrogen line emission very rapidly, after only a few hours. Vermeulen
{\it et al.\/} (1993a) found that the total ignition time for new clumps of gas
in the jets was 6--10 hours. Thus, the brightness profile of the SS433 jets
has been firmly established.

The geometrical and kinematical parameters of the jets were determined by
Borisov and Fabrika (1987) based on modeling of the profile of the moving
H$\alpha^-$ line (Fig.~7). Their model jet underwent precessional and
nutational motion and was filled with clouds of gas at its base, which
were radially distributed across the jet in accordance with a normal
distribution with standard deviation $\theta_j/2$. The gas moved along
ballistic trajectories at constant speed. Kopylov {\it et al.\/} (1986) suspected
that the gas in the jets was decelerated by $\Delta V_j/V_j \le 10^{-2}$,
corresponding to a shift of or no more   than 10 \AA\ over several days of
flight. This effect is weak, and even if it does exist, it does not
significantly affect the structure of the computed line profiles. The
opening angle of the jets was found to be $1\degdot0 < \theta_j < 1\degdot4$,
with both the natural opening angle of the   jet $\theta_j$ and the
parameters of the nutational trajectory making significant contributions
to the observed width of the moving line. The nutation angle is $\theta_n
= 2\degdot8 \pm 0\degdot3$. The typical structure of the line profiles
suggests that the number of clouds (gas clumps) in the jet is
$\sim 10^3 - 10^4$, or that the time to form one such cloud is $\sim 100$~s.
However, the rate at which gas enters the jet is also variable on time scales
of about $0\daydot3 - 0\daydot5$ (Borisov and Fabrika 1987; Vermeulen
  {\it et al.\/} 1993a). This sporadic activity (bullets)
creates large-scale structure in the line profiles.

\renewcommand{\thefigure}{7}
\begin{figure}[p]
\centerline{\psfig{figure=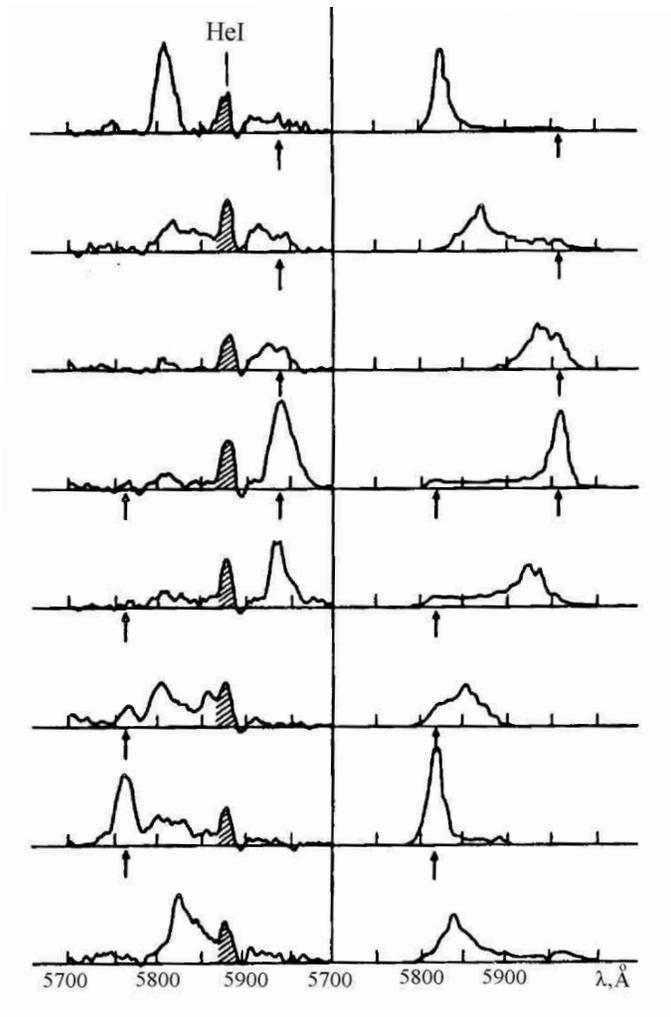,width=90mm,angle=-0.9}}
\caption{Observed (left) and modelled (right) profiles of the H$\alpha^-$
moving line, with the date increasing upward. The observational data were
obtained on June 21--28, 1981 by Kopylov {\it et al.\/} (1986, 1987). The stationary
He\,I\,$\lambda5876$ line is shaded. The arrows mark evolving
components at wavelengths corresponding to extrema of the
nutational radial-velocity curve.
}
\end{figure}

Below, when describing the physical state of the gas in the optical jets,
we discuss evidence that the gas in the jets is located in still smaller
clumps (in clouds or cloudlets) with sizes $l \sim 10^8$~cm, which form
as a result of thermal instabilities as the gas cools. Thus, we can speak
of a hierarchical structure for the jets: (i) small cloudlets, (ii)
larger clouds that form on time scales of $\sim 100$~s, (iii) large-scale
clumps of gas that form on time scales of $0\daydot3 - 0\daydot5$. As a rule,
the term ``bullets'' adopted in early studies of SS433 referred to bright
emission components of lines, which formed due to the effect of projection.
In the light of our current more detailed understanding of the jet
structure, this term may appear somewhat inappropriate, and we ascribe it
here to these last large-scale inhomogeneities. The time scale of $\sim 100$~s
coincides with the time that the jet propagates inside the funnel of the
accretion disk. This can be considered evidence that the jets are collimated there,
and that the observed inhomogeneities form due to thermal or hydrodynamical
instabilities during the time that the jets move inside the funnel. The
time scale of about   0.3--0.5~days could be comparable to the characteristic
time scale for instabilities in the outer parts of the accretion disk
associated with the formation of spiral shock waves, i.e., with processes
that modulate the rate of gas transfer in the central part of the accretion
disk.

\bigskip

\section{The Radio Jets and W50}

\subsection{The Uniqueness of the SS433 Radio Jets}

SS433 is a very bright radio star, whose central radio source radiates
at a level of about 1~Jy at centimeter wavelengths. Virtually all of the
emission of SS433 is non-thermal synchrotron radiation from the jets. The
structure of the precessing jets was directly visible in the first maps
obtained with the VLA (Hjellming and Johnston 1981). In spite of the fact
that the overall dimensions of the radio jets are a factor of a hundred
larger than the H$\alpha$ jets, the phase of the radio-jet precession
(orientation) is in good agreement with the kinematic model. Clearly, the
gas in which the radiating electrons are generated moves along the same
ballistic trajectories as the H$\alpha$ clouds, and is directly related
to them. This makes it possible to measure the distance to SS433 (5.0~kpc)
with an accuracy that is unprecedented in astronomy, about 5--10~$\%$. The
maximum of the radio emission occurs at a distance of $\sim 10^{15}$~cm
from the source (Hjellming and Johnston 1981; Romney {\it et al.\/} 1987; Vermeulen
{\it et al.\/} 1993b), from the same place where the maximum optical jet line
emission originates. The brightness of the radio jets gradually decreases
to a distance of $\sim 10^{17}$~cm, beyond which the jets are not visible
until a distance of $\sim 10^{20}$~cm. Here, the jets are decelerated and
the large-scale X-ray jets are observed together with an increased intensity
of the radio emission (Brinkmann {\it et al.\/} 1996). This forms the so-called
``ears'' of the W50 radio nebula, which contain regions of X-ray and
optical emission.

Figure~8 shows radio images of SS433 on various angular scales taken
from  Paragi {\it et al.\/} (2000). The figure shows (a) the W50 radio
nebula depicted as a mosaic of VLA images at 1.4~GHz (Dubner {\it et al.\/}
1998), (b) an image of the precessing jets on scales $\sim 10^{17}$~cm
obtained with the MERLIN interferometer and a global VLBI array (EVN+VLBA+Y1)
at 1.6~GHz, and (c) images of the inner jets obtained using the EVN,
VLBA, MERLIN, and the VLA at 1.6~GHz. In this last panel, both a region
of brightening of the radio emission at a distance of about 50~mas
along the jets and a weak radio structure perpendicular to the jets are
visible (see below).

\renewcommand{\thefigure}{8}
\begin{figure}[p]
\centerline{\psfig{figure=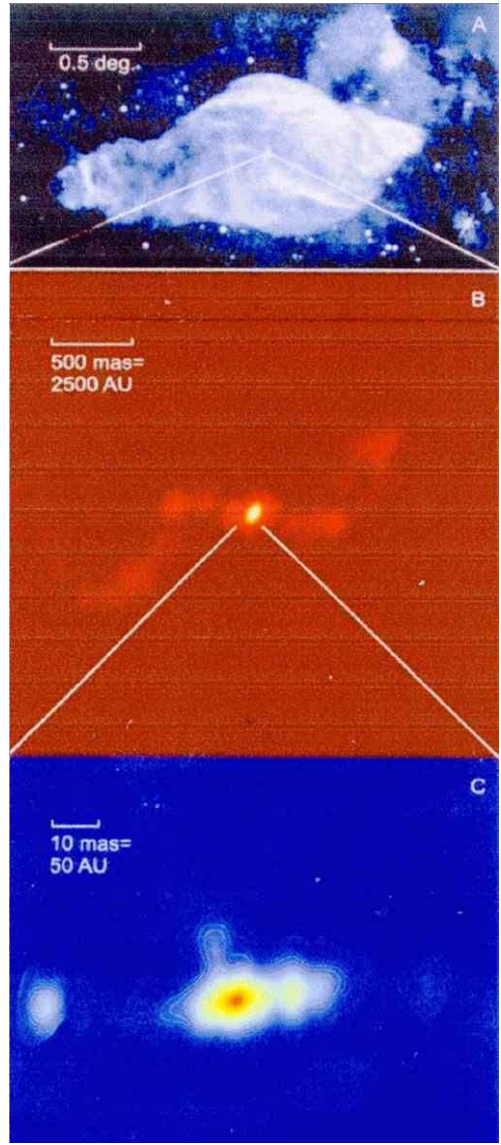,width=65mm}}
\caption{Radio images of SS433 on various angular scales (Paragi {\it et al.\/}
2000). (a) The W50 nebulosity (Dubner {\it et al.\/} 1998) at 1.4~GHz, (b) the
precessing radio jets,
(c) the inner jets, where the radio brightening zone about
50~mas along the jet and the weak radio emission approximately
perpendicular to the jets are both visible.
}
\end{figure}

The radio spectrum of SS433 (Seaquist {\it et al.\/} 1980) is a
typical synchrotron spectrum, with a spectral index $\alpha
\approx -0.6$ ($S_{\nu} \propto \nu^{\alpha}$) at frequencies of
0.3--22.5~GHz. A turnover or flattening of the spectrum is
observed at $\nu \sim 0.3$~GHz, which could be due to
synchrotron self absorption. However, it is no less likely that
it is the result of free--free absorption of the radio emission
by gas of the accretion-disk wind.

The radio emission is radiated by relativistic electrons that
travel with the jets and are continuously generated in them. The
jets of SS433 are ``heavy'', being composed of a $e^-p^+$
plasma, and propagate at a relatively low speed (compared to
known microquasars), $V_j = 0.26c$.  It is of fundamental
importance here that the SS433 jets are made up of dense clouds
of gas that are capable of propagating to appreciable distances
without appreciable deceleration. Therefore, these jets can be
observed spectrally in the optical and X-ray, and in images in
the radio and X-ray.  This is what determines the uniqueness of
the SS433 jets; more precisely, SS433 represents an evident and
undoubted example of heavy jets. The relativistic particles are
most likely accelerated in shock waves during interactions
between the jets and gas flowing from the accretion disk. It is
interesting that such interactions between the jets and the slow
disk wind must also be invoked to explain the longevity of the
optical radiation of the jets, as well as the six-day variations
in the radio~flux~of~SS433~(see~below).

The jet radio emission is significantly linearly polarized at  
10--20~$\%$ (Seaquist 1981;  Niell {\it et al.\/} 1981); the
polarization is variable and oriented along the
``instantaneous'' direction of the jets. Circular polarization
of the radio emission has also recently been discovered (Fender
{\it et al.\/} 2000). The degree of circular polarization is
0.3--0.6~$\%$ at 1--9~GHz, and its spectrum is $V \propto
\nu^{-0.9 \pm 0.1}$. Circular polarization can in principle
arise directly as a result of the synchrotron radiation of the
relativistic electrons in the jets. The inferred magnetic field
in this case is $\sim 50$~mG (Fender {\it et al.\/} 2000), i.e.
roughly the same order of magnitude as is required to explain
the low-frequency turnover of the spectrum (Seaquist 1981).
Fender {\it et al.\/} (2000) proposed that the observed circular
polarization arises via the conversion of linear to circular
polarization during the propagation of the radiation through a
plasma with elliptical (or linear) transmission modes. In this
case, the degree of circular polarization could be as high as
10~$\%$.

\enlargethispage{5mm}

To understand the processes occurring in the jets, it is also
important that a variable linear polarization in the UV has also
been detected (Dolan {\it et al.\/} 1997). The radiation is
polarized to 10--15~$\%$ near 2800 \AA, with the orientation of
the polarization coinciding with the jet direction, as in the
radio. The origin of this polarization is not clear (see the
section ``The Supercritical Accretion Disk and the Components
from Photometric Data'' below). If it is associated with the
jets, it arises at a distance from the source no larger than the
dimensions of the binary system, most likely in the places where
the jets appear above the wind photosphere. The maximum emission
  of SS433 is observed precisely in the UV, and the source of
this radiation is the accretion disk or a region directly above
the disk that is eclipsed by the secondary component in the
system.

\medskip

\subsection{Radio Variability}

Long-term monitoring of SS433 in the radio (Johnston {\it et
al.\/} 1981, 1984; Bonsignori-Facondi {\it et al.\/} 1986;
Fiedler {\it et al.\/} 1987; Bursov and Trushkin 1995) provides
an excellent demonstration of the active and quiescent states of
the object. The low-frequency radio variations lag the
variations at higher frequencies by on average several days. The
radio jets observed with VLBI are present in both active and
quiescent periods (Romney {\it et al.\/} 1987; Fejes {\it et
al.\/} 1988; Vermeulen {\it et al.\/} 1993b). In quiescent
phases, the radio flux of SS433 shows only moderate variations,
up to 10~$\%$, but powerful, often overlapping flares are
observed during active phases. The duration of active states
ranges from 30 to 90 days, with the duration of individual
flares being from one to several days.

There is no observed dependence of the times of ``switching on''
(i.\,e. transitions to active states) of the object on the
orbital or precessional phases. Likewise, there are no detected
variations of the radio flux itself with the orbital or
precessional periods. However, the radio flux does vary with the
6-day nutational period. Johnston {\it et al.\/} (1981) detected
such variability during an active period, when the radio
emission was dominated by flares.

Band and Grindlay (1984) suggested that, in the slaved-disk
model (in which the rotational axis of the optical star is
inclined to the orbital axis), flares should occur twice per
orbital period in a reference frame rotating with the equatorial
plane of the star; i.e., with a period of 6.06~days, due to
variations of the volume of the critical Roche surface with this
period. The decrease in the effective Roche surface should occur
when the relativistic star is in the node line; i.e., in the
equatorial plane of the donor star.

In the same slaved-disk model, twice during each orbital period,
there should be a perturbation of the accretion disk due to the
gravitational torque from the donor star acting on the outer
edge of the disk. The maximum perturbation of the disk arises
when the donor is perpendicular to the node line, here it is a
line of intersection of the disk and orbital planes.  This
well-known nodding mechanism for producing the motions of the
accretion disk (Katz {\it et al.\/} 1982) provides the most
natural explanation for the nutational motions of the jets.
Both mechanisms (Band and Grindlay 1984; Katz {\it et al.\/}
1982) can in principle modulate the mass transfer or the jet
activity in SS433. We will return to these mechanisms in the
next section, in connection with our description of
  flares in SS433.

Here, it is important to note that, if the six-day variations in
the radio flux (and the optical flux; see the section ``The
Supercritical Accretion Disk and the Components from Photometric
Data'') are associated with a restructuring of the accretion
structures or with real variations in the activity of the
object, we expect the variability to have the synodic period of
6.06~days:  $$f_{6.06}=2f_{orb}+2f_{pr},$$ however, if this
variability is associated with geometrical or projection effects
(as are the nutational motions of the jets), we expect the
variability to have a period of 6.28~days:
$$f_{6.28}=2f_{orb}+f_{pr}.$$

Trushkin {\it et al.\/} (2001) detected variations with a 6-day
period during a quiescent phase of the radio emission. These
variations may be associated with the nutational nodding of the
jets and the corresponding variations in the relativistic
beaming of the radiation. Although the amplitude of the
variation of the jet inclination to the line of sight due to the
precessional rotation is appreciably higher than that due to
nutation ($\pm 20^{\circ}$ as opposed to $\pm 3^{\circ}$), no
precessional variations have been detected, probably due to the
strong sporadic variability (the active states) of SS433 on time
scales comparable to the precessional period. This is consistent
with the fact that the 6-day variability has been detected only
in relatively short datasets or during the quiescent state of
the object. However, note that the conditions for interaction of
the jets and the slow disk wind, and consequently also the
conditions for the generation of relativistic electrons, should
also vary with the 6-day period (Panferov and Fabrika 1997). The
effect of the interaction is maximum at phases of the nutational
period when the nutational and precessional shifts of the jets
add. This mechanism for the 6-day modulations operates in the
immediate vicinity of the source, at $\sim 10^{14}$~cm, since
this is the distance travelled by the wind during six days of
revolution of the jets. In the case of either variations of the
relativistic beaming or variations of the conditions for
interaction of the jets and wind, we expect variability of the
radio flux with the synodic period of 6.28~days.

Thus, the nature of the six-day variations in the radio flux in
the active and quiescent states of SS433 remains at present
unclear, although there is a good basis to hope for progress in
this area. This will require accurate measurements of the period
of the variations (6.06 or 6.28~days) in both the active and the
quiescent states separately, and a comparison of the phases of
these variations with the known ephemerides of the orbital and
nutational periods. Possible origins of the variability include
both periodic variations in the accretion-disk structure (Band
and Grindlay 1984; Katz {\it et al.\/} 1982) and purely
geometrical effects. Depending on the mechanism that is
operating, we can expect various times delays of this
variability relative to the orbital or nutational photometric
variability.

\bigskip

\subsection{Flares}

During flares, the structure of the inner radio jets can undergo
dramatic variations, and one-sided jets are sometimes observed
(Romney {\it et al.\/} 1987).  It is likely that the specific
radio structure of flares depends not only on the asymmetry of
the ejection, but also on the interaction of the jets with the
surrounding gas from the disk wind of SS433 and the absorption
of radio emission in this gas. The radio spectrum changes
appreciably during flares. As a rule, it becomes flatter, and
the low-frequency turnover is shifted to 2--3~GHz. Detailed
analyses of individual flares (Seaquist {\it et al.\/} 1982;
Band and Grindlay 1986; Vermeulen {\it et al.\/} 1993c) shows
that there are at least two types of flares. In the first type
of flare, the flux at the flare maximum is approximately the
same at all frequencies, with the maximum first being reached at
higher frequencies and then gradually shifting toward lower
frequencies. The other type of flare is more complex, and they
have a threshold frequency near 1--3~GHz, below which their
behaviour is similar to flares of the first type. Above this
frequency, the maximum emission is reached simultaneously at all
frequencies, but the flux at the maximum decreases with
increasing frequency. It is possible that the first type of
flare is encountered either during the quiescent state of the
object or at the beginning of the active state. Neither the
first nor the second type of radio flare are in agreement
(Vermeulen {\it et al.\/} 1993c) with the standard model for a
single injection of relativistic electrons, followed by
adiabatic expansion of the electron cloud (Shklovskii 1960; van
der Laan 1966). The observed kinetics of the flares requires a
continuous generation of relativistic particles. This is in good
agreement with the idea that the interaction between the jets
and wind is sharply increased   during the flares.

It is interesting that the existence of two types of optical
flares has also been discussed (Kopylov {\it et al.\/} 1985;
Goranskii {\it et al.\/} 1998a). The first type of flares are
``white'' in terms of their $UBVR$ colours and have large
amplitudes, and the second are red flares. When an active period
begins, SS433 reddens (Irsmambetova 1997; Goranskii {\it et
al.\/} 1998ab), and a more powerful circumstellar gas envelope
develops at the activity maximum.  In spite of the numerous
observations of SS433 that have been obtained, studies of the
development of the flares in the radio or optical that include
spectroscopy are insufficient to enable firm conclusions. The
flares in SS433 are the result of a perturbation in the jet
activity, and spectral monitoring of the jets will be necessary
if we wish to understand their nature. Here, we present the main
regularities in the development of flares in SS433 based on two
sets of continuous, prolonged observations: the optical
spectroscopy and photometry of Kopylov {\it et al.\/} (1985) and
the radio monitoring and interferometry and optical spectroscopy
and photometry of Vermeulen {\it et al.\/} (1993abc). Some of
these regularities may not be confirmed by future observations,
but they are clearly visible in these data, and can be extremely
useful for improving our understanding of the flare mechanism.
In the quiescent state of the object, on the day on which an
optical flare (of the first type) develops or slightly earlier,
1--2 days before the flare is detected, the optical jets
``disappear''.  More precisely, the intensity of the jet lines
decreases substantially, and the positions of the lines deviate
sharply from the locations predicted by the ephemerides.  A dip
in the radio emission is also observed at this time. Further, an
optical and radio flare is observed, and the jet lines appear at
their expected locations with enhanced intensity. A new radio
``blob'' appears in the VLBI jet at the time of the radio flare.

This behavior suggests scenarios in which the initial flare
occurs due to a perturbation (for some reason) of the
inclination of the jets and the resulting conflict between the
jets and the disk wind in the immediate vicinity of the source.
It is possible that the interaction between the jets and wind
occurs directly in the funnel of the supercritical disk, if the
jets deviate by an angle that is larger than half the funnel
opening angle. The amplitude of the nutational nodding of the
jets is $\approx 3^{\circ}$, the amplitude of the jet jitter is
$2\degdot5 - 4^{\circ}$, and the amplitude of jet deviations
that could lead to flares is probably higher still. This flare
mechanism~-- large deviations of the jets and their interaction
with the walls of the funnel~-- places fairly rigid constraints
on the time scale for variations in the jet inclination. If the
size of the wind photosphere (see below) is $\sim 10^{12}$~cm
and the wind speed is $\sim 10^{8}$~cm/s, the funnel will be
refreshed over a time scale of only several hours. After the
initial flare, there could be perturbations of the outer parts
of the accretion disk, the atmosphere of the donor star and the
surrounding gas, leading to an active state of the object
accompanied by multiple flares.

The mass-exchange rate in SS433 is appreciably supercritical,
$\dot M \sim 10^{-4}\,M_{\sun}$/yr (Shklovskii 1981; van den
Heuvel 1981), so that flares are most likely not associated with
changes in the accretion rate, but instead with perturbations of
the accretion disk accompanied by perturbations in the jet
inclination. Possible mechanisms for such perturbations are
described above (Band and Grindlay 1984; Katz {\it et al.\/}
1982). In addition, it is possible that SS433 has a small
orbital eccentricity.  For example, purely due to the effect of
non-circularity of the orbit, the variations in the volume of
the critical Roche lobe can reach 2~$\%$ once per each orbital
period (Band and Grindlay 1984), even if the eccentricity is as
small as $e \approx 0.01$. When the components pass through
periastron and the node line of the equator of the star (Band
and Grindlay 1984) simultaneously coincides with the apsidal
line (or, alternatively, the node line of the accretion disk is
perpendicular to the apsidal line; Katz {\it et al.\/} 1982),
rather strong perturbations of the Roche lobe (or accretion
disk) are possible. Analysis of the intervals between eclipses
of the disk and star (Min\,I and Min\,II) shows that the orbit
of SS433 is close to circular, $e < 0.05$ (Fabrika {\it et
al.\/} 1990). It was recently found that the brightest optical
flares in SS433 occur in a distinct and very narrow interval of
orbital phases (Fabrika and Irsmambetova 2002).  This can be
taken as evidence of a modest eccentricity in the SS433 orbit,
since a non-circular orbit provides the only explanation for
flares that occur in distinct phases~of~the~orbital~period.

It follows directly from the observations that the region in
which the radio flares arise is $\sim 100$~AU, or 20~mas, from
the source (Vermeulen 1993bc). The time delay of the radio
flares relative to the optical flares is from several hours to
several days. It is likely that different types of flares
manifest the connection between optical and radio activity in
different ways.  Accordingly, not all radio flares are
accompanied by an optical flare.  There are relatively few data
on X-ray flares (Grindlay {\it et al.\/} 1984; Kotani {\it et
al.\/} 2002), since prolonged monitoring is required to analyse
the flares.  The X-ray emission of SS433 originates primarily in
the jets and the surrounding gas, in the immediate vicinity of
the source, at a distance of $\sim 10^{11 - 13}$~cm. We will
describe the X-ray data and parameters of the X-ray jets in more
detail in the next section.

\bigskip

\subsection{Radio Brightening Zone}

There is a radio ``brightening zone'' in the inner radio jets of
SS433, observed in VLBI images at distances of $\approx 50$~mas
from the central object (Romney {\it et al.\/} 1987; Vermeulen
{\it et al.\/} 1987, 1993b). After the appearance of new radio
blobs at the center, their strength appreciably weakens as they
move outward, but grows again as they pass through this
brightening zone, sometimes to values even exceeding the initial
flux density. The radio blobs rapidly fade beyond the
brightening zone. Figure~9 presents VLBI maps of SS433 obtained
at 4.99~GHz using the European VLBI Network (EVN) at intervals
of two days beginning on May 17, 1985 (Vermeulen {\it et al.\/}
1987). The dots mark two-day intervals in the trajectories of
the propagating jets. On average, the pattern is very
symmetrical, however deviations from symmetry in the two jets
and deviations from the trajectory corresponding to the
kinematic model are significant.

\renewcommand{\thefigure}{9}
\begin{figure}[p]
\centerline{\psfig{figure=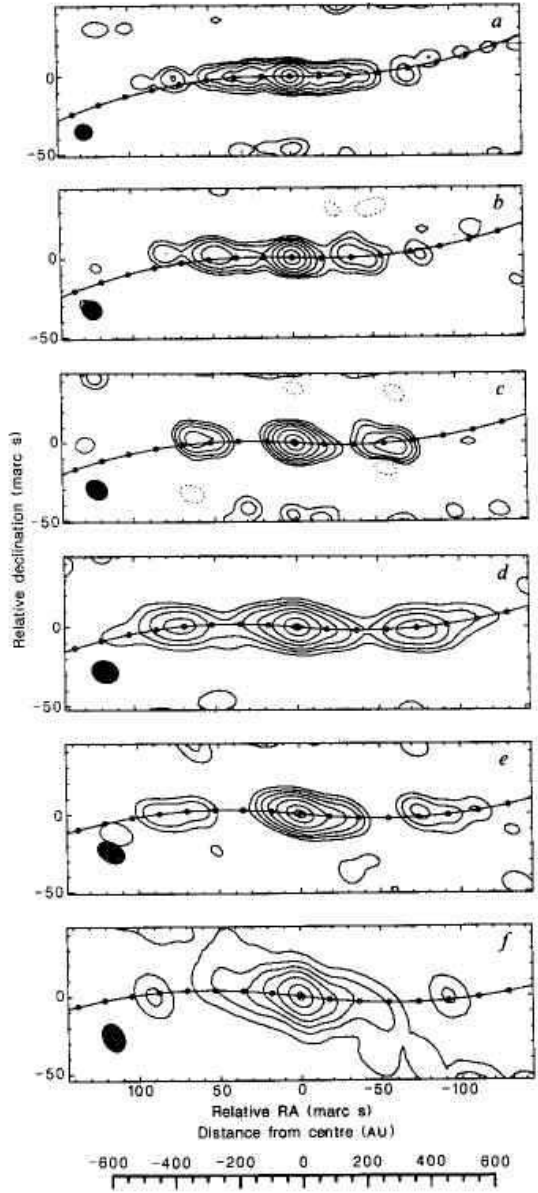,width=75mm}}
\caption{VLBI maps of SS433 (Vermeulen {\it et al.\/} 1987) obtained with an interval
of 2~days beginning on May 17, 1985. The shaded ellipse corresponds to the
angular resolution of the observations. The curves on each image show the
trajectory of the jets according to the kinematic model, and the points
mark two-day intervals on these trajectories for the propagating jets. The
radio brightening zone is located approximately 50~mas from the centre.
(Reproduced with permission from {\it Nature.\/})
}
\end{figure}

The radio emission along the VLBI jets is continuous, as is the
jet activity in the optical, and generally obeys the
regularities outlined above (weakening -- brightening --
weakening). However, the radio emission is also strongly
modulated by the individual blobs, with characteristic times for
their generation from one to several days. The modulation of the
brightness of the VLBI jets seems appreciably stronger than the
modulation of the H$\alpha$ jets. However, in the former case,
we observe modulations in the images of the jets, while in the
latter case, we have information about time variability of the
intensity of the line emission. It is quite possible that the
variability in the radio emission along the jets (in passive
states of the object) is associated not with variability in the
mass-loss rate in the jets, but with appreciable strengthening
of the particle-generation process at certain phases of the
nutational period; when the nutational and precessional motions
add, for example. A similar effect for the optical jets has been
well studied (see the previous section), but it is a projection
effect, when the line emission adds up at a single radial
velocity at phases of extrema in the nutation period, creating
bright emission lines in the spectrum, given the name
``bullets'' in early studies. The optical line emission along
the jets is essentially continuous, with the exception of
modulations with a characteristic time scale of $0\daydot3 -
0\daydot5$. Accordingly, the mass-loss rate in the jets is
likewise nearly continuous.

The ``active'' section of the VLBI jets begins about $1.5 \cdot
10^{14}$~cm   or 2~mas from the center (Paragi {\it et al.\/}
1999) and ends just beyond the brightening zone at $\approx 4
\cdot 10^{15}$~cm. The active section of the H$\alpha$ jets
begins $1.7 \cdot 10^{14}$~cm from the center (the maximum
emission arises at $4 \cdot 10^{14}$~cm; Borisov and Fabrika
1987) and ends just before the radio brightening zone. We can
see that the same region in the jets gives rise to the most
intense emission in both the radio and in the optical lines. It
is likely that a single mechanism acts in this region
(interaction of the jets with the slow wind), leading to the
generation of synchrotron electrons and maintaining the clouds
of relatively cool gas that radiate in the optical lines. This
mechanism will be discussed in detail in the section ``Structure
and formation of the jets'', and we describe it here only
qualitatively.

The radio brightening zone is located at a distance of about  
$3.7 \cdot 10^{15}$~cm, which corresponds to 5.6~days of flight.
Vermeulen {\it et al.\/} (1987) proposed that the interaction
between the jets and the gas of the optical star's wind ceases
at this distance from the center of SS433.  The jets sweep out
the gas of the slow wind on the surface of the precession cone,
and the wind refills sections of the precessional cone after the
passage of the jets. After the 164 days of the full precessional
revolution, the jets pass through new gas. The jets move through
new gas for only a few days, after which they enter a region
that is free of wind, not having had time to be filled, where
the jet gas clouds can expand freely and the radio emission can
become stronger.  This same scenario was discussed by Davidson
and McCray (1980) as an explanation for the length of the
optical jets. However, it follows from the observations that the
optical jet emission ends at a distance of $3 \cdot 10^{15}$~cm,
which corresponds to a radiation time of $\Delta t_j =
4.5$~days, one day less than the time of flight to the
brightening zone. In addition, the optical star of SS433
overfills its critical Roche surface, and is unlikely to possess
a powerful, isotropic wind.  It has been proposed (Panferov and
Fabrika 1997; Panferov 1999) that the jets interact with the
wind from the accretion disk, and that the wind speed in
near-polar regions of the disk in this scenario should be $V_w =
(\Delta t_j/P_{pr}) V_j \approx 2\,000$~km/s. Beyond the zone of
interaction between the jets and wind, there is a zone in which
the H$\alpha$ clouds expand (the zone of expansion  whose size
comprises one day of flight of the jets). The expansion of the
clouds leads to substantial variations in the jet structure and
an increase in the efficiency with which the relativistic
particles are generated. This mechanism is indeed able to
explain the appearance of the radio brightening zone.

\bigskip

\subsection{Equatorial Wind}

The most recent radio interferometric observations of SS433 with
high angular resolution (Paragi {\it et al.\/} 1999, 2000;
Blundell {\it et al.\/} 2001) revealed unusual structure in the
inner regions. Paragi {\it et al.\/} (1999) discovered a gap
$\sim 5$~mas ($3.7 \cdot 10^{14}$~cm) in size in the radio
emission at 1.6, 5 and 15~GHz. The size of the gap increases
toward lower frequencies, as it roughly should in a conical jet
geometry (Blandford and K\"onigl 1979; Hjellming and Johnston
1988) if the weakening of the intensity in the central region is
due to synchrotron self-absorption. The central source is
located along the line connecting the jets, but its position is
not symmetric relative to the two jets. The gap is larger on the
side of the receding (western) jet. These geometrical
properties, and also the different intensities and spectra of
the radio emission of the two inner jets, clearly indicate that
the inner radio jets are weakened by free--free absorption as
well as synchrotron self-absorption.  The size of the gap also
varies with time, and the geometry of the absorbing gas probably
depends on the precessional and orbital phases. The central gap
in the radio emission testifies to the presence of an equatorial
envelope. The absorbing gas surrounds the binary system in the
form of an inclined disk-like envelope, whose projection onto
the plane of the sky is roughly perpendicular to the direction
of the jets.

Paragi {\it et al.\/}~(1999) found clear evidence for the
existence of gas in the plane perpendicular to the jets of
SS433. Radio-emitting clouds were detected at 1.6~GHz at a
distance of 40--50 mas (200--250~AU) on both sides of the
central source, visible in Fig.~8c. Radio features in
significant disagreement with the kinematic model for the jets
(``anomalous ejections'') were also observed in SS433 earlier
(Romney {\it et al.\/} 1987; Spencer and Waggett 1984; Jowett
and Spencer 1995). However, deviations of the jets by more than
5--10$^{\circ}$ have never been observed in the optical
spectrum, in spite of the fact that the spectral observations
cover an interval tens of times longer than VLBI observations.
The equatorial regions of emission discovered by Paragi {\it et
al.\/} (1999) have a very high brightness temperature
($10^{7-8}$~K), which excludes the possibility that they are due
to thermal radio emission. These regions have different shapes
in different observing seasons (Paragi {\it et al.\/} 2000,
2002), i.e., they vary on characteristic time scales of at least
tens of days.  The available observations are insufficient to
reveal any periodicity in the variations of these regions.

In the data of Blundell {\it et al.\/} (2001, 2002), the
emission perpendicular to the jets does not form regions
separated from the source, and instead forms a smooth halo-like
structure. The radio spectra of these components are flat
($\alpha = -0.12 \pm 0.02$, $S_{\nu} \propto \nu^{\alpha}$), as
is characteristic of thermal emission.  However, the high
brightness temperature is in complete contradiction with a
thermal emission mechanism.

The speed of the equatorial wind in the immediate vicinity   of
SS433 has been derived from absorption-line radial velocities
(Fabrika {\it et al.\/} 1997a).  It depends on the angle above
the plane of the precessing disk, and varies from $\approx
100$~km/s to $\approx 1\,300$~km/s, with a mean speed of
$\approx 350$~km/s. The speed of expansion of this envelope has
also been estimated from the proper motion of clumps in   the
VLBI images (Paragi {\it et al.\/} 2002) to be $\sim
1\,200$~km/s.

The presence of powerful gas streams in SS433 flowing outward in the plane
of the binary system is confirmed by a whole series of independent
observations in the optical, X-ray, and radio. We will return to an
interpretation of the equatorial wind when describing other observations
below. We will describe the structure of these flows
in more detail when we discuss the
supercritical accretion disk of SS433. The existence of such flows follows
naturally from modern concepts about the formation of accretion disks in
binary systems in which the donor star overfills its critical Roche surface.

\bigskip

\subsection{W50}

The well known radio nebula W50 surrounds SS433 on scales of
tens of parsecs (Fig.~8a). A review of the results of studies of
this nebula is given by Margon (1984). The central position of
SS433, elongation of the nebula in the east--west direction
(PA$\,\approx 100^{\circ}$), along the axis of the jet
precession cone, and many X-ray and optical data leave no doubt
that W50 was formed (at least in this direction) as a result of
interactions between the jets and the interstellar gas. The
recent discovery of equatorial radio structure in SS433 (Paragi
{\it et al.\/} 1999), which probably formed as a result of gas
flowing from the system from the point L$_2$ beyond the
accretion disk (Fabrika 1993), suggests that W50 could be
excited by the constant activity of SS433 even in the direction
perpendicular to the jets (north--south). A dense equatorial
wind with velocity $\approx 300$~km/s would be easily able to
fill the body of W50 over $\sim 10^5$~yrs. Of course, this does
not mean that we should exclude the possibility that the
supernova explosion that gave birth to the relativistic star
SS433 played a role in   the formation of W50.

The image of W50 presented in Fig.~8a was obtained by Dubner
{\it et al.\/} (1998) on the VLA from 1465~MHz continuum
observations. It is reminiscent of a seashell. The central part
of the nebula forms a nearly ideal circle with radius
29$^\prime$ (42~pc for a distance of 5.0~kpc). This could be a
supernova remnant, however its size is not consistent with the
standard surface brightness--diameter relation for remnants
(Margon 1984). This may be associated with uncertainty in the
distance to SS433.

Based on the observed HI-line morphology and an analysis of the
interaction of W50 with the interstellar gas, Dubner {\it et
al.\/} (1998) found that the systemic radial velocity of the
nebulosity is 42~km/s, which in turn leads to a kinematic
distance of $3.0 \pm 0.2$~kpc. However, the accepted distance of
5~kpc was derived by comparing the pattern of motions in the
radio jets with the kinematic model, and there can be no doubt
about the legitimacy of the latter. Thus, the distance   to W50
derived both from the radial velocity of the nebula and from the
surface brightness--diameter relation is appreciably smaller
than the distance to SS433 derived from the known velocity of
propagation of the optical jets, $0.26 c$, which, in turn, is
measured using the transverse Doppler effect. This is a fairly
difficult problem.  We believe the solution lies in the unusual
properties of W50.

The velocity of W50 obtained from the HI lines is in reasonable
agreement with the systemic velocity of SS433 derived from the
radial-velocity curve for the He\,II\,$\lambda 4686$ line ($27
\pm 13$\,km/s, Crampton and Hutchings 1981a; $-13 \pm 12$\,km/s,
Fabrika and Bychkova 1990). The modest discrepancy between the
velocities of the nebula and the object could easily be
explained as a result of the kick given to the object and the
effect of uncompensated momentum during the supernova explosion
in the binary system. The eastern and western optical filaments
of W50 have radial velocities of 79 and 54~km/s, respectively,
and the formal mean velocity is $67 \pm 6$\,km/s (Mazeh {\it et
al.\/} 1983). However, these data are likewise in disagreement
with the long-baseline radio observations (and with the same
kinematic model), as well as with recent CHANDRA X-ray
observations of jet emission lines detected at a distance of
$\sim 10^{17}$~cm from SS433 (Migliari {\it et al.\/} 2002),
which demonstrate that the eastern jet is approaching the
observer while the western jet is receding.

The plane of the Galaxy passes nearly perpendicular to the axis
  of W50 from its western side, making the western part of the
  nebula shorter and brighter (Fig.~8a). The two sides of the
  nebula are asymmetric in many respects; for example, the
radio spectral index of the central region of W50 is $\alpha
\approx 0.5$, while the eastern and western parts of the nebula
have spectral indices of 0.8 and 0.4, respectively (Dubner {\it
et al.\/} 1998). Dubner {\it et al.\/} (1998) estimated the
total kinetic energy of expansion of the nebula to be $\sim 2
\cdot 10^{51}$~erg (for a distance of 3~kpc), which corresponds
to a total flux of kinetic energy of $3 \cdot 10^{39}$~erg/s for
a life time of $2 \cdot 10^4$~yrs for the nebula (Zealey {\it et
al.\/} 1980).

\bigskip

\subsection{The Extended Jets}

The eastern and western optical filaments of the W50 nebula (Zealey {\it et al.\/}
1980; Kirshner and Chevalier 1980; K\"onigl 1983; Mazeh {\it et al.\/}
1983) lie inside the projection of the jet precession cone at a
distance of ${\rm R_{W50} \approx 50}$~pc from SS433. The
filaments are oriented perpendicular to the jet. Extended X-ray
jets that can be traced out to the optical filaments have also
been detected, with the maximum X-ray emission ``fringing''
these filaments (Watson {\it et al.\/} 1983). The optical spectra of the
filaments show that the gas is heated by shock waves travelling
at a speed of 50--90~km/s. The line intensity   ratio
S\,[II]~$\lambda 6717/\lambda 6731$ has been used to estimate
the electron density, $n_e \approx 10^2$~cm$^{-3}$, and gas
pressure $P \approx 3\cdot 10^{-10}$~erg\,cm$^{-3}$ in the
filaments (K\"onigl 1983). Since the filaments are formed
by the sweeping up of the interstellar gas by the jets, the
pressure in the filaments should correspond to the dynamical
pressure of the jets. Based on this idea, K\"onigl (1983)
and Fabrika and Borisov (1987) estimated the mass-loss rate in
the SS433 jets, which, allowing for the real opening angle of
the jets $\theta_j \approx 1^\circ$, is $ \dot M_j \sim 5 \cdot
10^{-7} \,M_{\sun}$/yr (corresponding to a kinetic luminosity
$L_k \sim 10^{39}$ erg/s).

Knots of infrared emission were detected by the IRAS satellite
at the western wing of W50 (Band 1987), located along the
propagation axis of the western jet.  The eastern wing does not
have any appreciable infrared emission. The spectra of the knots
in the four IRAS bands (from $12\mu$ to $100\mu$) are fairly
steep. In her mapping of W50 with the ISOCAM camera of the ISO
observatory for studies of the processes decelerating the jets
and their interactions, Fuchs (2002) detected a number of knots
of emission in the 14--16$\mu$ band, some of which coincide with
regions of millimetre emission in the   CO (1--0) transition at
115~GHz. It is possible that the western SS433 jet collides with
and heats dusty regions, or it may be that the infrared emission
is synchrotron radiation.

The large-scale jets of SS433 observed in W50 represent a unique
laboratory for studies of the processes decelerating the jets
and their interaction with the interstellar gas. The diffuse
X-ray emission of~SS433 (the X-ray lobes) was first detected by
Seward {\it et al.\/} (1980). Watson {\it et al.\/} (1983)
mapped the X-ray emission in the vicinity of~SS433 using data
obtained with the Einstein Observatory. The extended X-ray lobes
or jets extend to the east and west from the source along the
precession axis of the radio jets, in full agreement with the
orientation and asymmetry of W50. The X-ray emission becomes
appreciable about 20~pc (15$^\prime$) from the source, reaches a
maximum at a distance of 50~pc (in the region of the optical
filaments) and disappears at distances of 60~pc. The diffuse
X-ray emission is much softer than the emission from the central
source. The luminosity of each jet is $\sim 6 \cdot
10^{34}$~erg/s (0.5--4.5~keV). Assuming that the X-ray emission
was thermal, Watson {\it et al.\/} (1983) estimated the total
thermal energy of the X-ray gas to be $\sim 1.2 \cdot
10^{51}$~erg, in reasonable agreement with the data of Dubner
{\it et al.\/} (1998).

One striking feature of the extended X-ray jets is that their
total opening angle is about 20$^{\circ}$ (Brinkmann {\it et
al.\/} 1996), much smaller than the opening angle for the
precession cone of the optical and inner radio jets
(40$^{\circ}$). The same geometry is shown by the outer radio
lobes of W50 (Fig.~8a), whose total opening angle is appreciably
smaller than that of the precession cone in the kinematic model.
It would be natural to expect the kinematic model to be in
agreement with the geometry of the extended structures in W50.
In addition, the surface brightness of the X-ray emission grows
with approach toward the precession axis, although, at first
glance, it seems that the X-ray jets should be hollow and have
their maximum radiation along the generating line of the
precessional cone ($\pm 20^{\circ}$).

To illustrate the extended jets, Fig.~10 presents an X-ray image
of SS433 constructed using ASCA GIS data (Kotani 1998).  This is
one of the deepest X-ray images of the vicinity of SS433. This
image was obtained by joining several images obtained with
different time exposures. The central bright source is SS433,
and the total size of the entire jet system is about one degree.

\renewcommand{\thefigure}{10}
\begin{figure}[t]
\vspace*{5mm}
\centerline{\psfig{figure=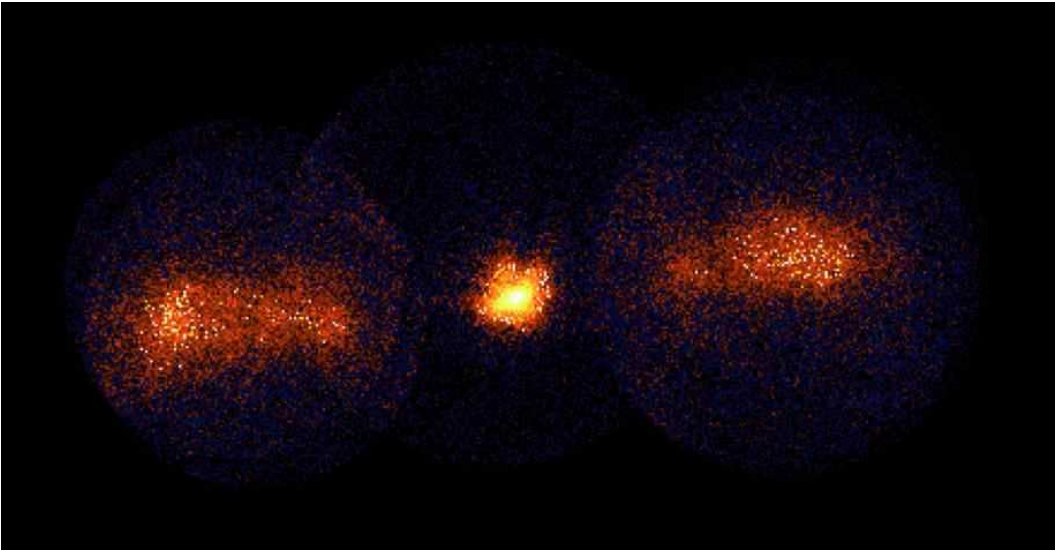,width=140mm}}
\caption{X-ray image of SS433 derived from ASCA GIS data (Kotani 1998).
The central bright source is SS433, and the total size of the entire
jet system is about one degree.
}
\end{figure}

Subsequent investigations of the X-ray emission in the vicinity
  of SS433 were carried out using the ROSAT, ASCA and RXTE
observatories (Yamauchi {\it et al.\/} 1994; Brinkmann {\it et
al.\/} 1996; Safi-Harb and Oegelman 1997; Safi-Harb and Petre
1999). The X-ray spectra of the two extended jets are different.
The spectrum of the eastern jet is non-thermal, with a power-law
photon index $\Gamma \approx 1.6$. There is a region of thermal
emission at the end of this jet ($T \approx 0.4$~keV), with the
X-ray and radio structures being very similar.  The emission of
the western jet is appreciably softer ($\Gamma \ge 2.3$), and
may even be thermal, but no additional thermal emission is
detected at the end of this jet (at the radio ``ear''). The
ROSAT data (Brinkmann {\it et al.\/} 1996) indicate that the
X-ray structure does not vary significantly along the extended
jets, although the contribution of the soft radiation
surrounding the jet grows outside. Namiki {\it et al.\/} (2000)
found that, with distance from the source, the X-ray spectrum
becomes softer, and the spectrum is non-thermal and continuous,
without emission lines. Analysis of RXTE observations (Safi-Harb
and Petre 1999) confirmed that, in a broader energy range (to
100~keV), the spectrum of the extended jets (the eastern jet) is
non-thermal, with $\Gamma \approx 1.45$, with the total X-ray
luminosity of the jet being $\sim 1.2 \cdot 10^{35}$~erg/s.  
No $\gamma$ radiation from W50 or SS433 has been detected
(Geldzahler {\it et al.\/} 1989; Rowell 2001).

The increase in the brightness of the X-ray jets with approach
toward the precession axis, as well as the decrease in the
opening angle of the jet-propagation cone at large distances --
the focusing of the extended jets -- could be associated with
hydrodynamical collimation of the precessing jets (Peter and
Eichler 1993) or with the interaction of the jets with material
from the supernova remnant and the formation of secondary
reflected shock waves propagating inside the precession cone
(Velazquez and Raga 2000). In this latter case, it is even
possible to explain the helical structure observed in the radio
ears of W50 (Dubner {\it et al.\/} 1998). The velocity of the
shocks in the direction of the symmetry axis in this case should
be no lower than $\sim 130$~km/s, so that a perturbation created
at the precession cone surface reaches the axis in $10^5$~yrs at
a distance from   the object of 40~pc.

However, it is not ruled out that the jet precession angle can
change with time. In the slaved disk (precession) model, there
is a beautiful mechanism that makes possible time variations in
the precession angle of the donor star. Matese and Whitmire
(1983, 1984) showed that a misalignment between the stellar
rotational axis and the orbital axis could be enhanced in the
mass transfer through the inner Lagrange point (the supernova
explosion serves as a suitable mechanism initially introducing
this misalignment). It is well-known that the orbit of the close
binary system will tend to be circularized due to tidal friction
(and mass transfer), and the misalignment between the rotational
and orbital axes decreases with time. However, during the flow
of gas through the inner Lagrange point, the specific angular
momenta of the lost and remaining gas are imbalanced, so that
the misalignment in the axes can be maintained (or even
increased) after the orbit approaches circularity.

The alignment of the stellar and orbital axes due to tidal
effects occurs over $\sim 10^5$~yrs (Papaloizou and Pringle
1983), which is less than the time for the star to evolve to the
stage of overfilling its critical Roche lobe. Consequently, for
the mechanism of Matese and Whitmire to operate, either the
relativistic object in SS433 must form after the star has filled
its Roche lobe, or this star should already be fairly evolved
before the formation of the relativistic object. This conclusion
is rather important. Indirect evidence that the mechanism of
Matese and Whitmire operates is provided by the fact that the
time for alignment of the stellar rotational and orbital axes in
a binary system such as SS433 is less than or roughly equal to
the time for circularization of the orbit (Papaloizou and
Pringle 1982), and the orbit in this system is nearly circular,
$e < 0.05$ (Fabrika {\it et al.\/} 1990). If the formation of a
circular orbit in SS433 as a result of the supernova explosion
is not purely coincidental (for example, the compensation by
chance of the orbital angular momentum by the asymmetrical
explosion of the supernova, which is unlikely), a mechanism that
prevents the rapid alignment of the axes in this system is
required.

\bigskip

\section{The X-ray Jets}

\subsection{Early Observations}

Early X-ray studies of the central source in SS433 are described in the
review of Margon (1984). Beginning in the mid-1980s, thanks to the
discovery of X-ray lines of the jets and observations during eclipses
of the accretion disk, it has become clear that the X-ray emission
is thermal, and is radiated primarily in inner regions, immediately
above the accretion disk, by cooling gas of the relativistic jets on
scales of $\sim 10^{12}$~cm. The total luminosity of the X-ray emission
  is $L_x \sim 3 \cdot 10^{35} - 10^{36}$~erg/s, substantially lower
than the bolometric luminosity of the accretion disk, $L_{bol} \sim
10^{40}$~erg/s. The X-ray emission is strongly variable, and its intensity
and spectrum (like those of the optical emission) depend on the activity
state (flares), orientation of the disk and jets (precessional phase),
and effects of eclipses of the optical star and absorption in the surrounding
gas (orbital phase).

The X-ray iron lines were first detected in an EXOSAT spectrum of SS433
(Watson {\it et al.\/} 1986; Stewart {\it et al.\/} 1987; Brinkmann {\it et al.\/} 1988). These
observations showed a relatively broad line that moved through the
spectrum. This movement was in good agreement with the kinematic model
if the line is emitted by highly ionised iron (Fe\,XXV, 6.7~keV) in the
blue (approaching) jet; the emission itself was obviously thermal. The
corresponding line from the receding jet had not been detected, and it
was thought that this could be the result of eclipsing of the receding jet
by the accretion disk
(in which case the X-ray jet must be comparatively short) or of substantial
weaking of the intensity of the receding jet due to the effects of
relativistic aberration. It was also concluded that the X-ray gas of the
jets had a low temperature ($kT \sim 2$~keV). The low emissivity of the
X-ray gas corresponding to this temperature led to the requirement for
a very high kinetic luminosity of the jets, $L_k \sim 10^{40-41}$~erg/s.

In subsequent GINGA observations of SS433 (Kawai {\it et al.\/} 1989; Brinkmann
{\it et al.\/} 1991), X-ray brightness decreases were reliably identified with
eclipses of the accretion disk. Some of these eclipses were in very good
agreement with optical eclipses, even according to simultaneous optical
observations (Goranskii {\it et al.\/} 1997). The eclipse light curves varied
substantially depending on the orientation of the disk (precession phase).

The X-ray line emission of the jets was not resolved in the GINGA observations,
but the complex behaviour of the broad shifted iron line at $E \approx 7$~keV
had already been noted. This was described as precessional motion of
``narrow'' iron emission against a background of broad line emission.
During eclipses, the intensity in the entire line
decreased in proportional to the total flux, indicating that most or all
of the X-ray emission was generated in the jets. The temperature of the
emitting gas derived from the GINGA data decreased sharply during eclipses,
from $k T \sim 30$~keV to $k T \sim 12$~keV at the centre of the eclipse,
from which it follows that the temperature of the jets falls off with
increasing distance from the source. In subsequent analyses
of the GINGA data (Yuan {\it et al.\/} 1995), a narrow
moving line component formed in the approaching jet was distinguished.
The intensity of this component was approximately constant in the rest
frame of the jet. It was noted that the intensity of the remaining broad
iron-line component (a weakly ionised iron line or blend of many lines)
varies in proportion to variations of the total X-ray flux,
in accordance with the precessional variations of the orientation of the
accretion disk. When the disk is maximally turned toward the observer,
the object becomes brighter.

\bigskip

\subsection{Localisation of the X-ray Source}

The X-ray emission outside the binary system is relatively weak, since
observations of the deepest eclipses in SS433 (Kotani 1998) indicate
that the fraction of external radiation is less than 30~$\%$. More
exact estimates are either not available, or they become model dependent:
What fraction of the X-ray emission is formed in the uneclipsed, cooling jets
($\sim 10^{13}$~cm)? What fraction is reflected or re-emitted in gas moving
in the wind? Is there an additional source of X-ray emission further from
the system, in the region of maximum emission of the optical and radio jets
($\sim 10^{15}$~cm) and the radio brightening zone?

Recent CHANDRA HETGS observations (Marshall {\it et al.\/} 2002) detected extended
X-ray emission around
the central source on scales from $1^{\prime\prime} - 2^{\prime\prime}$
($(0.7-1.5) \cdot 10^{17}$~cm) to $6^{\prime\prime}$. Unfortunately,
information about the structure of the central source itself was lost
due to the pileup effect during these observations. The extended
X-ray source is elongated in the direction of the jet precession axis,
with its intensity growing toward the centre. The total luminosity of this
source is $L_{x,ext} = 0.6\% \, L_x \approx 2 \cdot 10^{33}$~erg/s;
Marshall {\it et al.\/} (2002) did not detect any emission lines in its spectrum.

The CHANDRA, ASIS--S observations reported by Migliari {\it et al.\/} (2002)
confirm that the X-ray jets are resolved on scales of several arcseconds.
The direction of the jets is completely consistent with the direction of
the radio jets, and the maxima of the X-ray emission in the eastern and
western jets are observed at distances of $\ga 2 \cdot 10^{17}$~cm,
however the central source of SS433 was likewise affected by pileup
effects in these observations.
The 2--10~keV X-ray luminosity at these distances from the source
is $L_{x,ext} \approx 3-4 \cdot 10^{33}$~erg/s, which is
about $\sim 3~\%$ of the observed average X--ray luminosity   of
SS433.

Migliari {\it et al.\/} (2002) report the detection of emission lines shifted
in accordance with the kinematic model. A line at $\approx 7.3$~keV was
found in the spectrum of the eastern (approaching) jet, while a line  
at $\approx 6.4$~keV was found in the spectrum of the western (receding)
jet. Both emission lines may be
the Fe\,XXV\,K$\beta$ (7.06~keV) iron line, shifted because of the gas
motion in the jets with the velocity $0.26 c$. The relative
intensities of these two lines are also in agreement with the idea that
the lines are formed in the jets. The time for the jets to travel to the
emitting region is $\sim 200$~days. The emitting region is quite extended,
and covers no less than one whole precessional cycle of the jets (Migliari
{\it et al.\/} 2002). The continuum spectra (0.8--10~keV) of these two regions can be
fit by a bremsstrahlung spectrum with a temperature of $\sim 5$~keV, however
they can be equally well described with a power law with a photon index of
$2.1 \pm 0.2$.

The extended X-ray emission detected by Migliari {\it et al.\/} (2002) cannot be
radiation of the supercritical accretion disk scattered in external
gas, since the spectra of the western and eastern components should be the
same in that case. These data directly indicate reheating of the jets
at distances between $3 \cdot 10^{15}$~cm (the end of the optical jets
and the radio brightening zone) and $\sim 10^{17}$~cm.

It is likely that this X-ray emission on scales of arcseconds does not bear
any relation to the extended X-ray jets described above, since the emission
of the extended jets (deceleration of the jets) becomes appreciable at
distances from the centre that are hundreds of times larger ($\approx
15^{\prime}$). The X-ray emission on arcsecond scales could be
associated with interactions between the jets and the disk wind,
i.e., with the radio-emitting regions (the VLBI and VLA jets). Future X-ray
observations with arcsecond and subarcsecond angular resolutions will
provide answers to this question.

\bigskip

\subsection{ASCA Data: Lines and Spectrum of the Jets}

The ASCA observatory carried out about 30 observations of SS433 at various
phases of the orbital and precessional periods (Kotani {\it et al.\/} 1994, 1996,
1997ab; Kotani 1998). The X-ray eclipses have various depths
depending on the orientation of the accretion disk, which blocks from half
to two-thirds of the radiation. ASCA observations were able to resolve the
jet emission, and for the first time separate Fe\,XXV\,K$\alpha, \beta$,
Fe\,XXVI\,K$\alpha$ (appreciably weaker than the line of the helium-like
ion), and Ni\,XXVII\,K$\alpha$ lines detected for the approaching and
receding jets. Weaker K$\alpha$ lines of Mg\,XII, Si\,XII, Si\,XIV, S\,XV,
S\,XVI, and Ar\,XVII were detected only from the approaching jet; many
unresolved lines at 1--1.5~keV were also detected, as well as a fluorescent
stationary line
of neutral or weakly ionised iron Fe\,I--X
  at 6.4~keV (EW(Fe)$\approx 50$~eV),
which probably forms due to reprocessing of the radiation by the gas
surrounding the jets and making up the wind of the accretion disk. Thus,
a fundamentally new possibility arose of using the line intensities as a
diagnostic for the X-ray jets of SS433. Future observations of the X-ray
eclipses of the jets by the optical star of SS433 with better spectral
resolution (such as that provided by CHANDRA) will open rich opportunities
for direct investigations of the inner jets and the region above the
photosphere of the wind in which the jets appear.

The spectrum at hard energies of 5--9~keV where the iron lines are emitted
is in good agreement (taking into account both absorption and the
contribution of the main lines) with a
power-law with a photon index $\Gamma \approx 0.69$. At softer energies
(1--4~keV), where lines of less heavy elements are emitted, the power-law
spectral index   is
$\Gamma \approx 1.09$. The power-law approximations to
the spectrum in these two
intervals are consistent with the same absorption  
$N_H \approx 6.8 \cdot 10^{21}$~cm$^{-3}$ (Kotani {\it et al.\/} 1996).
The luminosity of SS433 at 2--8~keV is $L_x \approx 6.3 \cdot
10^{35}$~erg/s, with the luminosity in the brightest line of the
approaching jet being $L$(Fe\,XXV\,K$\alpha)^- \approx 2.3 \cdot
10^{34}$~erg/s.

In a simple model of adiabatically cooling jets, the ratio of the
intensities of the Fe\,XXV K$\alpha$ and Fe\,XXVI K$\alpha$ lines yield a
temperature for the base of the jets of $k T_0 \approx 22$~kev. The
temperature where the X-ray jets end~-- at a distance from the source of
$(2-3) \cdot 10^{13}$~cm, where the gas becomes thermally unstable~-- is
$\sim 1 \div 0.1$~keV. Beginning with a distance corresponding to a jet gas
temperature
  of 6--7~keV, there is absorption or obscuration of the
receding jet.

Estimates of the kinetic luminosity and mass-loss rate in the jets lead
to fairly high values. However, nearly all such estimates have assumed
improbably large opening angles for the jet, $\theta_j = 5^{\circ}$, while
the jets of SS433 are actually substantially better collimated than this,
$\theta_j \approx 1^{\circ}$ (Borisov and Fabrika 1987; Marshall {\it et al.\/} 2002).
In their more detailed jet model, Brinkmann and Kawai (2000) estimated the
kinetic energy flux to be $L_k \sim 5.7 \cdot 10^{39}$~erg/s. They demonstrated
on the basis of ASCA observations that the jet emission lines could be used
to study finer effects and to investigate the jet structure in detail.

Observations by the ASCA group (Kotani {\it et al.\/} 1997ab; Kotani 1998) led to
heavy-element abundances appreciably higher than their solar values. To
explain the spectra, the metal abundance must be enhanced by a factor of
1.5--2, and the abundance of Ni (Ni\,XXVII$^{\pm}$ at 7.3 and 7.7~keV) must
be increased by more than a factor of 20. This result could have far-reaching
implications, for example, about the occurrence of thermonuclear reactions
at the surface of a {\it neutron star} inside the supercritical accretion
disk. However, the observations of the optical jets are consistent with the
idea that they have a normal chemical composition. The cited nickel abundance
could be obtained if some additional effects were not taken into account
in the adopted approximation of the SS433 X-ray spectrum, since the
signal/noise ratio falls sharply at energies $>7$~keV (Kotani {\it et al.\/} 1997ab).
A model with ballistic jets cooling due to expansion and radiation was
used, but it is likely that additional sources of heating must be included
in the description of the X-ray jets (Brinkmann {\it et al.\/} 1988), for example,
reheating by collimated radiation of the supercritical disk
or shock processes
arising when the jet gas exits the nozzle (a channel in the wind). Similar
additional factors could affect the intensities of metal lines derived in
models. The CHANDRA observations (Marshall {\it et al.\/} 2002) are not consistent
with very high metal abundances.

\bigskip

\subsection{ASCA Data: Equatorial Wind}

The observations of the ASCA group revealed important behaviour that can
be most successfully interpreted in terms of a picture with absorption of
the radiation of the receding jet by a factor of 2--3, with the magnitude
of the absorption growing with distance from the source (Kotani {\it et al.\/} 1996).
The ratio of the intensities of the Fe\,XXV K$\alpha^+$ Fe\,XXV K$\alpha^-$ lines
from the two jets was found to $\approx 0.24 \pm 0.06$,
appreciably lower than the
value 0.66 expected for the given precessional phase. The lines in the
approaching jet should be brighter than those in the receding jet as a
consequence of relativistic boosting (see the following section for more
detail), but the lines of the receding jet proved to be systematically
weaker than expected even after allowing for this effect. Thus, it is
necessary to invoke additional absorption of the light from the receding
jet. However, the object screening this jet cannot be the accretion disk,
since the lines that are most subject to such screening would be the hottest
lines emitting in the jets closer to the source, which is not the case.
It was concluded that
the systematic relation $(I^+/I^-)_{soft} < (I^+/I^-)_{hard}$ was valid
for lines forming at various temperatures (Kotani {\it et al.\/} 1996, 1997ab;
Kotani 1998).
The observed line intensities required a weakening of the radiation from
the most distant regions of the receding jet by a factor of two to three.
This means that the ``accretion disk''
itself has relatively small dimensions,
and our view of the receding jet is unobstructed at distances of
$\sim 10^{12}$~cm from the source. Further, regions of the
receding jet radiating
at distances of $\sim 10^{13}$~cm (for comparison, the size of the system
is about $a \approx (4-5) \cdot 10^{12}$~cm) begin to experience appreciable
absorption.

Kotani {\it et al.\/} (1996) proposed that the absorption occurs in gas lost by
the system through the outer Lagrange point L2 (``sprinkling'' the disk).
Possible observational manifestations of an intense loss of gas in SS433
through the point L2 are discussed by Fabrika (1993). It is likely that
precisely such flows deform the orbital light curve in the optical
(Zwitter {\it et al.\/} 1991; Fabrika 1993), and that their effect is observed
on larger scales in the regions of radio emission perpendicular to the
jets detected in VLBI images (Paragi {\it et al.\/} 1999; Blundell {\it et al.\/} 2001).
It is possible that these equatorial regions can be detected most effectively
in H$\alpha$ line emission surrounding SS433 in the form of an extended
disk illuminated by the precessing accretion disk, in which case we should
expect variable extended H$\alpha$ emission at distances of $\sim
1^{\prime\prime}$ from SS433. However, to our knowledge, such observations
have not been carried out on the HST.

The results of ASCA X-ray observations are presented in the thesis of
Kotani (1998). At the precessional phase when the disk is maximally
facing the observer (near $T_3$ moment), the source is bright, the Fe\,K edge
is deep, the receding jet undergoes appreciable absorption, and the
temperature of the gas is higher in the approaching jet than in the
receding jet. At precessional phases when the disk is viewed edge-on
(near $T_{1,2}$), the X-ray source is weak, the Fe\,K edge is shallow,
and the intensities of the emission lines of both jets are the same.
This last circumstance is very important, and makes the conclusions that
can be drawn firm.

\bigskip

\subsection{CHANDRA Data: Narrow Multi-temperature Jets}

CHANDRA observations of SS433 (Marshall {\it et al.\/} 2002) have mainly
confirmed the conclusions reached earlier on the basis of the ASCA
observations. The CHANDRA observations are especially important for our
understanding of the X-ray jets of SS433, since the excellent spectral
resolution enables the direct detection and verification of the effects
described above. The spectrum of SS433 obtained with the CHANDRA HETG
spectrometer (Marshall {\it et al.\/} 2002) is shown
  in Fig.~11, where the range
of wavelengths in Angstroms corresponds to 1.08--8.3~keV. It was possible
to detect more than 20 emission lines from the approaching jet, 6 lines
from the receding jet and a line of neutral (or weakly ionised) iron at
6.42~keV in the spectrum. The strongest line was a helium-like line of
iron Fe\,XXV. The multi-temperature model for the
cooling jets was confirmed; lines of
the lighter elements Ne and Mg ($T \sim 10^7$~K) are observed together
with hot lines of Fe and Ni ($T \sim 10^8$~K). An appreciable number of
lines at low energies were not identified. The overlapping of these lines
could make a significant contribution to the continuum.

\renewcommand{\thefigure}{11}
\begin{figure}[p]
\vspace*{8mm}
\centerline{\psfig{figure=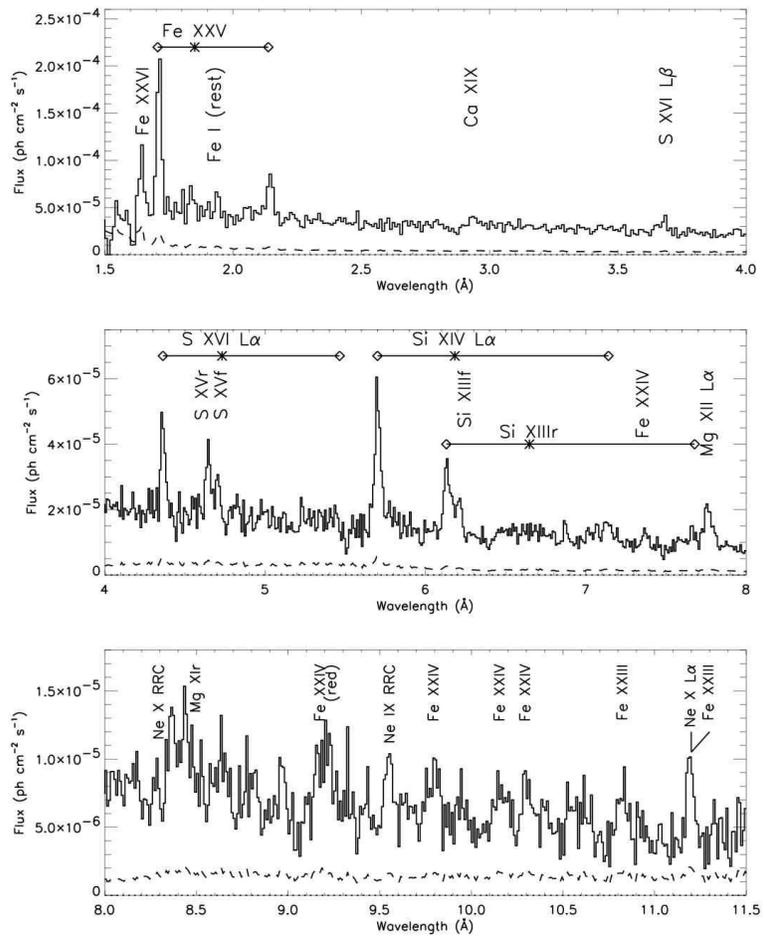,width=120mm}}
\caption{CHANDRA HETG spectrum of SS433 (Marshall {\it et al.\/}
2002). We can see primarily lines corresponding to the blue jet,
which are marked. The horizontal bars join the same lines
radiating in the blue and red jets (diamonds), while the
asterisks mark the unshifted positions of these lines. The
dotted line shows the statistical uncertainty.}
\end{figure}

The fact that CHANDRA observations were able to resolve the jet lines in
SS433 is a very important result. These lines proved to be appreciably
broadened, FWHM\,$\approx 1\,700$~km/s, with the line widths and radial
velocities having
approximately the same magnitude independent of the temperature
of the radiation. Marshall {\it et al.\/} (2002) found that the opening angle for
the X-ray jets is $\theta_{j,x} = 1\degdot23 \pm 0\degdot06$. Recall that
the opening angle of the optical jets found by Borisov and Fabrika (1987)
was $\theta_{j,opt} = 1\degdot0 - 1\degdot4$. When modeling the profiles
of the moving H$\alpha$ lines (Borisov and Fabrika 1987), the intensity
distribution across the jet was represented as a two-dimensional Gaussian
with $\sigma = \theta_j/2$. In contrast to the ``short'' X-ray jets, for
which fairly simple geometrical considerations suffice for estimates of
the opening angle, estimation of the opening angle of the optical jets
requires modeling of the line profiles, since nutational and precessional
shifts contribute to the total line widths. The coincidence of the opening
angles for the X-ray and optical jets is remarkable. This means that the
SS433 jets are indeed conical and move along strictly ballistical
trajectories, from the source itself at distances of $\sim 10^{11}$~cm
(where the jets emerge from below the wind photosphere and the
temperature of the jet gas is $\sim 10^8$~K) to the beginning of the zone
of expansion of the H$\alpha$ clouds, $\approx 3 \cdot 10^{15}$~cm
(where the temperature of the jet gas is $\approx (1-2) \cdot 10^4$~K).

Based on the positions of the lines, Marshall {\it et al.\/} (2002) deduced that
the velocity of the X-ray jets was $\beta = 0.2699 \pm 0.0007$, which is
$2920 \pm 440$~km/s higher than the velocity of the jets in the kinematic
model derived from optical data (Margon and Anderson 1989). However, if
we make the comparison using refined parameters of the kinematic
model (Eikenberry {\it et al.\/} 2001), the formal difference in the velocities
is reduced to $1560 \pm 340$~km/s. It is important to note here that the
temporal instability of the jet velocity (jitter) can reach $\pm 3\,000 -
5\,000$~km/s. Given the relatively short duration of the CHANDRA observations of
SS433 (29~ks), it seems premature to conclude that the X-ray and optical
jets propagate with different speeds. The complete coincidence of the
collimation angles for   the X-ray and optical jets, in turn,
suggests that they correspond to the same physical object
observed at different stages of its evolution.

The 0.8--8~keV CHANDRA X-ray spectrum is in good agreement with a power
law ($\Gamma = 1.35$, $N_H = 9.5 \cdot 10^{21}~cm^{-3}$) and with the
spectrum obtained by ASCA (Kotani {\it et al.\/} 1996); the mean luminosity of
SS433 at 2--10~keV is $L_x = 3.1 \cdot 10^{35}$~erg/s. The gas temperature
can be determined fairly reliably based on the ratios of the line fluxes of
hydrogen-like and helium-like ions. The observed jet emission lines form
at temperatures from $1 \cdot 10^8$~K to $5 \cdot 10^6$~K. The density-sensitive
Si\,XIII triplet can be used to derive the electron density,  $\sim
10^{14}$~cm$^{-3}$,  of the gas in the region of the jet where the temperature
is $1.3 \cdot 10^7$~K. Very weak lines (Ne\,X, Ne\,IX) arising as a consequence
of radiative recombination have been detected. If the strength of these lines
is due to photoionisation, this requires a luminosity of order $L_x \sim
10^{40}$~erg/s, which is, in principle, plausible if the X-ray radiation of
SS433 is collimated along the jets. However, based on the absence of other
strong lines that should arise during photoionisation, Marshall {\it et al.\/}
(2002) concluded that the gas was collisionally heated.

The emission measures for various ions have been determined in a model with
a conical, adiabatically cooling jet; optically thin thermal and collisional
plasma; and normal elemental abundances. A four-component (four-temperature)
model was also constructed, in which the temperature of the jet gas falls
from $1.1 \cdot 10^8$~K to $6 \cdot 10^6$~K and the electron density falls
from $2 \cdot 10^{15}$ to $4 \cdot 10^{13}$~cm$^{-3}$ at distances from the
base of the jet from $2 \cdot 10^{10}$ to $2 \cdot 10^{11}$~cm. The kinetic
luminosity of the jets derived in this model is $\L_k \sim 3 \cdot
10^{38}$~erg/s. Thus, the X-ray jet inferred by Marshall {\it et al.\/} (2002) proves
to be very short. This places substantial constraints on the distance from
the relativistic star at which the bases of the jets are located, such that
there are no instantaneous effects of eclipsing of the jets by the
  optical
star.

The X-ray iron line at 6.4~keV is not resolved in the CHANDRA spectra of
Marshall {\it et al.\/} (2002), FWHM\,$< 1\,000$~km/s. This line most likely arises
via fluorescence in the cool wind gas (Kotani 1998), or even in a cocoon
around the base of the jets (Fabrika 1997). Based on an analysis of the
X-ray eclipses and eclipses of the He\,II\,$\lambda 4686$ emission line and
the times of egress of the X-ray source and He\,II source from behind the
limb of the optical star, Goranskii {\it et al.\/} (1997) concluded that the region
of He\,II emission surrounds the X-ray source in the SS433 accretion disk.
We will consider the structure of the disk separately.

The CHANDRA data confirm the
screening of the receding jet detected by ASCA~--
the radiation of this jet is substantially weaker than that of the approaching
jet. However, the temperatures of the two jets proved to be approximately
equal. More prolonged observations are needed to test the hypothesis that
the radiation of the receding jet is absorbed in material flowing from the
system, since the gas flowing outward in the plane of the disk could be
appreciably non-uniform in the azimuthal direction.

Marshall {\it et al.\/} (2002) noted an interesting coincidence between the
expansion velocity of the jets in the transverse direction (more precisely,
the maximum possible expansion velocity) derived from the line widths and
the sound speed at a temperature of $\sim 10^8$~K derived from the line
intensities. If the width of the jets is determined by the free expansion
of the gas at the base of the jets, the opening angle should be equal to
$\theta_{j,x} = 2 c_s/V_j$, where $c_s$ is the sound speed for protons and
$V_j$ is the jet velocity. The jet opening angle that is obtained for a
gas temperature at the base of the jet of $T_0 = 1.1 \cdot 10^8$~K is
$\theta_{j,x} \approx 1\degdot4$, which virtually coincides with the
value derived from the observations. As the distance from the centre
increases, the gas cools, the sound speed falls, and the jets become
strictly ballistic. This coincidence of the opening angles represents
a weighty argument in support of the idea that the temperature $T_0$
is measured just at the point where the jets emerge from
under the photosphere of the cocoon surrounding their base.
On its own, this does not shed
light on the jet collimation mechanism, which is most likely hydrodynamic.
However, we can conclude that the jets should initially (in inner regions
hidden from the observer) be collimated no more poorly than is observed
in the X-ray and optical, and that the operation of the collimation
mechanism should end somewhere just before the emergence of the
jets from under the photosphere.

In recent observations with the CHANDRA HETGS published by
Namiki {\it et al.\/} (2003) SS433 was observed in precessional phase 
``disk edge--on''. The authors have found that a width of the iron line 
Fe\,XXV\,K$\alpha$ (FWHM(Fe) $\sim 4900$\,km/s) is considerably greater
than that of silicon line Si\,XIII\,K$\alpha$ (FWHM(Si) $\sim 2000$\,km/s).
Marshall {\it et al.\/} (2002) have also noted such a trend, that the
widths of lower energy lines are slightly less than average width of all
lines studied in the spectrum, however this trend was only marginal in 
their data. In the spectra of Namiki {\it et al.\/} (2003) the silicon line
width is in agreement with the line widths found by Marshall {\it et al.\/}
(2002), but the iron line is notably broader. The authors suggested that 
they detected a progressive jet collimation along its axis. These new data
show that it is necessary to accumulate more observations with high 
spectral resolution (CHANDRA) at different precessional phases and also
during the accretion disk eclipses to understand the structure of the 
SS433 X--ray jets and their distance of the source. Probably, a Compton
scattering in the jet gas or in surrounding medium may play an important
role in broadenning of the X--ray spectral lines. Furthermore at the disk
orientation ``edge--on'' its inner parts (the jet bases) are obscured
partly by outer rim of the disk (the Section ``The Supercritical Accretion
Disk and the Components from the Photometric Data''), that is the
geometrical effects have to influence the X--ray spectrum.

In very recent observations with the gamma--ray observatory INTEGRAL
Cherepashchuk {\it et al.\/} (2003) detected a hard X--ray radiation 
of SS433 in the energy band 20\,--\,100~keV. The hard X--ray spectrum 
appears relatively flat in this band with the power--law photon index
$\Gamma \sim 2$. The luminosity of SS433 in the hard X--rays is 
$L_x \sim 3 \cdot 10^{35}$~erg/s (25\,--\,50~keV) and 
$L_x \sim 1.2 \cdot 10^{35}$~erg/s (50\,--\,100~keV). Cherepashchuk {\it et
al.\/} (2003) found a precessional variability of the hard X--ray flux:
when the disk is maximally facing the observer, the flux in the 
25\,--\,50~keV band increases more than two times comparing with the
precessional orientation ``edge--on''. The precessional variability
confirms the notion that the outer rim of the disk blocks the inner 
region at the disk orientation ``edge--on''. Note that the relativistic
boosting effect changes the flux with about the same factor $\sim 2$ 
if we consider only one jet, and the effect is quite weak ($9\%$) for 
the both antiparallel jets.   

The presence of the hard power--law component in the X--ray spectrum 
means in turn a Comptonisation of soft X--ray photons generated in the
inner disk (in X--ray jets) on relativistic electrons. The relativistic 
particles may be accelerated in the same inner jets, where the jets
leave out the funnel of the supercritical disk (the next Section).  

\bigskip

\subsection{Inhomogeneity of the Jets and X-ray Variability}

As a rule, investigators analysing X-ray observations of the
jets have assumed that the jets are conical and completely
filled with gas. In contrast, the filling factor of the optical
jets must be quite small (Davidson and McCray 1980; Begelman
{\it et al.\/} 1980); the filling factor for the jet clouds at
the distance of the maximum H$\alpha$ emission
  ($\approx 4 \cdot 10^{14}$~cm) is $\sim 10^{-6}$ (Panferov
and Fabrika 1997). The jet gas is collected in clouds by the
action of thermal instabilities. It is quite likely that the
region of formation of these clumps (Bodo {\it et al.\/} 1988;
Brinkmann {\it et al.\/} 1988; Kotani {\it et al.\/} 1996) is
located at the end of the X-ray jets, where the gas cools to a
temperature of $\sim 0.1$~keV and should begin to fragment (see
following section).  The cloud sizes predicted by the
thermal-instability mechanism (Brinkmann {\it et al.\/} 1988)
and derived from relative hydrogen line intensities (Panferov
and Fabrika 1997) are~$\sim 10^8$~cm. There should be thousands
of such clouds even in the relatively short X-ray jets, and it
is very unlikely that it will be possible to detect their
presence in the X-ray, for example, from X-ray variability.

The structure of the H$\alpha$ line profiles in the optical jets
suggests the presence of $\sim 10^3$ larger-scale formations
that could be considered clusters of clouds (Borisov and Fabrika
1987; Panferov and Fabrika 1997), with the characteristic time
for the formation of these structures being $\sim 10^2$~s. This
approximately corresponds to the time for the motion of the gas
along the X-ray section of the jets.  Thus, the structure of the
optical jets and the relatively modest size of   the X-ray jets
suggest that the X-ray flux is very likely variable on time
scales corresponding to the formation times of these structures.

The time of $\sim 100$~s is also close to the time for the
propagation of the SS433 jet beneath the photosphere of the wind
of the supercritical accretion disk. For a mass-loss rate in the
SS433 wind of $\sim 10^{-4}\,M_{\sun}$/yr, the wind-photosphere
radius is $R_{ph} \sim 10^{12}$~cm (van den Heuvel 1981; Lipunov
and Shakura 1982; Fabrika 1997).  This size is in good agreement
with the dimensions of the source of optical and UV radiation
around the relativistic star derived from observations,
$R_{UV,opt} =(1.5-2) \cdot 10^{12}$~cm (Dolan {\it et al.\/}
1997). If we identify this object with the opaque part of the
wind or with a channel in the wind, the time for gas moving with
the velocity of the jets to cross this region is 200--300~s. The
jets should be accelerated and collimated on this, or even
shorter, time scales (if the geometry of the inner region is
complex and the photosphere where the X-ray emission emerges is
smaller than the photosphere where the UV radiation is
generated). We can also conclude based on the coincidence of the
time for the generation of large-scale inhomogeneities in the
jet and the time for the propagation of the jet beneath the
photosphere that we should expect X-ray variability with a
characteristic time scale of hundreds of seconds.

Kotani {\it et al.\/} (2002), Safi-Harb and Kotani (2002)
recently detected such variability in PCA/RXTE observations
obtained during the active state of SS433. Strong stochastic
variations of the brightness of SS433 are observed at energies
of 2--10 and 10--20~keV on time scales of $10^2- 10^3$~s, with
the minimum variability time scale being 50~s. Although the
accretion disk was viewed edge-on during these observations,
i.e., its orientation was not amenable to variability studies,
this variability was reliably detected. It is likely that the
X-ray flux of SS433 is also variable in quiescent periods,
possibly with a lower amplitude than in periods of activity.

Chakrabarti {\it et al.\/} (2002) discussed plausible mechanisms
for the production of bullet-like ejecta with a time scale of
50--100~s in the accretion disk of SS433 as a possible
explanation for the observed short-time-scale variability.  They
suggest non-linear oscillations of shocks in the accretion disk
or close to a sound barrier in the sub-Keplerian accretion flow
as one such mechanism. In this case, the accretion rate will
exhibit appreciable modulations with the time-scale needed.


Variability of SS433 on short time scales ($<100$~s) has been
searched for in the UV at 1400--3000 \AA\AA\ in HSP/HST
observations (Dolan {\it et al.\/} 1997). An upper limit to the
variability amplitude   of $< 1.2~\%$ was derived from these
observations. We can hope that short-timescale variability of
SS433 will eventually be detected in the blue and UV, where the
emission is radiated in regions of the wind close to the jets or
by gas surrounding the place where the jets emerge. Like in the
X-ray, the question of detecting rapid variability in the UV
probably reduces to the sensitivity of the detectors used and
the signal/noise ratio of the observations. The presence of
optical variability on time scales of several minutes is well
established (for example, Goranskii {\it et al.\/} 1987; Zwitter
{\it et al.\/} 1991); this is stochastic variability with an
amplitude of $\sim 0\magdot1$ that does not disappear even
during eclipses of the accretion disk.

\bigskip

\section{Structure and Formation of the Jets}

In the previous sections, we described the main observational
data on the SS433 jets, as well as constraints on the physical
conditions in the jets that can be derived directly from these
observations. Here, we describe the physical conditions and
state of the gas in the jets in more detail (some of these
results will naturally be model-dependent), as well as the main
published theories concerning mechanisms for the jet formation.
Our review is not aimed at a complete treatment of various
theoretical questions.

\bigskip

\subsection{The State of the Gas in the Optical Jets}

The precessional motion of the jets presents excellent
opportunities for studies of the physical state of the gas in
the jets. Comparing the intensities of the hydrogen emission
lines at various angles of the two jets to the line of sight and
assuming that the two jets are intrinsically identical and that
their internal properties do not depend on orientation
(precessional phase), we can characterise the gas heating
mechanism and elucidate the structure of the radiating clouds of
gas. Differing absorption in the stellar wind of the emission
from   the ``+'' and ``--'' optical jets (in contrast to the
X-ray jets) is ruled out, since the size of the jets is two
orders of magnitude larger than the dimensions of the SS433
binary system.

The observed intensity of the jet spectral lines ${\cal
J}_{obs}$ will depend on the relativistic aberration of the
emitted radiation. In a jet with a finite length or consisting
of individual evolving fragments with finite life times (Lind
and Blandford 1985; Begelman {\it et al.\/} 1984; Panferov and
Fabrika 1997), the intensity of the line radiation in the rest
frame of the jet is ${\cal J}_{co}=(1+z)^3{\cal J}_{obs}$, where
$z=\delta\lambda/\lambda_0$ is negative for the approaching jet
and, as before, $\lambda = \lambda_0 \gamma (1-\beta \cos
\eta)$. If we consider the continuum radiation, the exponent for
this dependence should be replaced by $2 + \alpha$, where
$\alpha$ is the spectral index ($I_{\nu} \propto \nu^{-\alpha}$).
At time $T_3$ ($\psi = 0$, the jet axis is inclined at the
minimum angle to the line of sight, $57^{\circ}$), the line
emission from the approaching jet should be brighter than that
from the receding jet, ${\cal J}_{obs}^-/{\cal J}_{obs}^+
\approx 2.4$. At precessional phases $T_{1,2}$ (when the jets
are located in the plane of the sky), the intensities of the two
jets should be the same.

Asadullaev and Cherepashchuk (1986) discovered that the jet radiation is
anisotropic. They found that the ratio of the intensities of the two jets
in the H$\alpha$ line in a frame co-moving with the jets is maximum at
precessional phase $\psi \approx 0$ and is equal to about 2--3. In this case,
the anisotropy can be interpreted in a model with clouds in the jets, with
the maximum radiation occurring in their frontal regions.

\renewcommand{\thefigure}{12}
\begin{figure}[p]
\vspace*{8mm}
\centerline{\hspace*{-5mm}\psfig{figure=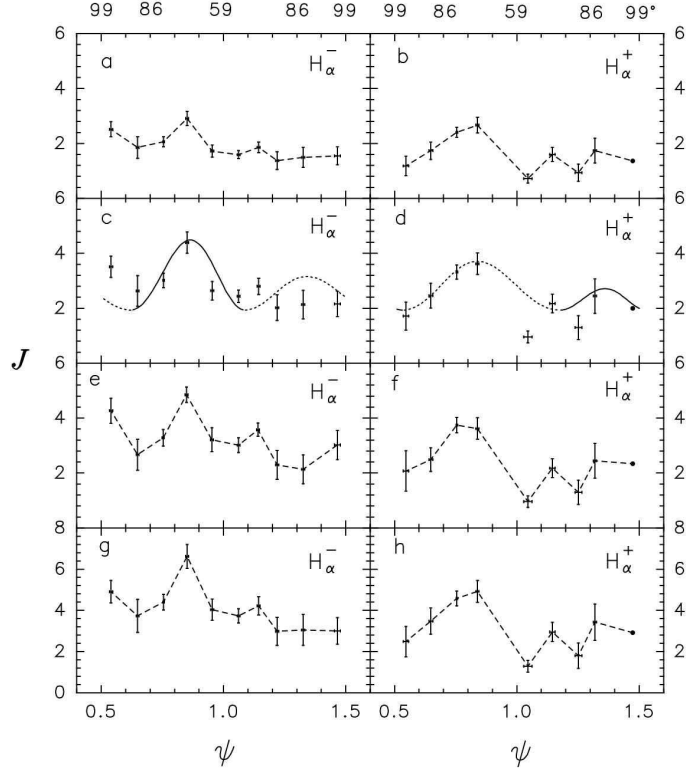,width=100mm}}
\caption{Precessional dependences of the intensities of the main components
of the moving H$\alpha^{\pm}$ lines in a comoving coordinate frame in units
of $10^{-10}~$erg/cm$^2\,$s (Panferov {\it et al.\/} 1997). The upper axis gives the
angle between the ``--'' jet and the line of sight. The intensities have
been corrected for interstellar absorption using the values $A_V$ = 7\magdot3
in (a) and (b), 7\magdot8 in (c)--(f), and 8\magdot3 in (g) and (h). Panels
(e) and (f) show the intensities obtained by summing all the components
of the moving-line profiles.
Model intensity curves are shown in panels (c) and (d), where
the solid curve corresponds to the radiation from the front
hemisphere of the emitter and the dashed curve to radiation from
the rear hemisphere.}
\end{figure}

In their study of the behaviour of the intensities of the moving H$\alpha$
lines based on a substantially larger amount of observational data,
Panferov {\it et al.\/} (1997) found that the angular distribution of the jet radiation
is rather complex. Figure~12 presents the precessional dependence of the
intensities of the main components of the moving H$\alpha^{\pm}$ lines in
a frame co-moving with the jets in units of~$10^{-10}~$erg/cm$^2\,$s. The
vertical axis plots the angle between
  the ``--'' jet and the line of sight
(the error in the values of this angle appearing in this same figure presented
by Panferov {\it et al.\/} (1997) has been corrected). The intensities have been
corrected for various values of the interstellar absorption ($A_V$ =
7\magdot3~-- 8\magdot3) near the true value
for~SS433, $A_V \approx 8$. The
data have been obtained for the main components of the moving-line profiles
(the ``young'' jet), which form near the source. This same figure (Figs.~12e,
f) shows the behaviour of the intensities obtained by summing all components
of the moving-line profiles. The figure presents the errors in the mean
intensities in intervals of the precessional phase $\Delta \psi = 0.1$. Each
phase interval contains the results of several authors obtained during various
precessional cycles during the ten years in which observations have been
made. Possible variations in the line intensities associated with active
periods are not reflected in the data in Fig.~12, since the mean data were
obtained over a long time interval. Possible inaccuracies in the adopted
magnitudes of the absorption and of the projection effects described above
(the nutational motion gives rise to an increase in the intensity of the
moving lines at extrema of the radial-velocity curves), likewise, do not
appreciably change the form of the dependences. In particular, the H$\alpha$
light curves of the jets derived from the main component or from the entire
profile are the same.

Figure~12 implies that the jet line radiation is anisotropic, and that the
directional beams for the radiation of both jets are similar; there are
maxima in the radiation both in the direction of motion and in the opposite
direction (forward and backward), with the axis of the beam not coinciding
with the jet-velocity vector. The maximum radiation is observed not at phase
$\psi = 0$, but shifted to phases 0.8--0.9. Model intensity curves obtained
taking relativistic aberration into account are shown in Figs.~12c and d,
where the solid and dashed curves correspond to the radiation of the front
and back hemispheres of the model radiator
(a gas cloud in the jet was specified
as a flattened spheroid). Panferov {\it et al.\/} (1997) found that the
front sides of the clouds in the SS433 jets are a factor of $\approx 1.7$
brighter than the back sides. The apparent equality of the intensities of
the ``+'' and ''--'' jets is the result of relativistic aberration of the
emitted light. The directions of the maximum radiation of the two jets
deviate significantly from the direction of the jet motion, by $\sim
30-40^{\circ}$, towards the direction of the precessional motion. The
dynamical interaction of the jets with the surrounding medium could lead
to such effects. Consequently, the dissipation of the kinetic energy of the
jets is the dominant process in heating the jets.

Anisotropic radiation by individual clouds is possible if their optical
depth in the hydrogen lines exceeds unity. Either these clouds are flattened
or they are transparent; i.e., the probability for line radiation
to exit is enhanced in a particular direction. This direction could
correspond to the direction of shocks in the jets that make the gas
transparent to line radiation due to the development of velocity gradients
in the gas clouds. The jets of SS433 do not propagate through a previously
established channel as they move through the surrounding medium, as would
jets propagating in a constant direction,
but must continually rebuild this channel, pushing it
toward the direction of the precessional motion. The jets interacting with
the wind create a cocoon and carry with them nearby pieces of the wind. The
transverse density and velocity profiles of the inter-cloud gas in
the jets will not be axially symmetric: larger gradients will develop in
the direction of the precessional rotation, and it is from that direction
that gas can flow into the jet. In this case, the direction of propagation
of shock waves in the jet is inclined to the velocity vector toward the
direction of the precessional rotation. The radiation of the jet gas in
optically thick lines is maximum along this axis. The vector of the maximum
H$\alpha$ radiation of the clouds is closest to the line of sight at
precessional phases $\psi \approx 0.8$. Either the inter-cloud gas flows
relative to the clouds inside the jet or weak shocks move through the
inter-cloud gas and clouds, but these perturbations propagate at an angle
to the jet axis and arise on the side in which the precessional motion is
directed.

Panferov and Fabrika (1997) studied the Balmer decrements in the SS433 jets.
The relative intensities of the H$\alpha^{\pm}$, H$\beta^{\pm}$ and
H$\gamma^{\pm}$ lines were found using a uniform dataset obtained over ten
years of observations, deriving the individual decrements from
spectra taken during a single night. The ratios of the hydrogen-line
intensities are the same in the two jets, but vary appreciably with the
precessional phase. At the phases of the maximum intensities of the moving
lines ($\psi = 0.7 - 0.9$), H$\alpha$/H$\beta= 1.6 \pm 0.2$ and
H$\gamma$/H$\beta =0.7 \pm 0.1$, while H$\alpha$/H$\beta= 0.9 \pm 0.2$
and H$\gamma$/H$\beta =0.8 \pm 0.1$ at the phases of the minimum intensities
($\psi = 0.0 - 0.2$). Such decrements are characteristic of high-density
gas, $n_e \ga 10^{12}$~cm$^{-3}$, when the populations of atomic levels
are determined primarily by collisional processes. In addition, the effects
of optical depth and projection are important during the formation
of the jet hydrogen lines. A comparison of the observed relative intensities
with the computational results of Drake and Ulrich (1980), which were
obtained for a uniform layer of gas within broad intervals of the physical
parameters took into account the influence of the Stark effect on the
probability for the photons to exit their emission region, demonstrates
that the gas in the SS433 jets should be dense and be formed of compact
clouds. The effective
size of individual gas clouds, or, more precisely, the optical depth in
the H$\alpha$ line, varies appreciably depending on the orientation of the
jets. The precessional phases when bright H$\alpha^{\pm}$ lines are
observed correspond to small optical depths of the layer. The following
parameters of the gas clouds and jets have been derived:
\begin{itemize}
\item mean particle density $ n \approx 10^{13}$~cm$^{-3}$,
\item gas temperature $T_e \approx 2\cdot 10^4$~K,
\item optical depth of the layer to line radiation $\tau$(H$\alpha)
\sim 10^2 - 10^4$, depending on the jet orientation,
\item size of the clouds $l \sim 10^8$~cm,
\item number of clouds in the jet $ \sim 10^{12}$,
\item volume filling factor for the jet clouds $\xi \sim 4 \cdot 10^{-6}$,
\item kinetic luminosity of the optical jets $L_k \approx 10^{39}$~erg/s.
\end{itemize}

Estimates of the kinetic luminosity for two intervals of the precessional
phase (bright and weak jet lines) are in reasonable agreement. These were
derived from the ratio $\epsilon_{H\beta}/n$~-- the ratio of the mean
efficiency of the radiation of a unit volume of gas in the H$\beta$ line
to the gas density (Drake and Ulrich 1980). The kinetic energy flux is
$L_k \approx L_{H\beta}\, m_p\, n\, V_j^3/2\,\epsilon_{H\beta}\, R_j$,
where the jet luminosity in this line is (on average) L$_{H\beta} \approx
7 \cdot 10^{35}$~erg/s, the mean quantity $\epsilon_{H\beta} \approx
1$~erg/cm$^3\,$s, and $R_j \approx 4 \cdot 10^{14}$~cm is the
length of the jets corresponding to the maximum line radiation.

\bigskip

\subsection{The Zones of Sweeping out and Expansion}

Clouds of the supposed dimensions freely expand on a characteristic time
scale of $ l/c_s \sim 100$~s, much shorter than the lifetimes of the
moving-line components (4~days). Consequently, the clouds are not in a
freely expanding state in the jets, and there must be some mechanism
preventing their expansion. The jets sweep up the wind gas along the
surface of the precessional cone; the length of the swept-out zone is
$P_{pr}\, V_w \sim 3 \cdot 10^{15}$~cm if the disk wind speed in
circumpolar regions of the disk
($\pm\, 2\,\theta_{pr}=40^{\circ}$) is $V_w \approx
1\,500-2\,000$~km/s. Within this region, the H$\alpha$ clouds are protected
from disruption by the dynamical pressure of the surrounding gas, and
they can slowly evolve. The H$\alpha$/H$\beta$ intensity ratio is appreciably
higher for the weaker secondary moving-line components that form in distant
sections of the jets than for the main components (Panferov and Fabrika
1997). This testifies to a decrease in the density of the clouds in the
jet with time (with increasing distance from the source).
If the wind density depends on distance as $\hat n \propto
R^{-2}$, the density of the clouds that are prevented from expanding by the
dynamical pressure of this gas falls by nearly two orders of magnitude by
the end of the swept-out zone. The dimensions of the clouds are $ l \propto
n^{-1/3} \propto \hat n^{-1/3} \propto  R^{2/3}$, and therefore increase
approximately fivefold. The volume emission measure in the gas of the clouds
$n^2\, l^3$ along the optical jets falls as $R^{-2}$, and the volume filling
factor of the jets falls as $\xi \propto R^{-1}$.

Beyond the zone in which the wind gas is swept up, the jets move in gas-free
space, since the gas there was swept up during previous precessional
passages of the jets, and the slow wind has not had time to refill this
space with gas. Here, the dynamical pressure of the gas on the clouds is
much lower, and the clouds are reheated and expand~-- the radiation in
the hydrogen lines ceases.

The zone of expansion of the clouds begins at
the boundary of the sweeping-out zone, at a distance of about
$3 \cdot 10^{15}$~cm (4.5~days of flight of the gas). Here, the density
of the incoming gas decreases sharply. The clouds, which are no longer
confined beyond the sweeping-out zone, expand and fill the entire volume
of the jets. This process probably leads to the appearance of numerous
shocks in the jets and to the rapid heating and expansion of the gas clouds,
and the efficiency of the generation of relativistic particles is increased.
Recalculating the cloud parameters to the outer boundary of the sweeping-out
zone, Panferov and Fabrika (1997) found that the freely expanding clouds
can fill the entire jet volume in approximately one day. This is consistent
with the location of the well known brightening zone in the radio jets of
SS433 (see the Section ``The Radio Jets and W50''), which is observed in
VLBI images (Romney {\it et al.\/} 1987; Vermeulen {\it et al.\/} 1987, 1993b) at a distance
of $3.7 \cdot 10^{15}$~cm (5.6~days of flight of the jet gas), i.e., at
a distance of $0.7 \cdot 10^{15}$~cm from the outer
boundary of the sweeping-out zone.

\renewcommand{\thefigure}{13}
\begin{figure}[h]
\centerline{\psfig{figure=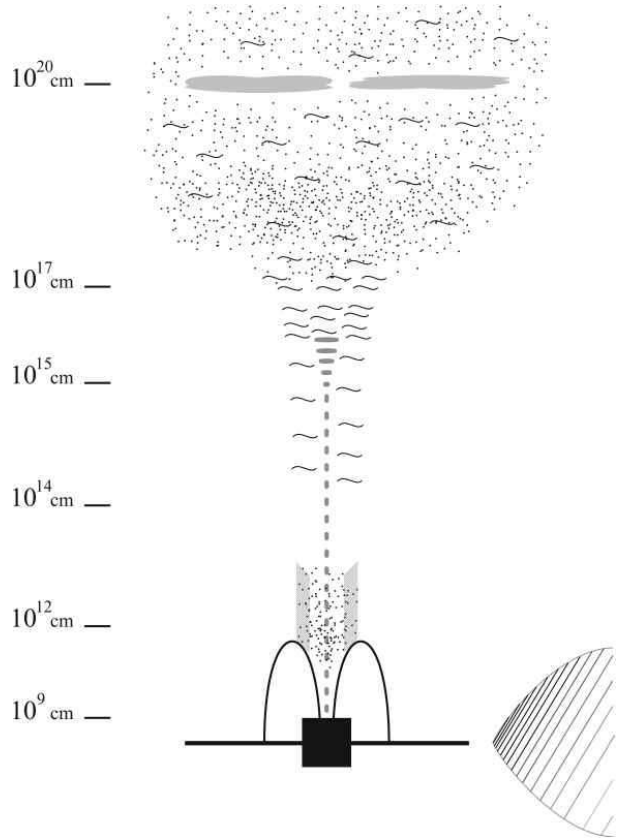,width=80mm}}
\caption{Schematic of the SS433 binary system and jets. The regions of
X-ray emission of the jets are shown by points, and the regions of radio
emission by wavy lines. The solid curves mark the photosphere of the
slow wind. The cocoons of hot gas above the wind photosphere are shaded.
The optical filaments at the ends of the large-scale jets ($\sim 10^{20}$~cm)
are also shown. Note that the jets on scales of $\sim 10^{12-13}$~cm are
the main source of the X-ray radiation. The black square shows a region
$\sim 100 R_g$ in size in the supercritical accretion disk. The outer edge
of the accretion disk and gaseous flows are not shown in this diagram.
}
\end{figure}

Figure~13 presents a schematic of the binary system and the jets
  of SS433.  The vertical scale indicated in the figure is
logarithmic, but the scale itself is not continuous, since the
purpose of the figure is to illustrate the main components of
the jets described above. Regions of X-ray emission of the jets
are shown by dots, and regions of radio emission by wavy lines.
The cocoons of hot gas above the wind photosphere are shaded
(see ``The Supercritical Accretion Disk from Spectroscopic
Data''). The optical filaments at the ends of the large-scale
jets ($\sim 10^{20}$~cm) are also depicted. The jets are not
observed below the wind photosphere, but they are the main
source   of X-ray emission on scales $\sim 10^{12-13}$~cm. The
filled square shows the   region $\sim 100 R_g$ in size in the
supercritical accretion disk that has been studied using
hydrodynamical simulations (see below).  The schematic does not
show either the outer edge of the accretion disk or gas flows,
which in reality have an appreciable impact on the brightness of
the system.

\bigskip

\subsection{Heating of the Jets}

The gas in the optical jet must be continuously reheated, since the
radiative cooling time for the gas in the clouds is several orders of
magnitude shorter than the time of their motion. The heating of the
H$\alpha$ clouds by UV and X-ray emission arising during the dissipation
of internal shocks (Fabian and Rees 1979) requires dispersion velocities
in the shocks of $\Delta V \sim 100$~km/s (Begelman {\it et al.\/} 1980) in order
to provide agreement with the observed relative intensities of the helium
lines in the spectra. However, this heating mechanism is improbable, as,
in addition, it should lead to the appearance of an appreciable X-ray flux
(that is not eclipsed, since the size of the jets is two to three orders
of magnitude larger than the size of the binary system). In particular,
if we recalculate the estimated X-ray luminosity obtained by Begelman
{\it et al.\/} (1980) taking into account the now known distance, jet opening angle,
and jet luminosity in the H$\alpha$ line, $L_{H\alpha} \approx 1 \cdot
10^{36}$~erg/s in a frame that is comoving with the jets (Panferov and
Fabrika 1997), the predicted jet luminosity at 2--10~keV is $\sim
10^{36}$~erg/s, which is approximately an order of magnitude higher than
is acceptable.

In principle, heating of the
jets by the collimated radiation coming from the accretion disk funnel
is possible (Bodo {\it et al.\/} 1985; Fabrika and Borisov 1987; Panferov
and Fabrika 1993), but this is always treated as being hypothetical, since
we cannot see the direct radiation from the funnel due to the orientation
of the SS433 system. Heating of the H$\alpha$ clouds by UV radiation is
unlikely, since the clouds should be uniformly heated, and the
side of the clouds nearest the source would not be brighter than the forward
side. Therefore, in the case of UV heating, the radial density of the clouds
must be $N_H < 10^{18}$~cm$^{-2}$, while analysis of the
hydrogen-line intensities implies hydrogen densities for the
H$\alpha$ clouds $N_H \sim 10^{21}$~cm$^{-2}$. For this same
reason, in the case of heating by collimated X-ray radiation,
the luminosity of the funnel at energies $\epsilon > 1.5$~keV
must be no lower than $L_{x,c} \ga 3 \cdot 10^{39}$~erg/s, in
order to provide the required H$\alpha$ flux (Panferov and
Fabrika 1993, 1997).

The possibility of heating of the jets by collimated radiation from the funnel
raises intriguing possibilities for investigating the directional beam of
the funnel radiation. The precessional and nutational motions of the axis
of the collimated radiation result in the ``slow'' jet (compared to the
speed of light) exiting from the light-cone directional beam at some distance
from the source. This will diminish the intensity of the heating, which
could be detected via analysis of the profiles of the moving lines. In
particular, the model profiles of
  the H$\alpha^-$ line shown in Fig.~7 were
obtained precisely under the assumption that the gas in the jet is heated
by collimated radiation (Panferov and Fabrika 1993). The beam of the
radiation was represented both as a two-dimensional Gaussian and as a flat
function (the radiation intensity does not depend on direction within the
cone). It was concluded that, in the case of heating by collimated radiation,
the total opening angle of the collimation cone was no less than
$\theta_c > 14^{\circ}$. However, the profiles of the moving lines proved
to be insensitive to the form of the directional beam for the collimated
radiation. It is known that the H$\alpha^{\pm}$ intensity falls off
exponentially along the jet (Borisov and Fabrika 1987), and that this is
probably determined by the gradual variation in the size of the clouds and in
the emissivity of the gas along the jet. Moreover, there are direct data
indicating that the dominant source of heating of the gas making up the
clouds is interactions of the jets with the gas of the slow wind. Therefore,
if the clouds are heated by collimated radiation as well, the contribution
of this heating to the overall heat balance of the gas is not the dominant
one.

Heating of the H$\alpha$ clouds via interactions of the jets with the
surrounding gas (Davidson and McCray 1980; Begelman {\it et al.\/} 1980; Brown
{\it et al.\/} 1991; Panferov and Fabrika 1997) is currently the most promising
mechanism. The parameters of the gas clouds in the jets will also be
determined by interactions of the jets and wind. This mechanism is in
good consistency with the observational data~-- the dependence of the
intensity of the jet line radiation on the precessional phase and the
anisotropy of the
cloud radiation in the hydrogen lines. This mechanism is based on the
necessary condition that the jets propagate through the accretion-disk
wind, and makes it possible to understand the length of the optical jets
and the formation of the radio brightening zone.

In their article based on the earliest spectroscopic data for SS433,
Davidson and McCray (1980) derived surprisingly precise values for the
main parameters of the jets and gas clouds: the electron density
($n_e \sim 10^{13}$~cm), temperature ($T_e \sim 1.5 \cdot 10^{4}$~K),
size ($l \sim 10^8$~cm), and volume filling factor for the
jet clouds. The requirement that the jet volume not be completely filled
with radiating gas follows from energetic considerations. In order for the
mass of the jets (and the kinetic-luminosity flux) to not be unreasonably
high, the emissivity of the radiating gas (which is $\propto n_e^2$) must
be rather high, and the clouds must not be too opaque to the line radiation
($\tau \propto l n$). The highest line emissivity of the gas is reached
when there are a large number of relatively small and dense clouds. Davidson
and McCray (1980), as well as Bodo {\it et al.\/} (1985), proposed that cool clouds
are dissipated due to their interaction with the hot ($\sim 10^{8}$~K)
intercloud gas that forms in the jets during their interaction with the wind.
Brown {\it et al.\/} (1991) considered various mechanisms for heating the gas in
detail, and concluded that heating via collisional interaction between the
jets and wind was most likely. In this case, the line radiation of the gas
should be linearly polarized (Brown and Fletcher 1992). The direction of
the polarization is orthogonal to the jets, and the expected degree of
polarization is $\approx 0.2~\%$.

It is improbable that each individual cloud radiating in the H$\alpha$
line experiences dynamical pressure from the gas incoming at the speed of
propagation of the jet. Interactions at a relative velocity of 0.26\,c
could easily lead to overheating and dissipation of the clouds. It appears that
a more complex interaction is realised, when the first set of clouds to
pass through a region sweep out the wind gas, possibly becoming disrupted
in the process. The jets create a channel, which moves in
the wind gas in accordance with the nutational and precessional motions.
The surrounding gas could become entrained in the jets and flow into them.
It is possible that shock waves or density waves propagate along the jets
in the reverse direction with relatively low speeds, heating the clouds
and preventing them from expanding. The parameters of the optical clouds,
probably, as well as those of the fragments into which the X-ray jets are divided
(Brinkmann {\it et al.\/} 1988), will be determined by complex processes occurring
over the course of the evolution of the jets.

Based on long series of spectral observations, Kopylov {\it et al.\/} (1986) derived
a limit on the deceleration of the jets of $\Delta V_j/ V_j \la 10^{-2}$
over 3--4 days in the lifetimes of the blobs. The loss of jet kinetic energy
corresponding to this deceleration is
$$
 \hat L_k = 2 L_k \Delta V_j/V_j
\la 2 \cdot 10^{37}\,L_{k39}~\mbox{erg/s},
$$
 where $ L_{k39}$ is the kinetic luminosity
in units
  of $10^{39}$~erg/s. Given the H$\alpha$ luminosity of the jets
$\sim 10^{36}$~erg/s, we find that the efficiency of cooling the gas via
H$\alpha$ radiation
  is $\ga 0.05$, which is quite acceptable (Begelman
{\it et al.\/} 1980; Panferov and Fabrika 1993).

Based on the condition that momentum be conserved,
$$\Delta\dot M/\dot M=\Delta V_j/V_j,$$
in order to satisfy the limit on the
jet deceleration, the density of the decelerating gas must be
$$\hat n = 8 \Delta V_j L_k / \pi m_p \sin^2 \theta_j R_j^2
V_j^4 \la  1.5 \cdot 10^{5}\,L_{k39}\,\mbox{cm}^{-3}.$$
It is
obvious that the source of this gas must be the wind flowing
from the accretion disk of SS433. The disk loses gas at a rate 
$\dot M_w \sim 10^{-4} \,M_{\sun}$/yr, and the wind speed in
circumpolar regions
  is $V_w \ga 1\,500$~km/s (Fabrika 1997). If
this flow were isotropic,   the
density of the gas through which
the jet passes would be
$$\hat n = \dot M_w / 4\pi R_{j}^{2} V_w
m_p \approx 4\cdot 10^6 \,\mbox{cm}^{-3}.$$
 Naturally, this outflow
from the SS433 disk should be anisotropic, with the wind in
circumpolar regions having higher speeds and lower densities.


\bigskip

\subsection{Ejection of Gas in the Jets}

In the first years after the discovery of SS433, a number of authors (see,
for example, Calvani and Nobili 1981) considered a model in which the gas
was accelerated by radiation pressure and the flow was collimated into
jets within the funnel of a thick accretion disk. Thermal instabilities
in the cooling gas of the jets can naturally explain the formation of clumps
of gas (Davidson and McCray 1980; Bodo {\it et al.\/} 1985), which are further
observed as H$\alpha$ clouds in the optical jets. Alternative approaches
to the problem of the non-uniformity of the optical jets are possible.
For example, Brown {\it et al.\/} (1995) proposed the existence of ``radiative
instabilities'' in the jets, when the cool gas radiating in the H$\alpha$
line does not form at all in isolated large-scale sections of the jets
due to variations in the balance of heating and cooling processes. Now,
in the light of X-ray studies of the jets, it is clear that the hot base
of the jets is as non-uniform as the optical jets, and the need for such
models is removed. However, this suggestion of Brown {\it et al.\/} (1995) remains
very pertinent, and some ``SS433-like'' objects may not show prominent jet
activity at all. Precisely the presence of a funnel in the supercritical
accretion disk, in which the gas is not only accelerated, but also compressed
into a thin stream, creates the conditions required for subsequent
fragmentation of the jets. The presence of a disk wind with which the jets
subsequently begin to interact due to their precessional and nutational
motions creates unique conditions enabling these fragments to survive
over the entire extent of the optical jets.

Bodo {\it et al.\/} (1985) considered the acceleration of the gas in the funnel
of a thick accretion disk (Jaroszynski {\it et al.\/} 1980; Rees {\it et al.\/} 1982;
Ferrari {\it et al.\/} 1985; Icke 1989). The funnel and disk are separated by
the walls of the funnel, which is a dynamical structure. Under the action
of a number of effects, the wall material can enter the funnel and be
efficiently accelerated outward by radiation pressure. The structure of
the accretion-disk funnel has been considered in a number of studies
(Lynden-Bell 1978; Sikora 1981; Narayan {\it et al.\/} 1983), as a rule,
in the context of accretion disks around supermassive black holes. The
funnel in a thick disk can naturally give rise to collimated radiation.
Taking into account the effects of reflection of the radiation off the
walls (Madau 1988) enhances the degree of collimation of the radiation
exiting along the funnel axis. If we allow for the scattering of the wall
radiation on the rarefied gas in the funnel ($\sigma_T \sim 1$), the degree
of collimation of the radiation could prove to be rather high, especially
if this gas moves outward along the funnel axis with the velocity
$v_j \sim 10^{10}$~cm/s.

Via hydrodynamical calculations of supercritical disk accretion
onto a black hole in a binary system assuming conditions close to those
for SS433, Eggum {\it et al.\/} (1985, 1988) concluded that the funnel
with its dynamical walls is formed in the inner regions, at
several tens of gravitational radii. The opening angle of the funnel
turns out to be $\theta_c \sim 30^{\circ}$. The optically thin gas in
the funnel is accelerated by radiation pressure to velocities $\sim
10^{10}$~cm/s. The walls consist of accreting gas, and bound the region
of the funnel (photocone of the collimated radiation) from the region of
convective plasma motions. Up to 80~$\%$ of the released gravitational
energy is accreted onto the black hole in the form of a kinetic energy
flux and radiation.
  About 1~$\%$ of the accreting material is ejected
in the funnel (in the jets), yielding a value that is close to the relative
mass-loss rate
  in SS433 jets, where $\dot M_j/\dot M_0 \sim 5 \cdot
10^{-7}/10^{-4} \sim 0.5~\%$ of the mass reprocessed by the disk is lost
in the jets. Computations have been done for accretion rates $\dot M/\dot
M_{crit} = 1-10$ (where $M_{crit}$ is the critical accretion rate required
to produce the Eddington luminosity),
while the accretion rate in SS433 reaches $\sim 10^3$.
A more precise comparison of computational results with the observed
picture for SS433 requires appropriate specification of the rate at which
mass enters the outer edge of the disk, the mass-loss rate on scales of
the spherisation radius, and, accordingly, the accretion rate in inner
parts of the disk, which are difficult to determine on the basis of
observational data. Overall, the computations of Eggum {\it et al.\/} (1985, 1988)
illustrate a basic scheme according to which we can understand the mechanism
for the ejection of gas and formation of jets in SS433.

The two-dimensional hydrodynamical computations of Okuda   (2002, and
references therein) support the mechanism of Eggum   {\it et al.\/} (1985,
1988) for the formation of the funnel and dynamical walls in inner regions
from several tens to hundreds of gravitational radii, where the radiation
pressure exceeds the gas pressure. The rarefied gas is accelerated in the
funnel to velocities of 0.1--0.2\,c. The structure of the funnel can be
very complex. The presence of convectively unstable zones located in a
broad torus-like region on the other side of the dynamical walls of the
funnel is also confirmed. Close to the black hole, there are advective
plasma motions. Okuda (2002) proposed that the acceleration of the gas
in the funnel could be stabilised by some mechanism which, under the
conditions leading to the given funnel structure, results in a constant
outflow speed of 0.26~c.

Comparatively recently, the significant role of convection has been
recognized in gas dynamics and energy transfer in the inner regions of accretion
disks (Stone {\it et al.\/} 1999; Abramowicz {\it et al.\/} 2002 and references therein).
The convective motions can form very powerful gas outflows from the
accretion disks, while, at the same time, the accretion flow is advective
very close to the black hole. A strong wind from the accretion disk could
also been formed because of standing shocks arising near a centrifugal barrier
in the disk (Molteni {\it et al.\/} 1994; Chattopadhyay and Chakrabarti 2002).
The simulations of Molteni {\it et al.\/} (1994) show that the disk wind
propagates not only in circumpolar regions of the disk, but also in
quite far directions from the disk axis ($\theta \sim 50^{\circ} - 60^{\circ}$).
In particular, the very same wind structure is observed in SS433
(Fabrika 1997).

Hydrodynamical computations of supercritical disk accretion are very
promising for improving our understanding of the mechanisms acting in SS433.
It is likely that, in the near future, computations extending to scales
$\sim 10^9 - 10^{10}$~cm, encompassing the spherisation radius of the disk,
will be carried out, enabling the elucidation of how the supercritical disk
wind forms. The spherisation radius of the accretion disk is (Shakura and
Sunyaev 1973) $R_{sp}= G M_x \dot M_{0} / L_{e}$, where $\dot M_{0}$ is
the rate at which gas enters the disk, and $M_x$ and $L_{e}$ are the mass
of the compact star and its corresponding critical luminosity. If the
accretion rate at the outer edge of the disk is $\sim 10^{-4}\,M_{\sun}$~/yr,
the spherisation radius is $R_{sp} \sim 10^{10}$~cm. Computations of the
funnel and wind on larger scales, $\sim 10^{11} - 10^{12}$~cm, may make
it possible to elucidate how the jets are collimated.

A number of models have been constructed for the formation of the jets
and radiation of SS433 both in the funnel of a thick accretion disk and
in a channel in the wind or gaseous envelope (Fukue 1987ab; Inoue
{\it et al.\/} 2001), where the channel has shapes varying from conical to
those in which the cross section of the channel grows with distance
more slowly, $S \propto r$. Models with a channel or funnel in the
accretion disk (or in material flowing outward from the disk) and collimated
radiation in this channel are very appropriate for SS433.

It is interesting that models with a supercritical accretion disk around
a neutron star in which the star does not possess a strong magnetic field
(Okuda and Fujita 2000) give approximately the same results, since the
funnel (as in the case of a black hole) forms around the rotational axis
of the disk. An obvious difference is
the accretion rate onto the surface of the neutron star and, accordingly,
the rate at which gas is ejected from the inner regions of the disk, since,
in contrast to a black hole, a neutron star cannot accept more material at
its surface than the amount determined by the critical
  accretion rate.

Several mechanisms for the ejection of gas and formation of jets for the
case of a neutron star with a strong magnetic field have been proposed.
Lipunov and Shakura (1982) considered supercritical accretion onto a
slowly rotating neutron star, where gas in the magnetosphere falls
onto the surface along force lines and is ejected in clumps along the
magnetic poles of the star with characteristic time intervals equal to the
free-fall time for blobs of gas arriving from the magnetosphere. In the
cauldron model of Begelman and Rees (1984), the jets are accelerated
when they pass through the magnetopause, as though through narrow de
Laval nozzles stabilised
by the pressure of the neutron star's magnetic field. As in the previous
model, an appreciable source of energy is required for the initial
acceleration of the gas in the magnetosphere, which could plausibly be
provided by the non-stationary accretion of blobs of gas or the rapid
rotation of a young neutron star with an oblique magnetic field.

Arav and Begelman (1992, 1993) developed a model for the
acceleration and collimation of gas in SS433 in which the jets
are ejected relatively close to the neutron-star surface and
propagate in a channel formed in the dense atmosphere created by
the accreting gas. The boundary layer of this
radiation-dominated channel in the atmosphere and the evolution
of the channel and jet with distance from the source were
considered in detail.  The boundary layer is represented by a
cocoon of low-density gas around the jets. It was demonstrated
that the channel and cocoon exhibit collimating properties and
efficiently collimate the radiation from the central regions.
This radiation-dominated channel model could prove to be very
useful for our understanding of the properties of channels in
supercritical accretion disks and jet flows within them. In
spite of the number of excellent models for SS433 as an object
containing a neutron star, the presence of this one in SS433 is
very improbable (see the next section).

\bigskip

\subsection{Acceleration, Collimation, Fragmentation}

The most promising models are those with the acceleration and collimation
of gas in the funnel of a supercritical accretion disk around a black hole.
Bodo {\it et al.\/} (1985) initially assumed that the radiation in the funnel was
collimated, the gas was optically thin and the main mechanism scattering
the radiation was Thomson scattering. Hot gas diffuses inside the funnel
and is accelerated by the existing collimated radiation. The accelerated
material acquires its essentially final speed while still deep inside the
channel, at distances $r \la 10^9$~cm, and a supersonic gas flow with a
temperature $T \ga 10^7$~K exits the funnel. Depending on the total
luminosity of the radiation coming out of the
funnel and the funnel opening angle,
the final speed of the material is $\beta = 0.1-0.6$.
In particular, for approximate values of the luminosity in units of the
Eddington luminosity and the
funnel opening angle appropriate for SS433,
$L_c/L_e = 10$ and $\theta_c = 30^{\circ}$   and $40^{\circ}$,
outflow speeds of $\beta = 0.35$ and $0.25$ are reached.

If the following mechanisms:
\begin{itemize}
\itemsep=-2pt
\item[(i)] finer tuning of the velocity of motion to the value
0.26~c,
\item[(ii)] collimation of the flow along the propagation axis,
\item[(iii)] the formation of dense clumps of gas along the axis
($\theta_j \sim 0.02$),
\end{itemize}
\noindent
act in an already accelerated flow of gas (for example, within the funnel in its upper part or in the region where the material exits the funnel), then, generally speaking, we will obtain jets like those
  in SS433. We consider each of these points in more detail below.

(i) Models for the acceleration of the gas in the SS433 jets by radiation
pressure propose the action of ``line-locking'' (Milgrom 1979b;
Pekarevich {\it et al.\/} 1984; Shapiro {\it et al.\/} 1986; Katz 1987). This mechanism
was actively discussed at the beginning of the 1970s in connection with
the acceleration of gas in envelopes surrounding active galactic nuclei
and the formation of absorption lines in quasars. In the case of SS433,
the observed jet velocity
of 0.26\,c corresponds to the Doppler shift of
the frequency of the main L$\alpha$ transition of the hydrogen atom to the
frequency of the ionisation threshold, Lc. If acceleration by radiation
pressure occurs via the absorption
of photons in L$\alpha$ frequencies and there is an
absorption edge Lc in the source spectrum, gas can be efficiently accelerated
only to velocities $\approx 0.26\,c$, since the source radiation intensity
falls off sharply beyond the Lc edge. This mechanism operates equally
effectively in locking the K$\alpha$--Kc radiation of hydrogen-like and
helium-like ions.

Investigations of the line-locking mechanism as applied to SS433 were
carried out primarily before the discovery of the X-ray jets, and,
accordingly, before it became clear that the main lines in these jets
belong to hydrogen-like and helium-line ions of heavy elements. Therefore,
consideration of heavy elements in these works had a somewhat speculative
nature, which was fully justified in subsequent years. It was concluded that
line-locking could be efficient in SS433, but with the additional
proviso that either there was an enhanced content of heavy elements
(such as iron) in the SS433 jets, or the jets were accelerated by collimated
radiation. This second hypothesis now appears very natural.

The efficiency of line-locking under the conditions in SS433 has been
placed under doubt more than once, but, as a rule, on the basis of
substantially overestimated values for the kinetic luminosity of the jets.
Nevertheless, it is difficult to imagine another acceleration mechanism
providing the striking constancy of the jet velocity in SS433. The jet
velocity does not depend on anything: the velocity remains constant, even
in the presence of variations in the luminosity of SS433 during flares
(by nearly an order of magnitude), in active states or in times of
appreciable brightness decreases. The jets sometimes disappear (see the
section ``The Radio Jets and W50''), but they always reappear with the
same velocity, 0.26\,c. More generally, one can say that, when the SS433 jets
contain cool gas clouds emitting in the hydrogen lines, the velocity of
the jets is always the same.

The surprising constancy of the jet velocity provides a very strong argument
in support of the line-locking mechanism. Even if the acceleration of the
jets is not entirely due to this mechanism, it seems very likely that at
least the tuning of the flow velocity to the observed value must be due to
this effect. If the limiting velocity
  of 0.26\,c is approached not from below,
via the acceleration of material in the funnel, but from above, via the
deceleration of material, line-locking can stop the deceleration at the
required velocity, since the gas begins to be accelerated by Lc photons when
the velocity is decreased to the critical value $V_j=0.26$\,c. If the flow
of gas in the funnel accelerated in the inner regions to velocities
$\sim 10^{10}$~cm/s begins to be decelerated by the wind coming in from
the funnel walls (the gas in this wind can adhere to inhomogeneities in
the main flow), this deceleration is sharply slowed at the threshold velocity
0.26\,c by the effect of line-locking. The mechanism for the formation of
instabilities in material accelerated by radiation pressure proposed by
Katz (1987) could sharply increase the efficiency of the formation of
inhomogeneities in the accelerated flow.

(ii) One candidate for the collimation mechanism could be hydrodynamical
collimation of the flow (Peter and Eichler 1996) due to interaction with
the walls in the upper part of the funnel or with the walls of the gas
cocoon surrounding the place where the jets emerge. Indicators of such a cocoon
in SS433 would be the fluorescence of a ``Fe\,I gas'' made up of weakly
ionised iron, manifest in the X-ray spectra (previous Section),
and evidence for a He\,II cocoon in the observed optical spectra
(``The Supercritical Accretion Disk from Spectroscopic Data'').
This same hydrodynamical collimation mechanism has been proposed to explain
the narrowness of the extended X-ray jets associated with the W50 nebulosity
(``The Radio Jets   and W50'').

The flow of gas in the funnel should interact with the walls, consisting
of gas from the slow, dense wind. A wind may blow from the walls and collide
with the main flow. In addition, due to the development of hydrodynamical
instabilities such as Kelvin--Helmholtz instabilities, inhomogeneities
(waves) should form in the walls and be carried away by the main flow.
Interaction with the inhomogeneities and with the wind from the walls
leads to the formation of axially symmetrical shocks in the flow, which
move along the main flow of gas toward the funnel axis and are carried by
the flow. The velocity of waves perpendicular to the motion of the main flow
is equal to or slightly higher than the sound speed, $c_s \sim 10^8$~cm/s
for a temperature $T \sim 10^8$~K (the temperature inferred for the base
of the jets using ASCA and CHANDRA data). In spite of the fact that the flow
itself is substantially supersonic, $V_j/c_s \sim 100$, a shock wave
propagating along the flow toward the axis could be weak. Thus, a ``soft''
oblique shock could propagate along the flow, compressing gas along the
axis of the motion. The collimation angle of this compressed region is
$2c_s/V_j \sim 0.02$ (or somewhat higher, since the temperature inside the
funnel should be higher than the temperature inferred from the observed
base of the X-ray jets), i.e., the gas is compressed into a narrow jet.
The question is whether thermal instabilities are capable of confining
the gas in this jet via the sharp decrease in the pressure due to the
radiative cooling, or whether the compressed gas on the funnel axis remains
too hot to form dense clouds. Nevertheless, the gas on the channel axis should
cool very efficiently due to its enhanced density, and its temperature should
fall. In any case, if the temperature at the exit from the funnel is
lowered to the measured value $T_{j,x} = 1 \cdot 10^8$~K, the jets cannot
expand for kinematic reasons, and are ``frozen'' at the level $\theta_j =
1\degdot4$, as discussed in the previous Section. Further, the thermal
instabilities can divide the jets into dense clumps, but this will occur
inside the cone $\theta_j$ and after the exit from the funnel, when the
gas temperature is lowered to $\sim 10^6$~K by adiabatic expansion and
radiative losses in the X-ray jets. As far as we are aware, no hydrodynamical
computations of the motion of the flow in the upper part of the funnel
have been carried out.

\renewcommand{\thefigure}{14}
\begin{figure}[t]
\centerline{\psfig{figure=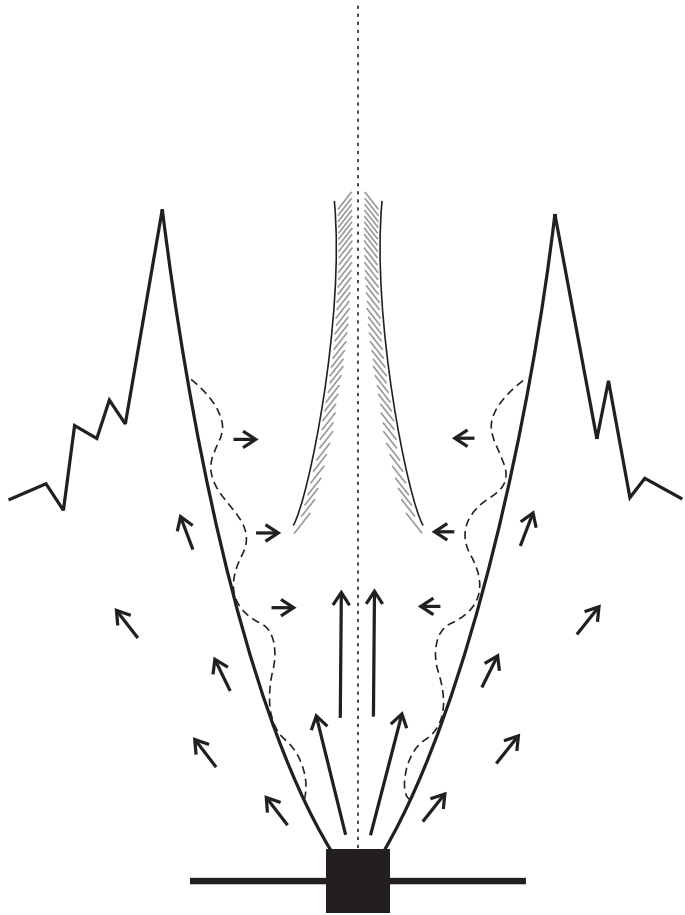,width=90mm}}
\caption{Schematic of the funnel in the disk wind. As in
Fig.~13, the black square denotes the region in which
hydrodynamical modelling of the supercritical disk has been
carried out (Eggum {\it et al.\/} 1985, 1988; Okuda 2002). The
bold curve denotes the photosphere of the slow wind. The dotted
curve shows inhomogeneities arising in the walls of the channel.
Proposed oblique shocks compressing gas into a thin ($2c_s/V_j
\sim 0.02$) stream are shaded.}
\end{figure}

Lebedev {\it et al.\/} (2002) presented results of laboratory 
experiments, where hypersonic plasma jets are generated in conically
convergent flows. The convergent flows are created by electrodynamic
acceleration of plasma in conically divergent array of fine metallic 
wires (Z--pinch array). Stagnation of plasma flow on the axis of 
symmetry forms a standing conical shock effectively collimating
the flow in the axial direction. In the experiments a hypersonic
(M~$\ga$20) well--collimated jet is generated. The jet collimation 
depends obviously on radiative cooling rate of the shocked plasma 
on the axis of symmetry. The experiments show high efficiency of the
discussed collimation mechanism, the geometry of the device (Lebedev 
{\it et al.\/} 2002) is very similar to that of the funnel in SS433, 
which we discuss. The jets formation by the convergence of supersonic 
conical flows was considered and developed by Cant\'{o} {\it et al.\/} 
(1988) in application to formation of interstellar jets. 

Figure~14 shows a schematic of the funnel in the disk wind illustrating
the features discussed above. As in the previous figure, the filled square
denotes the region in which hydrodynamical modeling of the supercritical
accretion disk has been performed (Eggum {\it et al.\/} 1985, 1988; Okuda 2002).
The bold curve denotes the photosphere of the slow wind. Inhomogeneities
arising in the funnel walls can appear on the boundary of the interaction
of the fast flow and slow wind. The proposed oblique shock waves compressing
gas into a narrow ($2c_s/V_j \sim 0.02$) jet are shaded.

At the gas compression on the axis of jet in the conically convergent
supersonic flows a generation of relativistic electrons is quite
probable. In principle, the mechanism of acceleration of the relativistic
particles may not differ on the particle acceleration mechanism in radio
jets (the Section ``The Radio Jets and W50''), where the electrons are
accelerated in shocks originating in interaction of the jets and the
disk wind. The relativistic electron energy needed for formation of the
hard X--ray spectrum in SS433 detected by INTEGRAL (Cherepashchuk 
{\it et al.\/} 2003) is $\gamma \sim 10$.

(iii) The fragmentation of the jets into dense clumps via the development
of thermal instabilities has been studied (Bodo {\it et al.\/} 1985; Brinkmann
{\it et al.\/} 1988). Bodo {\it et al.\/} (1985) found that, when it leaves the funnel
($r_{max} \la 10^{11}\,$cm), the gas begins to cool at distances $10^{11} \la
r \la 10^{12}$~cm. On short time scales much less than the time scale for
motion of the flow $r/V_j$, the thermal instabilities lead to the formation
of a two-phase medium: cool, dense fragments ($T\approx 10^4$~K,
$n\ga 10^{16}$~cm$^{-3}$) embedded in a hot medium ($T\ga 10^7$~K,
$n\approx 10^{14}$~cm$^{-3}$). The dimensions of the condensations
formed as a result of the instabilities are in the region of $l \approx
10^7 - 10^8$~cm, and are bounded from below by the thermal conductivity
and from above by the condition that a pressure equilibrium should be
established over the time for the development of the instability. Note that
these calculations were undertaken before the discovery of the X-ray jets
of~SS433 and the associated measurement of the parameters of the jet gas,
and they could be refined on the basis of the information now available.
The resulting scales for the clouds are very close to those derived from
analysis of the Balmer decrements of the jets (Panferov and Fabrika 1997),
$l \sim 10^8$~cm. Recall also that, according to the data of Marshall
{\it et al.\/} (2002), the temperature in the X-ray jets varies from $1.1 \cdot 10^8$~K
to $ 6 \cdot 10^6$~K from their bases to their ends (the accuracy of the
end values is limited by the weakness of the corresponding spectrum
and blending of lines at soft energies), and the electron density similarly
falls from $2 \cdot 10^{15}$ to $4 \cdot 10^{13}$~cm$^{-3}$.

In connection with the X-ray jets of SS433, Brinkmann {\it et al.\/} (1988) performed
a numerical study of the evolution of
a blob of material arising from initial density fluctuation
in a uniform, hot gas forming a conical jet. The jet
gas is in ionisation equilibrium via collisions and cools due to radiative
losses and expansion of the jets. In the linear stage of the thermal
instability, the blobs reach sizes similar to those obtained by Bodo {\it et al.\/}
(1985) based on a linear analysis. However, in the subsequent non-linear
stage, depending on a number of conditions, the blobs (clouds) can experience
both catastrophic compression and expansion. In this case, a simple stationary
state is not reached for any conditions, and the formation of clouds must
be considered as a dynamical phenomenon in the evolution of the jets. It is
certain that gas clouds with density contrasts $\ga 10^3$ form in the
cooling X-ray jets, and an additional mechanism to heat the gas in the
clouds is required in order to maintain them at temperatures $\sim 10^4$~K
over times $\sim 10^3-10^4$~s and to prevent collapse of the clouds. This
additional heating could be radiative (by the collimated radiation of the
funnel) or associated with the dissipation of shocks in the jet, radiative
heating by radiation of these shocks (Begelman {\it et al.\/} 1980; Fabian and
Rees 1979), or interaction of the precessing jets with the surrounding
medium (Davidson and McCray 1980).

Depending on the magnitude of this heating in the X-ray jet, there could
be substantial variations in estimates of both the physical conditions in
the gas and of the kinetic luminosity of the jets. In addition, the resulting
jet parameters depend strongly on assumptions about the jet structure. In
their model of the X-ray jet, Brinkmann and Kawai (2000) took into account
non-equilibrium effects and photoionisation of the gas, as well as the
non-uniform density distribution across the jet. In particular, in their model
with a narrow X-ray jet in which the density falls off with distance from
the jet axis in accordance with a Gaussian distribution ($\theta_j =
1\degdot4$) that is in best agreement with the observations, the estimated
kinetic luminosity
  is $L_k \sim 6 \cdot 10^{39}$~erg/s. This jet has a
rarefied, hot atmosphere.

 \bigskip

\section{The Supercritical Accretion Disk and the Components from Photometric
Data}

\subsection{The Light Curve of SS433: Precessional, Orbital and
Nutational Variability}

Extensive photometric observations have been obtained, and the results of
such observations can be found in various reviews (Margon 1984; Cherepashchuk
1989, 2002). However, we will be interested primarily in data that can be
used to draw conclusions about the SS433 accretion disk, or, more precisely,
about the shape and structure of the gaseous envelope that surrounds the
relativistic object. Much information about the structure of the disk and
gaseous flows has been obtained from spectral studies. Accordingly, we will
describe the available spectral data in more detail in the following section.

\renewcommand{\thefigure}{15}
\begin{figure}[t]
\vspace*{2mm}
\centerline{\psfig{figure=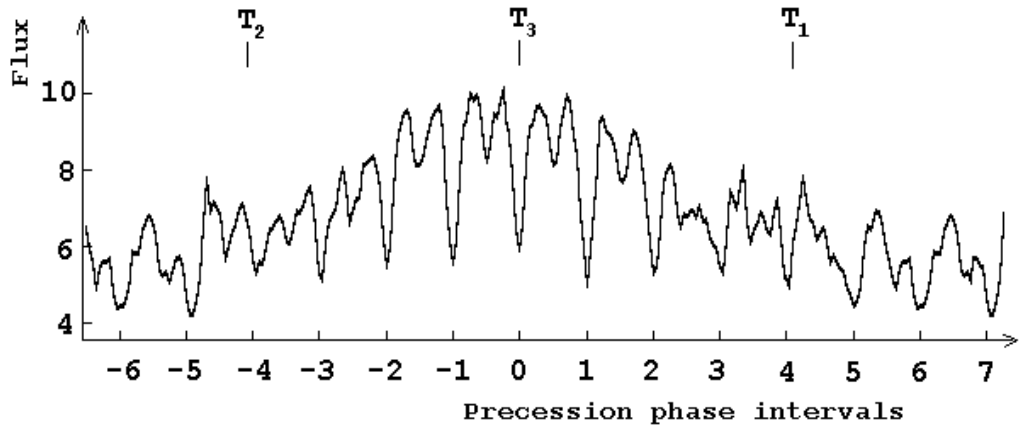,width=140mm}}
\caption{Flux density of SS433 in the $V$ filtre (mJy) over the precessional
cycle (Goranskii 2002). The precessional light curve is composed of mean
orbital light curves, and is presented in terms of fractions of the orbital
period.
}
\end{figure}

The light curve of SS433 is determined by three well-known periods~-- the
precessional, orbital and nutational periods~-- as well as by sporadic
variations~-- small-scale fluctuations, flares (described in the section
``The Radio Jets and W50''), and active periods (Leibowitz
{\it et al.\/} 1984; Kodaira {\it et al.\/} 1985; Kemp {\it et al.\/} 1986; Gladyshev {\it et al.\/} 1987;
Mazeh {\it et al.\/} 1987; Goranskii {\it et al.\/} 1987, 1997, 1998ab; Zwitter
{\it et al.\/} 1991; Fukue {\it et al.\/} 1997; Panferov {\it et al.\/} 1997;
Irsmambetova 1997, 2001).  Figure~15 shows the behaviour of the
$V$ flux density over a precessional cycle based on the data of
Goranskii {\it et al.\/} (1998b). These data were translated into
fluxes, averaged, and kindly presented to us by V.~P.~Goranskii (2002)
especially for this review. The precessional light curve is shown in fractions
of the orbital period; i.e., it is comprised of mean orbital curves obtained
at various precessional phases over all the years in which photometric studies
of SS433 have been undertaken. The orbital epochs when the accretion disk is
eclipsed and precessional epochs when the disk is maximally turned toward
the observer were
combined. In all, 2400 individual measurements were used.
When the accretion disk turns to face the observer during its precessional
rotation ($\psi =0$,
  time $T_3$), SS433 becomes brighter, the light curve
becomes more regular, and eclipses become deeper and more well-defined.
The depth of eclipses grows toward the blue, and appreciably decreases in
the red and infrared.


At precessional phases when the disk is viewed edge-on (Fig.~15) and the
jets lie in the plane of the sky, the light curve is very irregular, the
primary eclipses of the accretion disk by the star (Min\,I, $\phi=0$) become
shallow, and eclipses of the star by the disk (Min\,II, $\phi=0.5$) are
sometimes difficult to distinguish. At these precessional phases, the light
curve is very strongly distorted by flares (Goranskii {\it et al.\/} 1998b). Away
from eclipses in quadratures, the precessional variability amplitude in
the $V$ band is $\approx 0\magdot6$. At the centers of eclipses of the
accretion disk, the precession amplitude is lower, 0\magdot2--0\magdot3
(but is reliably determined);
SS433 also becomes brighter when the disk is turned toward the observer
and weaker when the disk is viewed
  edge-on. This led Goranskii {\it et al.\/}
(1998b) to conclude that, independent of the precessional orientation,
total eclipses of the disk
  in SS433 are never observed. The $B$--$V$ colour
varies little with the phases of the known periodic variations. When the
brightness decreases at precessional phases when the disk is viewed edge-on
or in eclipses, the object's $V$--$R$ colour reddens~-- the hot source is
eclipsed, while the red component of the emission (probably free--free
radiation of the gas surrounding the system) is not. Colour variations are
primarily observed during flares.

Such brightness variations with precessional phase could be associated with
(i) simple variations of the orientation of a geometrically flat source,
(ii) variations in the visibility of ``hot spots'' or funnels in the places
where the jets emerge in the photosphere of the outflowing wind (Lipunov
and Shakura 1982), or (iii) variations in the visibility of hot cocoons
surrounding the bases of the jets. In the last two cases, the ``disk''
could be a disk-like body whose thickness grows with distance from the center
and whose surface temperature grows toward the center. For example, roughly
this shape has been assigned to the accretion disk in some computations of
the SS433 light curves (Antokhina {\it et al.\/} 1992; Hirai and Fukue 2001). It is not
ruled out that the ``disk'' plays purely a screening function. The outer
edges of the disk can screen inner hot regions during the precession. In
addition, it is completely unclear whether the shape of this body (which
we traditionally call the disk) is determined by Keplerian rotation of
the disk material. It is more likely that the shape is determined
by the density, velocity and geometry of the wind, i.e., by the structure
of the wind photosphere. This means that the size of the bright body
surrounding the relativistic star could bear no relationship to the size of
the Roche lobe of this star (see below), which gives rise to serious
limitations when estimating the component mass ratio via modeling of the
light curves.

\renewcommand{\thefigure}{16}
\begin{figure}[p]
\vspace*{8mm}
\centerline{\hspace*{25mm}\psfig{figure=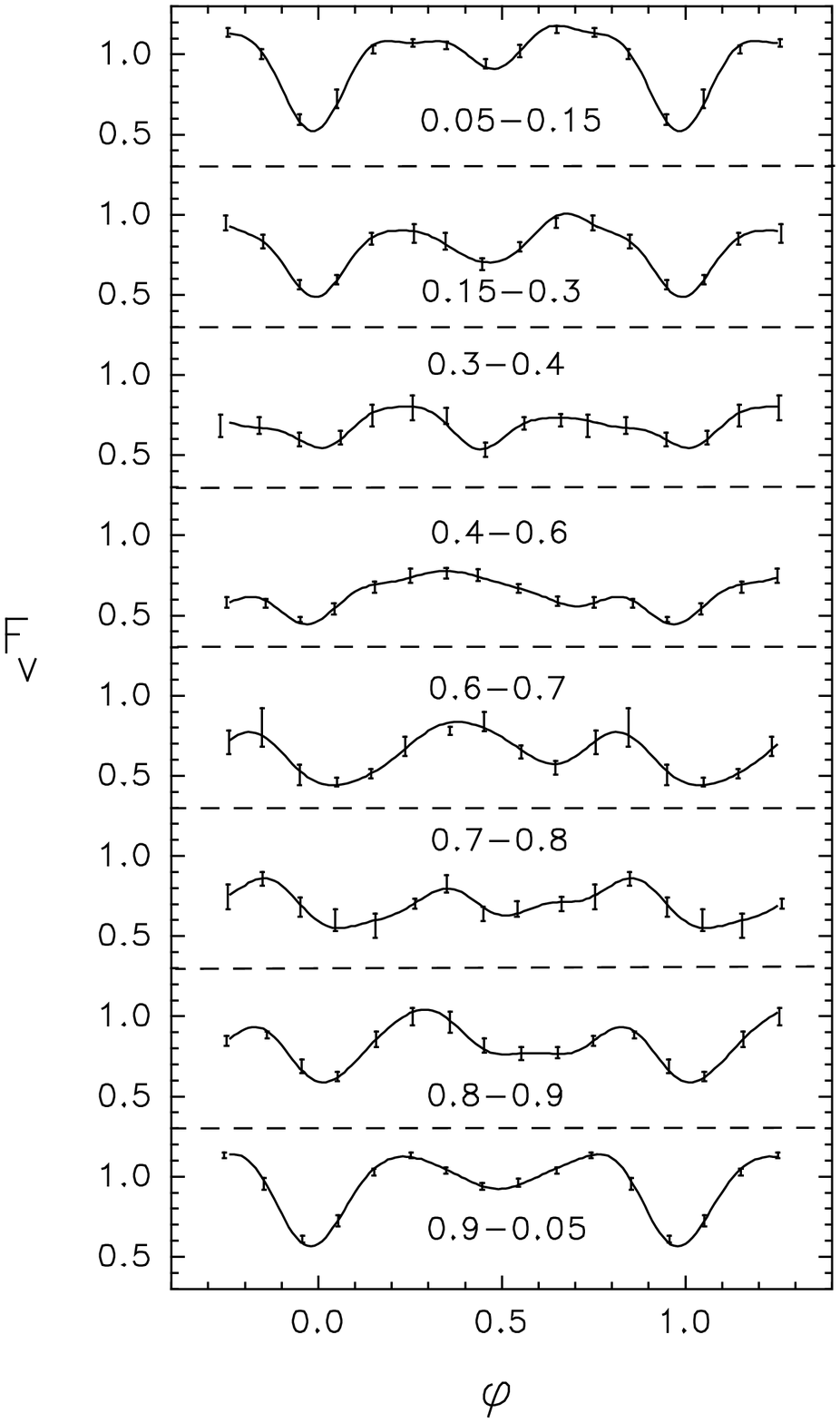,width=100mm}}
\vspace*{-5cm}
\caption{Orbital light curves of SS433 in the $V$ filtre averaged over
intervals of the precessional phase (Panferov {\it et al.\/} 1997).
}
\end{figure}

Figure~16 shows orbital light curves of SS433 in the $V$ band (Panferov
{\it et al.\/} 1997) obtained from all published data by averaging the brightness
measurements in intervals of the precessional phase. The flux $F_V=1$
corresponds to $V=14\magdot0$. The depth of the primary eclipses
Min\,I ($\phi = 0$)
varies appreciably as a function of the disk orientation; even their position
varies. If the orbital and precessional rotations have opposite directions,
as has been concluded in numerous studies and as should be true in the case
of driven precession of the optical star (a slaved disk), the primary eclipses
should lag the ephemerides before time $T_3$ ($\psi<0$) and lead the
ephemerides after time $T_3$ ($\psi>0$) in models with a bright source
coincident with the base of the approaching jet (models ii and iii above;
Fabrika 1984; Gladyshev {\it et al.\/} 1987).

The orbital light curves near the secondary eclipses Min\,II  
($\phi \approx 0.5$) are even more variable. If the effect of
reflection of the disk radiation from the surface of the donor
star or from densified regions in the wind near the donor
surface are sufficiently strong, then (again assuming the
directions of the orbital and precessional rotations are
opposite), at precessional phases before time $T_3$, the
brightness should be higher at elongation $\phi = 0.2-0.3$ than
at elongation $\phi = 0.7-0.8$, and Min\,II should lag somewhat
behind the computed value. At precessional phases after time
$T_3$, on the contrary, the brightness should be higher at
elongation $\phi = 0.7-0.8$ and~Min\,II should lead the computed
value. This is approximately the pattern observed in Fig.~16. In
addition, during the effect of heating, the maximum of the
precessional brightness measured separately at the phases of the
two elongations should also be shifted, due to the different
conditions for the heating and the visibility of the star from
the disk surface. More precisely, the precessional maximum
occurs earlier than the ephemerides time $T_3$ at elongation
$\phi = 0.2-0.3$, and occurs later than $T_3$ at elongation
$\phi = 0.7-0.8$.  This effect was noted by Gladyshev {\it et
al.\/} (1987).

The presence of eclipses Min\,II in the X-ray (Gies {\it et
al.\/} 2002a) may be associated with the reflection of light
from the donor star, but could also represent eclipses of a
region near the optical star if a modest fraction of the X-ray
radiation is formed there in shocks in the colliding winds
(Cherepashchuk {\it et al.\/} 1995). It seems reasonable to
suppose that the effect of heating of the donor surface (or of
perturbations in the wind) plays an important role in the
brightness variability of SS433, but modeling of the light
curves must be carried out to conclusively demonstrate that this
is the case.

Apart from heating,
two more effects that could distort the regular orbital variability are
known. The first of these is nutational variations of the brightness, which
distort the orbital variability in different ways at different precessional
phases (Goranskii {\it et al.\/} 1998b). The second effect is absorption in the
gas flowing out from the Lagrange point L2 behind the disk (Zwitter
{\it et al.\/} 1991; Fabrika 1993).
We will discuss these effects~below.


The third reliably established period in the optical brightness
variations is the nutational period of 6.28~days~-- the period
with which nodding of the jets is observed. The total amplitude
of the nutational brightness variations is $\Delta V =
0\magdot17$ (Mazeh {\it et al.\/} 1981, 1987; Goranskii {\it et
al.\/} 1998b). The additional source of red light (in the $R$
band) does not show nutational variations; i.e., the amplitude
of the variations of the blue source must be slightly higher
than   in the~$V$~band.

Here, we present ephemerides of the photometric variability with the
periods of the precession $P_{pr}$ (time of brightness maximum $T_3(phot)$),
orbital motion $P_{orb}$ (the center of the eclipse of the accretion disk
Min\,I) and nutation $P_{nut}$ (maximum brightness in the $V$ band) based
on the data of Goranskii {\it et al.\/} (1998b), which are the most recently
published. We repeat here spectral precession ephemerides (maximum separation
of the relativistic emission lines $T_3(spec)$).  
These ephemerides may prove useful in computations and planning
of observations.
$$
T_3(phot)=\mbox{JD}\,2450000+162\daydot15\cdot\mbox{E}
$$
$$
T_3(spec)=\mbox{JD}\,2443507\daydot47 + 162\daydot375 \cdot\mbox{E}
$$
$$
\mbox{Min\,I}=\mbox{JD\,}2450023\daydot62\pm0\daydot26+
(13\daydot08211\pm 0\daydot00008)\cdot\,\mbox{E}
\vphantom{X_{X_X}}
$$
$$
\vphantom{X^{X^X}}
\mbox{Max}=\mbox{JD\,}2450000\daydot94+6\daydot2877
\cdot\,\mbox{E}
$$

The orbital elements are the most accurate at the end of 2001, and were
confirmed after publication (Goranskii {\it et al.\/} 1998b) by recent observations
of eclipses (Goranskii 2002). When the ptotometric elements of the precessional 
period are compared with the spectral precessional elements 
(Eikenberry {\it et al.\/} 2001; ``The Optical Jets''), they are in reasonable
agreement given at the instabilities and inaccuracies that are usual for
precessional clocks. The accuracy of the spectral precessional period
should naturally be higher than that of the photometric precessional period,
since the amplitude of the shifts of the moving lines is much larger than
the uncertainties in the positions of lines in the spectrum (by a factor of
$\sim 10^3$). In photometric observations, the amplitude of the precessional
variability does not exceed the typical measurement errors by such a large
factor. Nevertheless, we study the precession of the accretion disk via
photometry, while we observe the motion of the jets via spectroscopy, so
that a comparison of the precessional motions of the jets and disk may have
fundamental value.

\bigskip

\subsection{The Light Curve in Active and Quiescent States}

The light curve in Fig.~15 was obtained by averaging all data in both
active and quiescent states of the object. An analysis of the precessional
and orbital variability separately in active and quiescent states was carried
out by Fabrika and Irsmambetova (2002), using the same photometric database
of the Sternberg Astronomical Institute (1979--1996). Active states of
SS433 (see the Introduction and the section ``The Radio Jets and W50'')
were identified using the results of the Green Bank Interferometer
monitoring program (http://www.gb.nrao.edu/fgdoss/gbi/gbint.html), as well
as on the basis of the optical data themselves. The precessional light curves
in active and quiescent states are appreciably different. In quiescent states,
the mean brightness of SS433 (of the accretion disk) outside of eclipse
depends strongly on the precessional phase. The object becomes weaker in
the $V$ band by approximately a factor of 2.2 when the accretion disk turns
from being maximally turned to face the observer to being viewed edge-on.
However, in active states
(excluding obvious flares), the brightness of SS433 depends only very
weakly on the disk orientation. It is especially important that, when the
disk is turned to maximally face the observer (time $T_3$), the brightnesses
in active and quiescent states are the same. When the disk is viewed edge-on
(times $T_1, \,T_2$), the object is significantly weaker in quiescent states
than in active states.

The orbital light curves in the active and quiescent states are approximately
the same~-- primary and secondary eclipses are also observed. At the center
of eclipses of the accretion disk by the star (primary minima,
$\phi \approx 0$), SS433 brightens by a factor of $\approx 1.6$ in active
states compared to quiescent states. The precessional modulation at the
center of these eclipses is roughly the same in active and quiescent
states: the brightness is slightly higher when the disk is turned to face
the observer than when it is edge-on. There exists some source of
radiation in SS433 that is not eclipsed by the star even in blue light. This
may be associated with radiation scattered in the outer gas of the wind.

The photometric behaviour of SS433 in active and quiescent states suggests
a model (Fabrika and Irsmambetova 2002), in which the optical brightness of
the object is primarily determined by some hot body in the central part of the
accretion disk, which could be two hot gaseous cocoons surrounding the
inner (X-ray) jets. In active states, the size of these cocoons increases,
and they are not blocked from the observer by the rim of the accretion
disk when it is viewed edge-on, or by the donor star during
primary eclipses.  The approximate equality of the amplitudes
of the precessional modulation and of the primary eclipses
indicate that the size of the donor is approximately equal to
the size of the outer rim of the disk in projection onto
the plane of the sky.


When the disk is maximally facing the observer, the mean
brightness at elongations is the same in active and quiescent
states. This may indicate that the cocoon surrounding the
approaching jet is not blocked by the outer rim of the disk at
precessional phases when the disk is maximally facing the
observer, so that its radiation can be observed right down to
its base, and that the luminosity of the cocoon does not depend
appreciably on its size (or the activity state of SS433). This
may be the case if the cocoon gas scatters radiation ($\tau_T
\sim 1$) arriving from inner regions~-- from a funnel in the
accretion disk or the wind. These cocoons may be identified with
a source   of UV radiation in which Dolan {\it et al.\/} (1997)
detected strong linear polarization directed along the jets, or
with the source of the two-peaked He\,II\,$\lambda4686$ line in
the spectrum of SS433 (see next section).

\vspace*{0.3cm}

\bigskip

\subsection{The Nutational Clock and Time for the Passage of
Material through the Disk}

Due to the conjugacy of the nutational period of 6.28~days with the orbital
and precessional periods, the distortions of the orbital light curves due to
the nutational oscillations will depend on the precessional phase
(Goranskii {\it et al.\/} 1998b). In particular, as can be found from the ephemerides
presented above, the nutational maxima at time $T_3$ occur at elongations.
The nutational variability leads to some shifts in the positions of
eclipses. The magnitude and sign of these shifts depend on the precessional
phase. Model fitting of the orbital light curves of SS433 should take these
nutational oscillations into account.

Information about the phases of the spectral and photometric variability
with the precessional and nutational periods is very important for our
understanding of the precession mechanism operating in SS433. As was already
mentioned, the most successful scenario for the precession is driven
precession of the donor star, whose rotational axis is not coincident with
the orbital axis, with a drifting or ``slaved'' accretion disk (Shakura 1972;
Roberts 1974; van den Heuvel {\it et al.\/} 1980; Whitmire and Matese 1980; Hut
and van den Heuvel 1981; Matese and Whitmire 1982). The massive donor star
can undergo driven precession (Papaloizou and Pringle 1982; Collins 1985).
In their analysis of periodic perturbations of the disk, which does not
lie in the orbital plane, by the gravitational field of the donor
star, Katz {\it et al.\/} (1982) concluded that the precessional and nutational
motions of SS433 can best be explained by a slaved-disk model. The analysis
of nutational synodic phenomena is an accurate tool, and the amplitudes
of these motions and possible variations of the period depend directly on
many parameters of the binary system. Collins (1985), Collins and Newsom
(1986) and Collins and Newsom (1988) developed a dynamical model for SS433
(see also the revised dynamical model in the recent paper by Collins and Scher
(2002)) in which it was possible to deduce the properties of the precessing
star, orbital eccentricity, and apsidal motions.

In terms of frequencies, the nutational period of 6.28~days looks like
$f_{nut} = 2 f_{orb} + f_{pr}$. In practice, the perturbations of the
accretion disk occur with a period of 6.06~days; this is the nodding of
the disk motion (Katz {\it et al.\/} 1982), $f_{nod} = 2 f_{orb} + 2 f_{pr}$.
As could be expected (``The Radio Jets and W50''), flares in the SS433
system should ``feel'' precisely this nodding period; i.e., the beating
period in a coordinate system rotating with the accretion disk or the
star. However, in the observer's system, where photometric and spectral
effects also depend on the angle between the line of sight and the
disk or jet axis, this period becomes 6.28~days.

Perturbations of the disk (or accretion flow) of SS433 lead to corresponding
variations in the jet inclination if the time for the material to pass through
the disk is not too long, and does not greatly exceed the perturbation period.
Information about variations of the inclination of the rotational moment
of the outer parts of the disk should reach the inner regions (the source of
the jets) without being appreciably distorted or smoothed. We will set aside
questions concerning the structure of the inclined disk and the interactions
of various harmonics in the perturbations of the outer edge of the disk
produced by the gravitational field of the donor star (Katz {\it et al.\/} 1982).

Qualitatively, the model is such that, at times of elongations, the
perturbation of the disk by the donor star leads to a shift of the disk
rotational axis in the plane of the sky (further, this perturbation affects
the direction of propagation of the jets), but this does not change the
inclination of the jets to the line of sight. Therefore, perturbations
applied to the disk at times of elongations do not lead to shifts in the
moving lines. Perturbations of the edge of the disk at times of
conjunctions are directed perpendicular to the plane of the
sky, and will therefore change the inclination of the jets. In particular,
at precessional phases close to $T_3$ (the inclination of the jets and
of the disk axis to the line of sight is $\approx 60^{\circ}$), perturbations
at conjunctions tend to align the disk in the orbital plane. Since the
inclination of the orbital axis to the line of sight is $\approx 78^{\circ}$,
the effect is to cause the H$\alpha^{\pm}$ lines to approach each other,
while the brightness of the system weakens. A quarter orbital period later,
at elongation, the disk and jets return to their initial positions,
the H$\alpha^{\pm}$ lines move apart, and the brightness grows. Of course, we
will see the reaction of the disk and jets to the gravitational perturbations
by the star only after the time required for the material to pass through
the disk and for the jets to move to the region of efficient line radiation.
In reality, in addition to perturbations of the outer parts of the disk,
we should bear in mind that the conditions for the formation of the disk
(the locations of the relativistic star relative to the equator of the donor)
change with the nutational phase. The heating of the donor surface is
inhomogeneous due to shadowing by the disk and flows. All these effects
change the geometry for the mass transport, and depend
  on the phase of the beating between the precessional and
orbital periods.

Mazeh {\it et al.\/} (1987) found that the jet nutational phases derived
  from
spectral and photometric data are not quite coincident: the photometric
maxima lead the nutational radial-velocity extrema by about one day.
Goranskii {\it et al.\/} (1998b) confirmed this result by showing based on 16 years
of data (950 nutational periods) that the nutational increase in the
brightness of SS433 coincides with the maximum shifts of the H$\alpha^-$
lines toward the blue, with a small but significant phase shift
$\Delta \phi_{nut} = 0.10 \pm 0.02$ (Goranskii {\it et al.\/}, 1998b). The nutational
deviations of the jets lag the optical variability. This phase
shift corresponds to a delay of 0.6~days, or a distance travelled by the
jets of $\approx 4 \cdot 10^{14}$~cm. This is precisely the distance
at which the maximum radiation of the jets in the moving H$\alpha$ lines is
achieved (Fabrika and Borisov 1987; Vermeulen {\it et al.\/} 1993a). This leads us
to conclude that the nutational oscillations of the brightness are associated
with the bases of the jets. There are also nutational variations in the
X-ray (Gies {\it et al.\/} 2002a), with the phase of the maximum X-ray flux
roughly coinciding with the phase of the maximum   in the
optical.

However, the energetics of the jets are insufficient to explain
  the 6-day
variations in the optical. It is natural to propose that the entire central
engine of the disk--jets system participates in the nutational variability.
The amplitude of these variations is about $\sim 10^{39}$~erg/s, comparable
to the kinetic luminosity of the jets. Even the X-ray luminosity of the
jets is roughly a factor of $\sim 10^4$ lower than the bolometric
luminosity of the accretion disk, and the optical continuum luminosity of
the jets should be lower than the X-ray luminosity, so that the optical
radiation of the jets cannot explain the nutational brightness variations
($\Delta V \approx 0\magdot17$). These variations may occur in inner regions
where the bases of the jets are located and which, like the jets, should
execute precessional and nutational motions.

If the brightness variations with the nutational period are caused by
nodding of the outer parts of the disk or the donor star, the time for
the passage of the material across the disk is either approximately ``zero''
(e.g., equal to the free-fall time, which is several tenths of a day), or is
equal to a multiple of the nutational period. However, the problem of the short
implied time for the passage of material across the SS433 disk arises for
other reasons as well, that are not associated with photometric variability.

Using the model of Katz {\it et al.\/} (1982) for the nodding motions of the accretion
disk, Gies {\it et al.\/} (2002a) found based on new observations of the moving jet
lines that the nutational deviations of the jets occur 1.0 day later than
the primary perturbations at elongations (or at conjunctions; the important
thing here is the fixing of the time in terms of the orbital clock). In this
way, Gies {\it et al.\/} (2002a) confirm the delay of oscillations of the jet
relative to the photometric variations. While 0.6~days (the delay of the
jet nutation relative to the photometric variability found by Goranskii
{\it et al.\/} (1998b)) agrees well with the required time for the motion of the
jet gas to the region of H$\alpha$ radiation at a distance $R_j \approx
4 \cdot 10^{14}$~cm, it seems that the remaining time of 0.4~days must
be the time for the passage of the signal across the accretion disk? It is
certain that such analyses can provide information about the structure of
the SS433 disk. If the result of Gies {\it et al.\/} (2002a) indicating a time delay
of 1.0 day is confirmed by future observations (or by new analyses of
existing data using more accurate, modern ephemerides), this may mean that
the material in the accretion disk reaches inner regions over the free-fall
time. The tilted disk could consist, for example, of flows of material that
lose momentum in shock waves.

In fact, a time delay of about one day was found in early studies, before
the nutational photometric variability had been discovered. Collins and
Newsom (1986) found in the framework of a dynamical model that the jets
deviate 0\daydot83 $\pm$ 0\daydot2 after the times of perturbation of the
precessing body. They proposed that the time for the passage of material
across the disk and further to the region of optical jet emission was
0\daydot8 plus a time equal to a multiple of the nutational period. Katz
{\it et al.\/} (1982) also found a time delay of 0\daydot9 in a model for the nodding
motion of the disk (the delay of the deviation of the jets after the
perturbations at elongations), but they used data on the radial velocities
of the He\,II\,$\lambda 4686$ line obtained by Crampton and Hutchings (1981a)
to determine the orbital phase. Thus, the delay of the times
of noddings of the jets relative to the times of perturbations of the accretion
disk of $\approx 1$~day can be considered to be reasonably well established.

One test of the possibility that the time for the passage of material across
the disk is a multiple of the nutational period would be a comparison of
the precessional phases determined photometrically and spectroscopically.
It is thought that the known precessional brightness variability is associated
with precession of the disk. However, the precessional brightness fluctuations
(Gladyshev {\it et al.\/} 1987; Goranskii {\it et al.\/} 1998b) do not lead the precessional
motions of the jets, as they should; in contrast, a delay of the precessional
photometric wave relative to the jet precession is observed. This delay
comprises no more than three to four days (Goranskii {\it et al.\/} 1998b), and could
easily be associated with inaccuracy of the photometric ephemerides for the
precession. Nevertheless, the approximate phase coincidence of the photometric
and spectral precessional periods does not support the idea that the time
for the passage of material in the disk is a multiple of the nutational
period. However, it is not possible to rule out this idea if the precessional
variability is in no way related to precession of outer parts of the accretion
disk.

In summary, we can conclude that the nutational brightness variations are
associated with variations in the orientation of hot inner regions in the
places where the jets emerge (for example, cocoons of hot gas). The
precessional variability most likely has the same origin. Independent of
the specific interpretation of the photometric variability, the problem
exists of the small time inferred for the passage of material across the
disk, which is close to the free-fall time. However, we cannot rule out the
possibility that the time for the passage of material across the disk is
a multiple of the nutational period.

\bigskip

\subsection{Outflows in the Disk Plane and Gaseous Flows}

As soon as the disk is oriented edge-on (beginning with precessional phase
$\approx0.3$), the gas trail beyond the disk can cross the line of sight at
orbital phases $\phi > 0.5$. Absorption in this flow of gas lost by
the system could plausibly give rise to an appreciable dimming of the
brightness, roughly as is observed in Fig.~16, where we see a significant
weakening of the brightness after Min\,II at precessional phases $\psi=0.3-0.7$.
Absorption in such a flow could also be significant even when the accretion
disk is not edge-on. Due to the precessional shift of the point L2, the orbital
phase where we could expect absorption will also shift, and this will shift
the position of Min\,II in approximately the same direction as in the case of
heating of the surface of the optical star. It is not possible to determine
which of these effects distorts the light curve more strongly without
specialised modeling of the precessional and orbital brightness modulations.

If absorption in flows of gas lost by the system is indeed that substantial,
the optical depth of the outflowing material to Thomson scattering will be
$\sim 1$, and the radial density will be $N_H \sim 10^{24}$~cm$^{-3}$. In
addition to its influence on the optical light curves (Zwitter {\it et al.\/} 1991;
Fabrika 1993), it is believed that precisely this material absorbs the
radiation of the receding X-ray jet (Kotani {\it et al.\/} 1996) and gives rise to
the radiation of the equatorial VLBI disk further from the system (Paragi
{\it et al.\/} 1999, 2000; Blundell {\it et al.\/} 2001), and may even be observed on larger
scales (seconds--minutes of arc) in optical line emission (Fabrika 1993).
There is direct evidence for the presence of spreading flows of material
in the plane of the accretion disk, which we will describe when we discuss
the spectra of SS433.

The depth and shape of X-ray eclipses of the accretion disk vary substantially,
depending on the precessional phase (Kawai {\it et al.\/} 1989; Brinkmann {\it et al.\/} 1991;
Yuan {\it et al.\/} 1995; Kotani {\it et al.\/} 1997b). Some of these X-ray eclipses are in
reasonable agreement with optical eclipses. Thanks to the ability to obtain
nearly continuous observations, observations of X-ray eclipses make it
possible to directly study the structure of the accretion disk, and to map
this region and its near environment.

\vspace*{0.1cm}

In addition to eclipses by the optical
star, other substantial decreases in the X-ray brightness have been detected
(with amplitudes as large as those during the eclipses, or even larger), which
it has not been possible to reconcile with the canonical picture of eclipses in
the binary system (Brinkmann {\it et al.\/} 1989). These ``unforeseen'' eclipses in
SS433 are not fully understood, and these brightness dips are most likely
due to absorption in both accretion flows in the system and flows of gas
out of the system roughly in the plane of the accretion disk.
The orbital phases and structure of these brightness
decreases are consistent with them being associated with a turning point of
a flow near the Roche lobe (Brinkmann {\it et al.\/} 1991; Lubow and Shu 1975).
Absorption could also arise in gaseous clouds above the plane of the disk
and in gaseous outflows from the system. Hydrodynamical computations of the
formation of the accretion disk when the donor overfills its critical Roche
lobe (Sawada {\it et al.\/} 1986; Chakrabarti and Matsuda 1992) show that a complex
structure with spiral shocks should exist in the disk, with powerful outflows
of gas beyond the disk. Computations of the formation of an inclined slaved
accretion disk (Bisikalo {\it et al.\/} 1999) also indicate a very complex structure
for the gaseous clouds outside the plane of the disk and the presence of
a shock wave along the stream of accreting material.

\bigskip

\subsection{Sharp Brightness Decreases}

The sharp and deep dips in the optical brightness of the object that are
observed at various precessional and orbital phases (Henson {\it et al.\/} 1982;
Kemp {\it et al.\/} 1986; Gladyshev {\it et al.\/} 1987; Goranskii {\it et al.\/} 1998b) may prove
important for our understanding of the nature of the components of SS433.
The deepest observed dip, to V$= 17\magdot3$, occurred during a primary
eclipse when the disk was oriented edge-on (Henson {\it et al.\/} 1982); the
brightness of SS433 dropped by 2\magdot5 compared to its normal level for
the corresponding orbital and precessional phases. It is interesting that
the object was in an active state at that time, and its brightness was
high 0\daydot5 before and 0\daydot5 after this record brightness dip
(Goranskii {\it et al.\/} 1998b). Other brightness decreases by 1\magdot9 and
1\magdot1 relative to the normal brightness level have been observed outside
of eclipse. These dips imply that the optical star in SS433 is at least
a factor of 20 weaker than the accretion disk in the $V$ band, since the
usual brightness of SS433 is V\,$=14\magdot0$. Adopting a distance to the
object of 5.0~kpc and an absorption of $A_V=8\magdot0$, we find that the
luminosity of the donor is $M_V > -4\magdot5$, if, of course, this star
itself is not subject to eclipses during these brightness dips.

These sharp decreases in the disk radiation are surprising, all the more so
because the strongest dip occurred in an active state. Is it possible that
the mass-transfer process is sufficiently non-stationary during times of
activity that material does not reach the disk for some time? It may be
that, as in the case of the ``unforeseen'' X-ray eclipses, we must
think about eclipses of an optical source in the accretion disk by gaseous
flows. An alternative scenario in which the $V$ brightness dips are associated,
on the contrary, with sharp increases in the rate with which material
is supplied to the disk is also possible. Roughly speaking, the luminosity
of the supercritical disk does not depend on the accretion rate $\dot M_0$.
The size of the photosphere of the supercritical disk's wind is $R_{ph}
\propto \dot M_0^{3/2}$, and the temperature of the photosphere is $T_{ph}
\propto \dot M_0^{-3/4}$. When there is a sharp increase in $\dot M_0$, for
example by a factor of 10--100, the photosphere temperature falls to several
thousand Kelvin, the size of the wind photosphere grows by a factor of
several tens (or even more, since the main source of opacity will become
free--free transitions and molecules rather than Thomson scattering), and
the entire binary system will end up being deep beneath the wind photosphere.

Observations of the sharp brightness dips in other spectral regions (or at
least in two different optical bands) would help elucidate the nature
of these strange brightness decreases. Note that such dips argue in favour
of a short time for the passage of material~across~the~disk.

\bigskip

\subsection{The Spectral Energy Distribution and Parameters of
the Components}

Outside of eclipses and at precessional phases when the accretion disk is
turned to face the observer, the luminosity and temperature of the object
are substantially enhanced. It is difficult to accurately estimate the
temperature of the radiation from optical photometric data, since it is
necessary to work in the far Jeans region of the spectrum. Based on
$WBVR$ photometry, Cherepashchuk {\it et al.\/} (1982) found that the spectrum
of SS433 at brightness maximum was consistent with that of a black body
with a radiation temperature of $T \ga 50\,000$~K subject to absorption
$A_V=7\magdot4 - 8\magdot3$, with the hot body having a radius
$R \approx 2 \cdot 10^{12}$~cm and bolometric luminosity $L_{bol} \ga
10^{40}$~erg/s. Murdin {\it et al.\/} (1980) arrived at essentially the same
conclusion in one of the earliest studies of SS433 ($A_V \approx8^{\rm m}$,
$T \sim 30\,000$~K, $L_{bol} \sim 3 \cdot 10^{39}$~erg/s). Wagner (1986)
drew similar conclusions based on spectrophotometric data ($A_V =7\magdot8
\pm 0\magdot5$, $T \sim 32\,500$~K, $R \sim 30 R_{\sun}$, $L_{bol} \sim 4.4
\cdot 10^{39}$~erg/s), and confirmed that the source becomes hotter when
the bright, precessing body is observed nearer to the pole. Formally, in the
presence of absorption $A_V > 8\magdot2$, the temperature of the source derived
from optical photometric data approaches infinity.

\renewcommand{\thefigure}{17}
\begin{figure}[t]
\centerline{\hspace*{-10mm}\psfig{figure=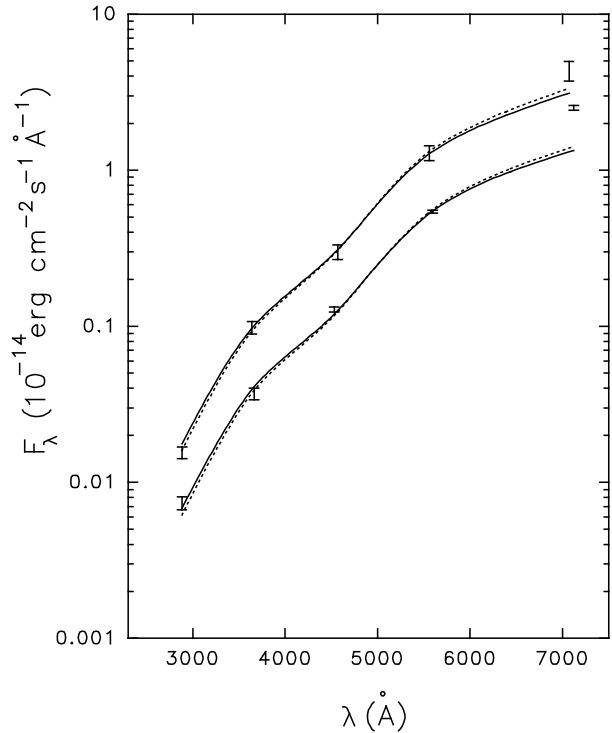,width=80mm}}
\caption{Observed fluxes of SS433 in the F227M and $UBVR$ bands
(Dolan {\it et al.\/} 1997). The upper panel shows the bright
precessional state and the phases of orbital elongations, and
the lower panel shows the edge-on precessional state and the
phase of eclipse of the accretion disk. Approximations using a
single blackbody spectrum taking into account interstellar
absorption at the isophotal wavelengths of the SS433 spectrum
are shown. The solid curves show the best solution for the $UBV$
data only, while the dotted curves show the solution for
$UBV$~+~F227M. The intensity of the additional radiation in the
$R$ band is independent of the precessional and orbital phases,
and is equal to $\Delta F_R=1.2\cdot10^{-14}$\,erg/cm$^2$\,s\,\AA.}
\end{figure}

Observations with the HST/HSP confirmed the high temperature of the radiation
of SS433 derived from optical photometry. Figure~17 presents the observed fluxes
in the F227M and $UBVR$ bands based on the data of Dolan {\it et al.\/} (1997).
The ultraviolet observations were conducted in a band centered
on 2270 \AA\ during both bright (precessional phase close to
zero and orbital phase close to elongation) and dim (edge-on
precessional orientation and phases of eclipses of the accretion
disk) states. The mean $UBVR$ fluxes for these two states were
derived from the optical photometry database. Approximations
using a single blackbody spectrum (without using the $R$ data) show that
the temperature of the object's radiation in the bright state is $T =72\,000
\pm 20\,000$~, with absorption $A_V = 8.4 \, (+0.2,\,-0.3)$,
  source radius
$R = 2.0 \cdot 10^{12}$~cm and bolometric luminosity $L_{bol} =
  (4 - 8) \cdot
10^{40}$~erg/s. It was not possible to obtain a satisfactory approximation to
the spectrum using a single blackbody emitter in the dim state; Fig.~17
shows the spectrum for $T = 4.9 \,(+2.9\, -1.5) \cdot 10^4$~K, $A_V =
8.2 \pm0.5$, $R = 1.5 \cdot 10^{12}$~cm and $L_{bol} \approx 1 \cdot
10^{40}$~erg/s.

These data also confirm the presence of a cool envelope, as
follows from optical and infrared photometry. There is an
additional   source of radiation in the $R$ band (Fig.~17),
independent of the
  precessional and orbital phases, equal to
$\Delta F_R = 1.2 \cdot 10^{-14}$~erg/cm\,s\,\AA. This
additional ``red light'' has now also been distinguished in
analyses of flares of SS433 (Irsmambetova 1997, 2000; Goranskii
  {\it et al.\/} 1998a). As was already mentioned above (``The
Radio Jets   and W50''), SS433 displays both ``white'' (i.e.,
hot) and ``red'' optical flares. Red flares are due to the
excitation of free--free emitting gas located around the system,
and show delays in the $R$   and $I$ bands by several hours
relative to the corresponding flares in   the $V$ band.  In
addition, as a rule, SS433 reddens in the transition to the
active state.

Observations in the far UV were conducted by Gies {\it et al.\/} (2002a) using
the HST/STIS at 1150--1700 \AA. Unfortunately, SS433 was not detected
in these observations, and it was only possible to place an upper limit
on the flux. The precessional phase at the observing epoch corresponded
to the dim state (edge-on). Using these upper limits, the data of Dolan
{\it et al.\/} (1997), archival IUE observations and the optical spectrophotometric
data of Wagner (1986), Gies
  {\it et al.\/} (2002a) found that the temperature of
the source derived by Wagner (1986) based on optical data in the dim state
($T = 21\,000$~K
  for $A_V = 7\magdot8$) was consistent with the upper limits
for the far-UV flux obtained from the STIS observations. The implied radius
of the hot source was $R = 2.3 \cdot 10^{12}$~cm.

The uncertainties in estimates of the temperature of the accretion disk (more
precisely, of the photosphere of the disk wind) depend on the photometric
data used, assumptions about the absorption law, and the complexity of the
source. It is usually assumed that this is a single object radiating as a
blackbody. Note also that the spectral energy distribution of the supercritical
disk should not correspond exactly to a blackbody distribution (Lipunova
1999; Okuda   and Fujita 2000).

Fuchs {\it et al.\/} (2002) detected He\,I and He\,II emission in an infrared spectrum
obtained with ISOPHOT (2.4--4.8 and 6--12 $\mu$m), and concluded that the
spectrum was similar to a WN spectrum. However, it is known from the
radial-velocity curves that the He\,I and He\,II lines in SS433 are radiated
in gas flows, in the accretion disk and in the accretion disk wind rather
than in the donor wind. The observed
spectral energy distribution at 4--12 $\mu$m agrees well with that expected
for free--free radiation in an optically thick envelope,
  with $\alpha = 0.6$
in the power-law spectrum $F_{\nu} \propto \nu^{\alpha}$. Optically thin
free--free radiation is observed between 2.4 and 4 $\mu$m with $\alpha = - 0.1$
(see also Giles {\it et al.\/} 1979; McAlary and McLaren 1980); and the spectrum is
well described by blackbody radiation by dust with $T \sim 150$~K at the
longer wavelengths, 12--60 $\mu$m. Fuchs {\it et al.\/} (2002) estimated the mass-loss
rate in the wind to be $\dot M_w \approx 1.0 \cdot 10^{-4}\,M_{\sun}$/yr.

Cherepashchuk {\it et al.\/} (1982) estimated the parameters of the optical star using
four-band photometric data obtained during eclipses of the accretion disk.
They found that the temperature of this star is in the range $T_s = 13\,000 -
43\,000$~K, while its radius is $R_s = 28 - 47\,R_{\sun}$. No spectral lines
from the star were detected. It was suggested
  (Fabrika 1998) that optical
lines should be searched for among high-order terms of the Balmer series,
since the Balmer emission decrement is very steep while the stellar
absorption decrement is flat; however, blue spectra of SS433 with the required
quality were not obtained.

Recently, Gies {\it et al.\/} (2002ab) presented promising
data. They proposed that the best phases for searches for absorption lines
of the star were during eclipses of the disk and precessional phases
$\psi \approx 0$, when outflows of gas from the system do not cross the line
of sight. They detected an He\,I\,$\lambda 6678$ absorption component at
these phases, which moved over three successive nights in agreement with
the orbital motion of the donor. Gies {\it et al.\/} (2002b) detected absorption
lines in the blue, which are probably associated with the donor photosphere.
We will describe these results in more detail in the   next
section.

\bigskip

\subsection{Polarization of the Optical and UV Radiation}

The results of linear polarization measurements could prove important for
our understanding of the accretion disk of SS433. The optical polarization
in the $BVRI$ bands is about 2~\% and is variable (McLean and Tapia 1980;
Efimov {\it et al.\/} 1984; Dolan {\it et al.\/} 1997), with the polarization angle directed
along the plane of the accretion disk (perpendicular to the jets). This is
in good agreement with the expectations for scattering of the disk radiation
by free electrons located above the disk. The degree of polarization grows
slightly toward the blue. However, Dolan {\it et al.\/} (1997) found that the
character of the polarization changes sharply in the UV. The degree of
polarization in the $U$ band is $8.9\pm 0.9~\%$, and has already grown to
$13.7 \pm 3.0~\%$ in the 2770 \AA\ band (FWHM~$= 340$~\AA, HST/HSP). In
addition, the orientation of the polarization changes by $90^{\circ}$,
so that it is directed along the jets in the UV. The UV polarization is very
variable from observation to observation, and can sometimes reach $20~\%$
in   the 2770 \AA\ band.

Dolan {\it et al.\/} (1997) proposed that the UV polarization arises due to
Rayleigh scattering on neutral hydrogen atoms located in the plane
of the accretion disk beyond the binary system. This imposes strong
constraints on geometrical models for the sources of the optical
  and UV
radiation in the SS433 accretion disk, since it remains unclear why the
UV radiation efficiently scatters on a gas in this plane, while the
optical radiation does not.

In the context of the most recent X-ray
observations and optical spectral observations during eclipses (see
below), a picture is sketched out for the inner regions of SS433, where the
bases of the jets are shrouded in hot gas immediately above the disk. The
temperature of this cocoon falls sharply from $\sim 10^8$~K on its axis
(the region of the jet motions) to $\sim 5 \cdot 10^4$~K at the edges (the
He\,II cocoon). The region of stationary X-ray Fe-line fluorescence is
probably located there as well. The UV radiation also forms in inner regions,
since the effects of eclipses by the optical star are visible in this
radiation (and in its polarization). If the UV radiation forms in the gas
surrounding the bases of the jets or in cocoons, this could lead to its
polarization. The emerging radiation will be Thomson scattered by the cocoon
gas, with the plane of polarization of the scattered radiation oriented
along the jets, in agreement with the data of Dolan {\it et al.\/} (1997). A more
detailed quantitative model for the region of UV radiation and computations
of the polarization of the emerging radiation are very desirable.

 \bigskip

\section{The Supercritical Disk from Spectroscopic Data}

\subsection{The ``Stationary'' Spectrum of SS433}

The ``stationary'' lines in the spectrum of SS433 include the strongest
hydrogen lines, as well as emission lines of He\,I, He\,II, C\,III, N\,III, and
weaker Fe\,II emission (Murdin {\it et al.\/} 1980; Crampton and Hutchings 1981ab;
Dopita and Cherepashchuk 1981; Falomo {\it et al.\/} 1987; Filippenko {\it et al.\/} 1988;
Kopylov {\it et al.\/} 1989; Fabrika {\it et al.\/} 1997a; Gies {\it et al.\/} 2002a;
Fuchs {\it et al.\/} 2002). The spectrum is indeed reminiscent of the spectrum of a WR
star, or more precisely, of a member of the recently identified class
of late WN stars, of type WN10--WN11 (Crowther and Smith 1997; Bohannan
and Crowther 1999). These stars, in turn, are close relatives of the
Luminous Blue Variables
 (LBVs; Humphreys and Davidson 1994), which display
evidence for hot outflowing atmospheres and late-WN spectra when their
brightness is weak. The similarity between the spectrum of SS433 and~WNL
spectra is probably not a coincidence. The conditions for the formation
of the SS433 spectrum and lines in the wind of the supercritical disk are
probably close to those in the winds of late-WN stars, and the chemical
compositions might also be similar if the donor
 in SS433 is a fairly evolved
star.

It is primarily the variability and behaviour of the stationary H$\alpha$
that has been studied in detail, since it is the strongest line in the
spectrum and is a convenient line for observational studies, since the
object is very bright in the red part of the spectrum. The H$\alpha$
equivalent width is strongly variable. It is usually in the range
100--300~\AA, but sometimes reaches 1000~\AA. The profiles of powerful
emission lines (FWHM~$\sim 1\,000$~km/s, full width at the base 3\,000--5\,000~km/s)
vary with the precessional phase, becoming more structured and often broader
when the accretion disk is oriented edge-on. In contrast, the intensity of
the lines grows appreciably toward phases near $T_3$. For example, on
average, the H$\alpha$ luminosity varies by a factor of three with the
precessional phase (Asadullaev and Cherepashchuk 1986; Fabrika {\it et al.\/} 1997a),
and is $L(H\alpha) \approx 7 \cdot 10^{36}$~erg/s when the disk is maximally
turned to face the observer. The structure of the hydrogen-line profiles is
determined by gas motions and has a large-scale nature; strong emission
components appear, often in the blue and red
 wings ($\pm 1\,000$~km/s). It is
not known, however, whether some of the profile structure arises due to
line absorption from the blue
(and red) side. The hydrogen and He\,I lines
also show ``regular'' P\,Cygni profiles, which arise in the wind, at
precessional phases near edge-on orientation of the disk. At such times,
the Fe\,II lines can have such strong blue absorption components that the
emission components disappear. This indicates strong non-isotropy of the wind.
All the lines in the SS433 spectrum form either in various places in the wind
flowing out from the disk or in gaseous flows in the system. The line
profiles and fluxes vary strongly during flares.

\bigskip

\subsection{He\,II Radial-velocity Curves and the Mass Function}

Studies of the line radial velocities and the behaviour of the lines during
eclipses of the accretion disk have provided much information about the
gaseous flows in SS433 and the structure of the disk wind, both in inner
regions where the bases of the jets are located and just beyond the limits
of the binary system. The observational manifestations of the flows and wind
vary strongly with the precessional and orbital phases. The inclination
of the system's axis to the line of sight is well known from analyses based
on the kinematic model ($i \approx 78^{\circ}$), so that it is not an unknown
parameter, as is usually the case with studies of binary stars.
Here, we will describe investigations of the behaviour of the He\,II\,$\lambda
4686$ line, which yield, in particular, the mass function of SS433, and then
describe studies of gaseous flows and the wind in the system.

The He\,II line possesses the highest excitation potential of all the lines
in the SS433 spectrum, and is also the only line whose orbital motion reflects
the motion of the relativistic star (Crampton and Hutchings 1981a). SS433 is
probably the only star for which so many contradictory reports of its mass
have been published. This is associated with objective difficulties. The
mass ratio can be obtained from the duration of eclipses, but the size of the
body surrounding the relativistic star turns out to depend on the precessional
phase and the observing wavelength. In addition, the size of the wind
photosphere is in no way related to the size of the Roche lobe of the
relativistic component. The mass function can be derived from the orbital
variability of the line radial velocities, but the line radiation occurs
primarily in gas flows, so that the radial velocities vary appreciably
with the precessional phase.

Crampton and Hutchings (1981a) found that the half amplitude of the
He\,II\,$\lambda 4686$ radial velocities is $K=195 \pm 19$~km/s, with a mean
velocity (the $\gamma$ velocity) for the line of $V_0=+27 \pm 13$~km/s,
while the orbital phase of the transition of the radial velocity through
the $\gamma$ velocity from the positive to the negative region is (recalculated
to the modern, precise ephemerides) $\phi_0 = -0.01 \pm 0.02$. This is the
phase of the upper conjunction of the source, and coincides with the time of
the upper conjunction of the accretion disk. The mass function derived from
these data is $f(M) = M_o^3\,/\,(M_x + M_o)^2 \approx 10.8\,M_{\sun}$, where
$M_o$ and $M_x$ are the masses of the optical and relativistic star,
respectively. The scatter of points about the radial-velocity curve is
appreciable (as in later studies), and is not associated with measurement
errors or insufficient spectral resolution, but with real variability of the
radial velocity and the structure of the line profile. In particular, the
neighbouring H$\beta$ line is substantially brighter and displays a sharp
one-peaked profile (showing that the measurement accuracy for this line
is high), but Crampton and Hutchings (1981a) found that the orbital curve
of the radial velocities of this line (and also of the~He\,I and Fe\,II lines)
varies substantially with the season. When it is exhibiting regular behaviour,
the H$\beta$ radial velocity varies with a half-amplitude of $\approx 80$~km/s,
and has a mean velocity of $\approx +220$~km/s and $\phi_0 = 0.25$. These
last two values indicate that the H$\beta$ radial velocity is appreciably
distorted by effects such as absorption in the wind, and the line itself
forms in a gaseous flow.

Fabrika and Bychkova (1990) found that the radial-velocity
orbital curve of
the He\,II line depends on the precessional phase: when the accretion disk is
maximally turned to face the observer ($T_3$, $0.9 \la \psi \la 0.1$), the
main contribution to the emission of this line is made by the accretion disk
(more precisely, by regions near the relativistic object), while the main
contribution at other precessional phases is made by a gaseous flow outside
the accretion-disk region. We will describe the structure of this flow in
more detail below. Near the phase of $T_3$, the He\,II radial-velocity curve is
similar to that obtained by Crampton and Hutchings (1981a):  $K=175 \pm
20$~km/s, $V_0=-13 \pm 12$~km/s, and $\phi_0 = 0.03 \pm 0.01$, corresponding to
the mass function $f(M) \approx 7.8\,M_{\sun}$. At other precessional phases,
when a stream dominates, $K \approx 80$~km/s and $\phi_0 = 0.12 \pm 0.04$. The
half-amplitude of the He\,II radial velocity based on all data without regard to
the precessional phase is $\approx 120$~km/s. The gaseous stream that radiates
the He\,II line lags the relativistic object in orbital phase by~$\Delta \phi
\approx 0.1$, and is probably directed from the star toward the disk. Later,
Fabrika {\it et al.\/} (1997a) confirmed using additional observations that the
half-amplitude of the orbital variability of the He\,II line at precessional
phases near $T_3$ is $K=176 \pm 13$~km/s. In their analysis of eclipses of
this line profile based on coordinated observations near primary eclipses,
Goranskii {\it et al.\/} (1997) and Fabrika {\it et al.\/} (1997b) were able to distinguish
a broad component of the line forming in the disk (it is fully eclipsed at
phase $\phi = 0.0$) and a narrow component that forms in the flow (it is
partially eclipsed at phase $\phi = 0.1$, and its radial velocity corresponds
to the radial-velocity  curve of the flow).

D'Odorico {\it et al.\/} (1991) reported a low mass function for SS433, with a
half-amplitude for the He\,II variability of $K \approx 112$~km/s and a
phase for upper conjunction of the source based on their radial-velocity
curve of $\phi_0 = 0.08$. This half-amplitude for the radial velocities
leads to low values for the mass function, $f(M) \propto K^3$. D'Odorico
{\it et al.\/} (1991) obtained their spectra over 120~days, and this interval did
not include precessional phases corresponding to the maximum turning
of the disk to face the observer, $0.9 < \psi < 0.1$. Therefore, their
radial-velocity curve corresponds to the gaseous flow rather than the
relativistic component, as is also clear from its parameters. The data of
D'Odorico {\it et al.\/} (1991) is in full agreement with the results of~Fabrika
and Bychkova (1990) for precessional phases that correspond to times when
the disk is viewed nearly edge-on.

It is interesting that Crampton and Hutchings (1981a) did not divide their
observations according to precessional phase, but nevertheless obtained a
large amplitude for the He\,II line shifts. Their spectra were obtained in
two observing seasons, both of which included the precessional-phase interval
$0.9 < \psi < 0.1$, when the contribution of the accretion disk to the He\,II
line emission is maximum (we do not know the specific selection of spectra
used by these authors for their radial-velocity analysis). It is possible
that this meant that Crampton and Hutchings (1981a) were able to record
the real radial-velocity curve of the accretion disk. They note that the
intrinsic half-amplitude of the shifts of this line may be somewhat smaller,
but it cannot be less than 150~km/s.

Thus, it has been fairly firmly established (Crampton and Hutchings 1981a;
Fabrika and Bychkova 1990; Fabrika {\it et al.\/} 1997a) that the mass function for
SS433 derived from the orbital motion of the accretion disk (the He\,II line)
is in the range 7--10\,$M_{\sun}$. Recently, Gies {\it et al.\/} (2002a) discovered
that the phase shift in the orbital variability of the radial velocities of
the C\,II\,$\lambda 7231, \, 7236$ emission lines in the spectrum of SS433
corresponds to the orbital position of the accretion disk, with the
half-amplitude of the variations being $K \approx 160$~km/s. This confirms
the results obtained for the He\,II line. It is possible that other lines
radiating near the relativistic star will be discovered in the near infrared,
where the object is fairly bright.

\bigskip

\subsection{The Component Mass Ratio}

The component mass ratio of SS433 $q=M_x/M_o$ has been estimated many times.
Based on modeling of the optical eclipses, Antokhina and Cherepashchuk (1987)
concluded that the mass ratio was
 $q \approx 0.25$. However, there is a fairly
broad residual minimum allowing larger mass ratios, in particular,
$q \sim 0.4$, and possibly even higher values. High values of $q$ were derived
by Leibowitz (1984), $\ga 0.8$, and Hirai and Fukue (2001), $\approx 1-1.5$.
These last authors fit light curves using a thick-accretion-disk model.
Similar studies of the shape of X-ray eclipses yield appreciably lower mass
ratios. Antokhina {\it et al.\/} (1992) concluded that $q = 0.15 - 0.2$, with values
up to 0.3 possibly being allowed. Kotani {\it et al.\/} (1998) found the ratio
$q \approx 0.22$ based on an analysis of eclipses observed by ASCA. If a
point source is eclipsed by the star, which fills its critical Roche lobe,
the duration of the X-ray eclipses leads to the value $q = 0.15$ (Goranskii
{\it et al.\/} 1998b).

X-ray and optical models for the eclipses yield systematically
different mass ratios, with an ``informal'' average of the values being near
$\approx 0.3$. With this $q$, the mass function $f(M) =7-10\,M_{\sun}$ yields
a mass for the relativistic star of $M_x=3.5-5.1\,M_{\sun}$. However, it is
certainly not correct to simply average the various mass-ratio estimates, and
we must understand the origins of the differences and devise more complex
models for the regions of the optical and X-ray emission in order to explain
them. It is also important to learn how to take into account additional
absorption in gas flows in the system, which distorts the shape of the X-ray
eclipses. Other limitations in this method for determining the mass ratio
are imposed by the following two assumptions: (1) the source cannot be larger
than the Roche lobe of the compact star (an assumption that we must discard)
and (2) the size of the donor corresponds fully to the size of its Roche lobe.
A star losing mass at a rate of $\sim 10^{-4}\,M_{\sun}$/yr could have a very
dense and extended atmosphere. Failure to take this into account could lead
to underestimation of the mass ratio.

Studies of variations of the orbital period could prove to be an effective
method for determining the mass ratio. Goranskii {\it et al.\/} (1998b) found that
the period did not vary to within 0\daydot00008
 over 17 years of intense
observations. Analysis of eclipses using archival material indicates that
the period has remained virtually constant for 34 years. This corresponds to
an upper limit for the rate of variation of the period of $\dot P_{orb}
\la 2 \cdot 10^{-7}$. Fabrika {\it et al.\/} (1990) investigated variations of the
period as a function of the
rate of the mass transfer and mass loss in SS433 under
the assumption that
all the material up to a rate of $\sim
10^{-4}\,M_{\sun}$/yr reaches the accretion disk, and that, further, some
portion of the gas is lost through the
 Lagrange point L2 beyond the
relativistic component, while the remaining gas is lost from the system
in the form of a wind from the inner regions of the disk. They based this
analysis on an incorrect determination of the rate of variation of the
period (they took large-scale fluctuations in the O--C diagram to be variations
in the period). It is now clear that the period of SS433 is surprisingly
stable (Goranskii {\it et al.\/} 1998b). Nevertheless, if we suppose that $\dot
P_{orb} = 0$,
 the analysis of Fabrika {\it et al.\/} (1990) yields a component-mass
ratio of $q = 0.7 - 0.8$.

It is likely that progress in estimating the masses of the stars
 in SS433
will come from spectroscopic measurements. If the absorption component in
the He\,I\,$\lambda 6678$ emission line detected by Gies
 {\it et al.\/} (2002a) arises
in the atmosphere of the donor star, the observed shift of the radial velocity
of this component yields the  estimate
$K_o=126 \pm 26$~km/s, which, in turn,
leads to the mass  ratio
$q = 0.72 \pm 0.17$ (Gies {\it et al.\/} 2002a). If the mass
function is $f(M) \approx 7.8\,M_{\sun}$, a mass ratio of
$\approx 0.7$ yields a mass for the  relativistic star of $M_x
\approx 16\,M_{\sun}$.

Of course, a number of indirect arguments, one of which is the huge luminosity
of SS433, suggest that there is a black hole with a mass of $\sim
10-20\,M_{\sun}$ in the system. However, without direct measurements of the
mass ratio, the compact star must remain only a very likely black-hole
candidate. The detection of lines of the secondary component of the system
would be an optimal solution to the problem of determining the mass ratio.

Such lines were detected in the recent observations of Gies {\it et al.\/} (2002b),
which were conducted during three successive nights that encompassed an
eclipse of the accretion disk by the donor star at the precessional phase
when the disk is maximally turned toward the observer ($\psi \approx 0$).
This last circumstance means that material spreading in the plane of the
disk does not screen the donor, and we have the best chance of seeing the
photospheric spectrum of this star. Weak absorption lines of Ti\,II, Fe\,II,
Cr\,II, Si\,II, Sr\,II, Ca\,I, and Fe\,I were detected in the blue part of
the spectrum (4000--4600~\AA), which resembles the spectrum of an
evolved star~-- an A-type supergiant ($T_{eff} \sim 8\,000$~K). These lines
became stronger when the disk was maximally eclipsed. The radial velocities
of the absorption lines showed an orbital shift that was opposite to the
shift shown by the lines emitted by the disk surrounding the compact star.
Gies {\it et al.\/} (2002b) present weighty evidence that they have detected the
spectrum of the donor star in SS433. Based on the amplitude of the
absorption-line shifts and taking into account the mass function derived
from the He\,II line (Fabrika and Bychkova 1990), $f(M) = 7.8\,M_{\sun}$,
Gies {\it et al.\/} (2002b) estimate the mass ratio to be $M_x/M_o = 0.57 \pm 0.11$
and the masses $M_o = 19 \pm 7$ and $M_x = 11 \pm 5$. It is obvious that
further more detailed observations in the blue should make it possible to
refine the parameters of the components of SS433, but the results of
Gies {\it et al.\/} (2002b) have already demonstrated that there is a black hole
 in this system.

\bigskip

\subsection{Properties of the Gas Flow from the He\,II and
H$\beta$ Lines}

In coordinated spectral and photometric observations of SS433 during eclipses,
Goranskii {\it et al.\/} (1997) were able to distinguish three components in the
He\,II\,$\lambda 4686$ emission. Components in the He\,II profile are also clearly
visible in the spectra of D'Odorico {\it et al.\/} (1991), however continuous
observations from night to night near the primary eclipse are required in
order to identify these components and trace their variations. In addition,
at different precessional phases, the He\,II components have different
appearances and display different behaviour during eclipses (Fabrika
{\it et al.\/}~1997b). The He\,II emission consists of a ``narrow'', essentially Gaussian
profile with FWHM~$\approx950 \pm 20$~km/s and two components forming a
broad two-peaked profile.

The narrow He\,II component is not eclipsed at the center of the photometric
eclipses. The region in which this component is emitted experiences partial
eclipses ($\approx 30-40~\%$) at orbital phase 0.1.
 The H$\beta$ line consists
only of one narrow component (FWHM~$=840 \pm 40$~km/s), which also
undergoes partial eclipses with amplitudes of about $15~\%$, but at
orbital phases $\phi=0.1-0.25$. The profile and width of the H$\beta$
line and the narrow
He\,II component do not vary during the course of the eclipses. Fabrika
{\it et al.\/} (1997c) also distinguished eclipses in the~H$\alpha$ line using
observations in a narrow filter and in the neighbouring continuum. The
eclipses occur at orbital phase $\phi\approx0.2$, and the depth of the
H$\alpha$ eclipse is about $15~\%$.

The narrow He\,II component and the hydrogen lines form in a {\it gaseous
flow directed toward the accretion disk}. The half-amplitude of the radial
velocities of emission lines emitted in the flow is  $K\approx80$~km/s.
The orbital phases at which the regions of line radiation in the flow are
at upper conjunction grow from $\phi_0 \approx 0.1$ for He\,II (the narrow
component near precessional phase $T_3$ or the entire line at other precessional
phases) to $\phi_0 = 0.1-0.2$ for the He\,I lines and~$\phi_0\approx0.25$ for
the hydrogen lines (Crampton {\it et al.\/}~1980; Kopylov {\it et al.\/}~1989;
Goranskii {\it et al.\/}~1997; Fabrika~1997; Fabrika {\it et al.\/}~1997abc; 
Gies {\it et al.\/}~2002a). We can conclude from this information that the observed flow is
very extended. The phase and duration of the He\,II and H$\beta$ eclipses
(Goranskii {\it et al.\/}~1997) indicate that the size of the flow is no less than
$0.4\,a$ and that, on average, the flow itself is separated from the accretion
disk by a distance of $\sim 0.6\,a$, where $a$ is the distance between the
components. The flow is directed toward the accretion disk, and the gas
temperature in the flow falls with distance from the disk from
$(3-5) \cdot 10^4$~K to $(1-2) \cdot 10^4$~K. It is possible that the gas
is heated by shock waves arising when the flow comes into contact with the
accretion disk.

Essentially, all the main emission lines in the spectrum except for the
He\,II lines are radiated in the flow. The profiles of lines emitted in the
flow are very strongly distorted by absorption in their blue wings (in the
outer wind), and are shifted toward the red by up
 to $\Delta V_r \sim
200$~km/s, depending on the optical depth of the line and the wind velocity
along the line of sight; the magnitude of this shift also strongly depends
on the precessional phase (Fabrika 1997; Fabrika {\it et al.\/} 1997a; Gies
{\it et al.\/} 2002a). The widths of the flow lines are much larger than the virial
velocity of the system. Kopylov {\it et al.\/} (1989) proposed that the flow
is optically thick to electron scattering ($\tau \sim 150$), and that the
lines are broadened by scattering of the emerging radiation.

\bigskip

\subsection{The Structure of the Disk and the Central Region
from the He\,II Line}

The precession of the disk and its eclipses by the optical star create
the rare possibility of directly studying the disk itself and the region
where the relativistic jets appear. Here, we must recall that the object
surrounding the relativistic star in SS433 is traditionally called a ``disk'',
although it resembles a disk less and less with each new investigation.
The wind of the supercritical accretion disk is observed around the
relativistic star, and its structure is certainly complex.

The broad, two-peaked He\,II component is completely eclipsed at
 the centre
of the primary minimum. Based on observations of
 an eclipse near precessional
phase $\psi = 0.95$, Goranskii {\it et al.\/} (1997) established that the blue wing
of the broad profile appeared first
 during the egress of the disk from eclipse
($\phi =0.1$), with the red wing appearing the following night. In another
eclipse at precessional phase $\psi=0.0$, two peaks with approximately equal
intensities appeared simultaneously, when the disk emerged from behind the
limb of the star. The distance between the maxima of the two-peaked profile
is $\Delta V \approx 1\,500$~km/s. Such a profile cannot belong to lines
radiated in the disk. For a Keplerian velocity $\Delta V\,/\,2 \approx
750$~km/s and reasonable masses for the compact star $\sim 5 \,\,M_{\sun}$,
the size of the disk in the He\,II line would be $\sim 10^{11}$~cm.
The time for egress from eclipse of such a disk would not be more than two
hours, while the observed egress from eclipse of the broad He\,II component
lasts no less than a day.

Goranskii {\it et al.\/} (1997) proposed that the two-peaked He\,II profile forms in
hot, gaseous cocoons surrounding the bases of the approaching and receding
jets. The sequence in which the blue and red components appear from behind
the limb of the star is consistent with the geometry for the positions
of the jets in the standard model in which the precessional and orbital
motions are in opposite directions. The velocity of the outflow
of gas in these ``He\,II cocoons'' was estimated from the known inclination
of the jets at the observation epoch
 to be $V_w(\mbox{He\,II}) \approx 1\,500$~km/s.
If we indeed see the hot bases of the jets in the He\,II emission, the distant
cocoon radiating the red He\,II wing is not screened by the accretion disk; i.e.,
in projection onto the plane of the sky, the distance between the relativistic
star and the cocoon is larger than the radius of the disk. Essentially the
same situation is observed in the X-ray (see the section ``The X-ray
Jets'')~-- the body of the disk does not block the receding jet. A full
eclipse of both components of the broad He\,II profile by the star means that
the size of the star exceeds the projection of the accretion disk onto the
plane of the sky.

Based on a comparison of the times for the emergence of the He\,II region and
the X-ray source in Fe\,XXV line from behind the limb of the star, Goranskii
{\it et al.\/} (1997) found that the size of the He\,II emission region ($0.25-0.30\,a$)
exceeds the size of the X-ray emission region ($\approx 0.20\,a$). The
observations of the X-ray eclipse of the Fe\,XXV line were taken from Ginga
data (Kawai {\it et al.\/} 1989). In those observations, the jet X-ray line and the
stationary line of weakly ionized iron were not well resolved. It is possible
that the He\,II region surrounds not only the X-ray jets but also the region
in which the fluorescent iron line is emitted. It was also noted that the
blend of the C\,III, N\,III\,$\lambda_{eff} \,4644$ lines ($T \sim 30\,000$ K)
showed behaviour during eclipses that was similar to that shown by the He\,II
line (narrow line peaks at the centre of eclipses and an overall broadening
of the blend during the egress from eclipse).

All these data suggest that the base of the jets can be represented by
a cocoon of hot gas enshrouding the region through which the jet passes,
with the temperature in the cocoon falling from $\sim 10^8$~K at the axis
to $\sim (3-5) \cdot 10^4$~K at the edges. It is likely that this same
region is the source of the polarized UV radiation (Dolan {\it et al.\/} 1997).
Observations in several
other eclipses (Fabrika {\it et al.\/} 1997b) confirmed that only the narrow He\,II
component remains at the centre of eclipses, with the two-peaked, broad
component of the profile appearing during the egress of the disk from eclipse.
The pattern for the appearance of the blue and red components of the two-peaked
profile varies with the precessional phase, and is not always evident due
to missing observations because of variable observing conditions.

The profiles of lines arising in the supercritical disk (the ``superdisk'')
were calculated by Fukue (2000), taking into account the effects of
reradiation and screening of the radiation of some parts of the disk by
others. The lines are two-peaked, as in an ordinary disk; but if advective
motions, in which the velocity of the radial flow toward the centre grows
sharply, are taken into account in the dynamics of the disk, the blue
component of the profile becomes brighter than the red component. This is
a projection effect associated with the fact that part of the hot, inner
surface of such a disk is blocked by the edge of the disk. Overall, the
orientation of such a superdisk relative to the observer appreciably
influences the observed radiation. While the edges of the disk can be
essentially dark, the near-polar radiation is strengthened by reradiation
effects. The luminosity of the superdisk is $L\sim 9(H/r)L_e$, where $H/r$
is the ratio of the thickness to the radius of the disk, i.e., it depends
on the degree of opening of the disk. It is interesting that, for a mass
of the compact object $10\,M_{\sun}$, the luminosity of the superdisk is
approximately equal to the observed bolometric luminosity  of
SS433, $\sim 10^{40}$~erg/s.

It would be tempting to associate the two-peaked He\,II profile with the
superdisk (Fukue 2000), especially since the blue peak is often observed
to be brighter than the red peak. However, as is noted above, the size of
the region in which this line is radiated appreciable exceeds the Keplerian
radius, so that the He\,II peaks cannot
be formed in the disk.

Filippenko {\it et al.\/} (1988) detected Paschen lines ($P_{11} - P_{15}$) with
two-peaked profiles in the spectrum of SS433, with the distance between the
peaks being $\Delta V \approx 290$~km/s. These observations were carried out
at precessional phases that are close to those when the disk is edge-on
and outside of eclipses. In addition to the Paschen lines, the Fe\,II and
H$\beta$ lines also had two peaks. The ratio of the intensities in the two
peaks changed slightly over three consecutive nights of observations.
Filippenko {\it et al.\/} (1988) proposed that the two-peaked lines are manifestations
of the accretion disk. The time for such a disk to be eclipsed by the optical
star (1--1.5~days) is quite reasonable from an observational point of view,
so that it should be possible to test the hypothesis that the double lines
arise in the disk by searching for eclipses in the lines at the epochs of
primary minima. Variations in the emission-line profiles or in the radial
velocities at the level of $\sim 100$~km/s during eclipses should be
relatively easy to detect, but no reports of such effects appear in the
literature. The two-peaked (``multi-component'') nature of the lines is
observed at precessional phases when the disk is edge-on (Crampton and
Hutchings 1981b), and it is quite likely that it is associated with the
flow of gas from the system through the outer Lagrange point (an
excretion disk) in the plane of the accretion disk. Filippenko {\it et al.\/} (1988)
put forth this hypothesis as an alternative explanation of the observed
two-peaked emission lines. In this case, the total mass of the system should
be rather high ($M \ga 40\, M_{\sun}$). This is in agreement with the
most recent estimates of the component masses based on spectroscopic data
(Gies {\it et al.\/} 2002b).

\medskip

\subsection{Precessional Modulation of the Stationary Lines}

Crampton and Hutchings (1981ab) noted that the radial velocities
of the emission and absorption lines depend on the precessional
phase. In their study of the precessional variability of the
radial velocities, Fabrika {\it et al.\/} (1997a) concluded that
it reduces to a variable absorption component in the blue wing
of the emission lines.  At precessional phases when the
accretion disk is edge-on, the absorption in the blue emission
wings increases sharply and shifts toward the centre of the
emission line (a P\,Cygni profile is visible); the remaining
line emission turns out to be shifted toward the red. When the
disk begins to turn to face the observer (the precessional phase
approaches time $T_3$), the absorption moves away from the
emission toward the blue, and the intensity of the absorption
decreases, so that the radial velocity of the emission lines is
decreased. The radial velocities of some He\,I lines and of the
He\,II line approach the normal value for the system ($V_r \sim
0$~km/s) as the time $T_3$ is approached. This is directly
related to the effect described above that the real orbital
variability of the He\,II line can be measured only at
precessional phases $0.9 < \psi < 0.1$.  Accordingly, the
intensity of the emission lines grows when the disk faces the
observer and decreases when the disk is observed edge-on
(Crampton and Hutchings 1981b; Asadullaev and Cherepashchuk
1986; Fabrika {\it et al.\/} 1997a), which is also a consequence
of the variable absorption. Thus, at precessional phases when
the disk is edge-on, the mean radial velocities of emission
lines grow and their intensity decreases, while, on the
contrary, the emission-line radial velocities decrease and their
intensities grow when the disk partially faces~the~observer.

The orbital and precessional variability of the radial
velocities distort each other. The orbital variability varies
appreciably with precessional phase. On the other hand, after
correcting for the orbital variability, the scatter of line
radial velocities in the precessional curves is appreciably
smaller. Certain regularities in the precessional variability of
various lines appear, in particular, for the He\,I, H$\beta$ and
He\,II emission lines: the higher the amplitude of the
precessional variability (from $50$ to $120$~km/s), the lower
the mean radial velocity of the lines (from $100$ to
$200$~km/s). These effects also provide evidence that the
precessional variability is not associated with real variations
of the regions of line emission, but instead with a variable
contribution of the absorption in the blue wings of the lines.
This is possible if the regions of emission and absorption are
spatially~separated.

\enlargethispage{5mm}

Apart from the precessional and orbital
periods in the variations of the emission-line radial velocities, new
periods are observed (Fabrika {\it et al.\/} 1997a), the strongest of which is
23.22~days ($K_{23} \approx 115$~km/s). In contrast to the precessional
modulation, the variations with these periods are not associated with
corresponding variations in the underlying absorption. It may be that this
periodicity is a consequence of nodding motions in the accretion flow
or spiral shocks in the disk.

\medskip

\subsection{Variability of Absorption Lines. Profile of the
Velocity of the Disk Wind}

The absorption lines in the spectrum of SS433 behave in a
surprising fashion.  An example of the absorption lines can be
seen in Fig.~1, which shows the weak blue absorption components
in emission lines that create P\,Cygni profiles. The strength of
the absorption lines grows sharply when the disk is edge-on;
there are two such times during the precessional cycle of SS433,
corresponding to the  times $T_1$ ($\psi = 0.34$) and $T_2$
($\psi = 0.66$). Accordingly, the absorption lines become
stronger twice during each precessional cycle (Crampton and
Hutchings 1981b). This should obviously be associated with an
increased density of gas lost by the system in the plane of the
accretion disk, since the line of sight lies in the plane of the
disk at times $T_1$ and $T_2$.

The intensity of the absorption lines is also increased at orbital phases
$\phi \sim 0.1$, i.e. immediately after eclipses of the accretion disk
(Fabrika {\it et al.\/} 1997b; Fabrika 1997). This effect can be seen in Fig.~1.
The upper spectrum was obtained nearly at the centre of the primary
eclipse, while the lower spectrum was obtained at orbital phase
 $\phi =
0.096$. The increase in the absorption intensity during the egress from
eclipse (when the bright source emerges from behind the limb of the star)
is associated with an increase in the gas density along the line of sight
in a zone of perturbed wind. The wind from the disk blows at the donor
star, and the density of the wind should be enhanced at the interface where
there is an interaction between the wind and star and perturbation of the
wind. At precessional phases when the disk maximally faces the observer, the
absorption lines are generally barely discernible, but they still become
substantially stronger during eclipses and immediately afterwards, at
orbital  phases $\phi = 0.0 - 0.2$. The higher the wind speed along the
line of sight (the closer to precessional phase 0), the earlier the
strengthening of the absorption lines begins and ends. The geometry for
the perturbations in the flow of gas around the star should indeed depend
on the wind velocity. The higher the wind velocity around the star (compared
to the constant velocity of the orbital motion), the less should be the
curving of the wake in the perturbed wind. Precisely this relation follows
from observations. We can see in Fig.~1 that the absorption lines are
easily discernible even at the centre of the primary minimum. When these
observations were obtained, the wind velocity along the line of sight was
$\sim 1\,200$~km/s.

The precession of the accretion disk makes it possible to use the absorption
lines to measure the wind velocity in SS433 (Fabrika {\it et al.\/} 1997a) as a
function of the polar angle $\alpha$ measured from the disk axis. According
to the kinematic model, we are able to study the wind only in the polar-angle
interval $60^{\circ} < \alpha < 90^{\circ}$. When the disk is edge-on
($\alpha =90^{\circ}$), a dense and slow wind is observed ($V_w \sim 100$~km/s),
while the wind velocity sharply increases as the angular distance from the
plane of the disk increases, reaching values $V_w \sim 1\,300$~km/s. When the
disk is maximally turned to face the observer, the H$\beta$ and He\,I
absorption lines become very weak. Figure~18 shows the velocity of the wind
from the accretion disk as a function of the polar angle measured from
absorption lines of various elements in the polar-angle interval $60^{\circ} <
\alpha < 90^{\circ}$. The radial velocities of the absorption lines were
measured using data for many precessional cycles. While the hydrogen and He\,I
lines show essentially the same dependence, the iron line (the unblended
Fe\,II\,$\lambda$5169 line) follows this dependence only for velocities
$\sim 600$~km/s, beyond which its radial velocity again begins to decrease,
reaching values $V_w \approx 340$~km/s when $\alpha \approx 60^{\circ}$.
The figure also shows the wind velocity indicated by the He\,II line near angles
$\alpha \sim 10^{\circ} - 20^{\circ}$ under the assumption that the two-peaked
He\,II profile forms in cocoons enshrouding the bases of the jets. In contrast
to the H$\beta$, He\,I and Fe\,II data, the wind velocity indicated by the
He\,II line is not derived from direct measurements.

\renewcommand{\thefigure}{18}
\begin{figure}[t]
\centerline{\hspace*{-15mm}\psfig{figure=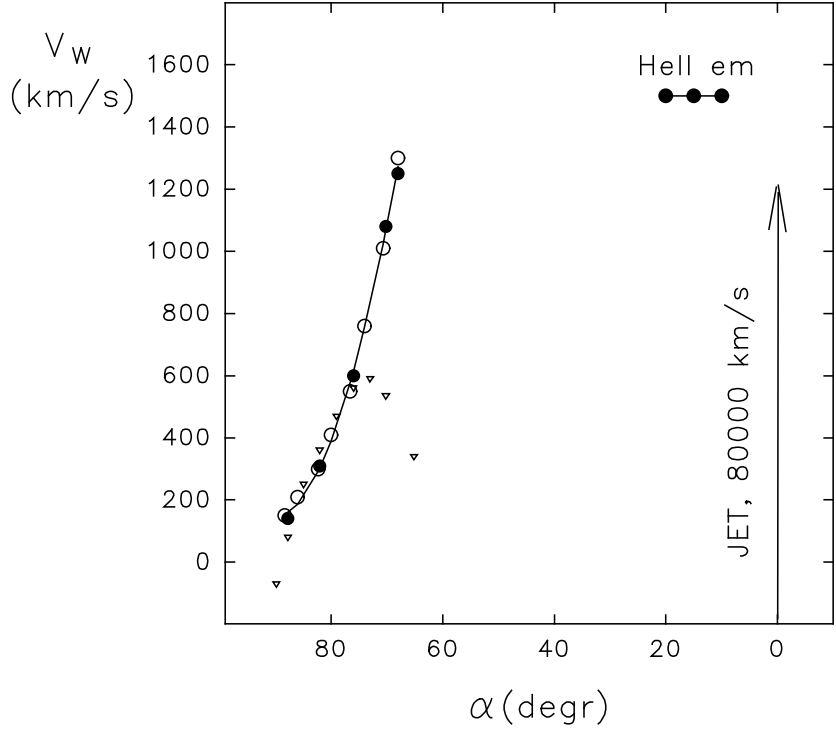,width=110mm}}
\caption{Velocity of the wind outflow from the SS433 accretion
disk as a function of the disk polar angle (Fabrika {\it et al.\/}
1997a), derived from observations of absorption lines at
different orientations of the disk. The open and filled
circles on the left show the data for the H$\beta$ and
He\,I\,$\lambda5015$ absorption lines. The triangles show the
outflow velocity measured using the Fe\,II\,$\lambda$5169
absorption line. The reverse behaviour of the velocity derived
from the iron line indicates that the fast wind overtakes the
slow wind at larger distances from SS433. The terminal mean wind
velocity along the line of sight is about $V_w \approx
340$~km/s. The data on the He\,II emission (the He\,II cocoon) are
model-dependent.}
\end{figure}

As the polar angle decreases, the velocity of the gas flowing out from the
disk grows sharply from 100--150~km/s for $\alpha = 90^{\circ}$ to  
$V_w \ga 1\,300$~km/s for $\alpha \approx 60^\circ$ (Fabrika {\it et al.\/} 1997a;
Fabrika 1997). In this range of $\alpha$, the wind velocity is well
approximated by the relation
$$
V_w=(8\,000\pm100\,\mbox{km/s})\cdot\cos^2\!\alpha+150\pm10\,\mbox{km/s}.
$$
The available data on the wind are in very
good agreement with the picture of gas
flowing out from a supercritical accretion
disk, for which a model was first described by Shakura and Sunyaev (1973).
In this model (see also van den Heuvel 1981; Seifina {\it et al.\/} 1991), the
final wind velocity is
$V_w \sim (2 G M_x / R_{sp})^{1/2}$, where $R_{sp}$
is the spherisation radius of the accretion disk. Adopting $V_w = 1\,500$~km/s,
we find that the spherisation radius in SS433 is $R_{sp} \sim 7
\cdot 10^{10} \, m_6$~cm if the mass of the relativistic star is taken to
be $m_6 = M_x / 6 \,M_{\sun}$. At distances from the relativistic star
exceeding 
$R_{sp}$, the SS433 accretion disk is
``normal'', i.e. it may not differ strongly from the accretion disks
of cataclysmic variables. Continuing with the same simple relationships,
the rate at which gas is supplied to the SS433 accretion disk 
 $\dot M_a =
L_e  R_{sp} / G M_x$ turns out to be
$\dot M_a = 2 \, L_e / V_w^2 \sim
10^{-3} \, m_6 \, M_{\sun}$/yr. The critical or Eddington luminosity
is $L_e \sim 8 \cdot 10^{38} \,m_6$~erg/s for the same mass of the compact
star. The observed bolometric luminosity of SS433 exceeds this critical
luminosity by an order of magnitude.

The rate at which gas flows from SS433 is $\dot M_e \sim 10^{-4}
\,M_{\sun}/$yr (Shklovskii 1981; van den Heuvel 1981). The size of the
wind photosphere $R_{ph}$ is determined by the velocity of the outflow, the
mass-loss rate and the temperature of the gas (the absorption coefficient).
As is noted above, the observed radius of the bright source around the
relativistic object in SS433 is $R_{ph} \approx (1.5 - 2) \cdot 10^{12}$~cm,
and the observed blackbody temperature of the source is $T_{ph} \ga
5 \cdot 10^4$~K. If the outflow from the disk is spherically symmetrical
(which is a crude approximation), the mass-loss rate in the wind will be
$$
\dot M_e\approx4\pi m_p R_{ph}V_w/\sigma_T\sim10^{-4}\,
M_{\sun}/\mbox{yr}.
$$
This is close to many other estimates of $\dot M_e$ based on independent
data, such as IR data (Shklovskii 1981) and radio observations (Blundell
{\it et al.\/} 2001). It turns out that the rate at which gas reaches the outer edge
of the disk appreciably exceeds the rate at which gas flows from SS433:
$\dot M_a / (\dot M_e + \dot M_j) \sim 10$. It is possible that this
provides support for models with advective supercritical accretion disks
(Eggum {\it et al.\/} 1985, 1988; Okuda 2002), in which an appreciable fraction of
the accreted material and radiation is absorbed by the black hole. This
would also indirectly confirm that the relativistic star in this system is
a black hole.

\bigskip

\subsection{Structure of Equatorial Outflows in SS433}

The absorption lines appear and are sharply enhanced approximately at the
times when the disk is viewed edge-on $T_1$ and $T_2$ (Crampton and Hutchings
1981b), but in fact, the times when the absorption intensity is maximum
are delayed somewhat relative to the precise times $T_1$ and $T_2$ by an
amount $\Delta \psi (I_{abs}) = 0.15$. Fabrika {\it et al.\/} (1997a) proposed
that this delay was associated with the time necessary to accumulate
sufficient line-of-sight optical depth in the outflowing gas to form the
absorption lines. The emission-line radial-velocity maxima show roughly
the same delay. Absorption in the wind distorts the blue side of the
emission-line profiles, and the maximum emission-line radial velocities are
observed when the line profiles are maximally distorted (at ``edge-on''
phases). However, the maximum emission-line radial velocities do not occur
at phase $\psi = 0.5$ (half way between $T_1$ and $T_2$), as might be
expected, and show a delay by $\Delta \psi (V_{r,\,em}) \approx 0.12 \pm 0.07$
averaged over all lines. The minimum intensity due to the precessional
variability of the emission lines should be observed when there is maximum
absorption in the wind, at phase $\psi = 0.5$, with the intensity maximum
at phase $\psi = 0.0$. The delay is also observed here: the H$\alpha$ intensity
minimum lags phase $\psi = 0.5$ by $\Delta \psi (I_{H\alpha}) = 0.19$,
according to the data of Asadullaev and Cherepashchuk (1986); the maximum
H$\alpha$ intensity lags phase $\psi = 0.0$ by $\Delta \psi (I_{H\alpha})
\approx 0.1$ (Fabrika {\it et al.\/} 1997c), while the minimum H$\beta$ intensity
lags $\psi = 0.5$ by $\Delta \psi (I_{H\beta}) = 0.13$ (Fabrika {\it et al.\/} 1997a).
Finally, roughly the same delay is observed for the behaviour of the
absorption-line radial velocities with varying precessional phase. When we
observe SS433 when the accretion disk is oriented edge-on, the absorption-line
radial velocities are maximum ($-100$~km/s); after the second edge-on position
($T_2$), the disk turns toward the observer and the absorption radial velocity
decreases appreciably. The maximum absorption radial velocity is observed not
at time $T_2$ ($\psi = 0.66$), but later by $\Delta \psi (V_{r,\,abs})
\approx 0.11$ averaged over all the lines. The phase of the maximum radial
velocity is somewhat different for each absorption line, which was taken into
account when constructing Fig.~18.

All these delays are very similar. It is important that they have been
measured for different lines and using different parameters of the lines,
but all these delay effects have one origin~-- variability of the line
absorption. The delay is determined by the time required for the accumulation
of sufficient optical depth in the gas flowing out in the plane of the
accretion disk to form the absorption lines. In the plane of the disk, the
gas flows out with the velocity $V_w \approx 100-150$~km/s, as follows from the maximum
(but negative) H$\beta$ and He\,I absorption-line radial velocities, but
the maximum radial velocity measured from the Fe\,II absorption lines varies
in the range $V_r = +50$ to $-150$~km/s. The Fe\,II lines have a very weak
emission component, so that their radial velocities can be measured most
accurately, without significant systematic errors. On the other hand, the
orbital variability could introduce appreciable distortion, and we believe
for this reason that the flow velocity in the plane of the accretion disk has
been estimated only very approximately, $V_w \sim 100$~km/s.

The behaviour of the radial velocities and intensities of the absorption
lines with precessional phase indicates that the outer parts of the SS433
accretion disk participate in the precessional motion. Indeed the outflow of
gas from the outer edge of the disk, and therefore the outer edge of the disk
itself, participate in the precessional motion. This means that the angular
momentum of the material flowing from the donor star also precesses,
providing independent support for slaved precession of the SS433 accretion
disk and driven precession of the donor star (Shakura 1972; Roberts 1974;
van den Heuvel {\it et al.\/} 1980; Whitmire and Matese 1980; Katz 1980; Hut and
van den Heuvel 1981). Accordingly, the inner parts of the disk where the
rapid wind and jets form also precess.

In the plane of the disk, the system can lose gas most efficiently through
the libration point L2, and this loss of gas is associated with the removal
of angular momentum during the formation of the disk. Through the L2 point the system can lose
at least half of the total gas supplied by the donor star overfilling its
Roche lobe  (Sawada {\it et al.\/} 1986). An additional source of
loss of angular momentum appears above the plane of the disk~-- the
supercritical wind. The gas lost by the system through L2 leaves the system
along a winding spiral. It is likely that this is the outflow inferred by
Filippenko {\it et al.\/} (1988) from the two-peaked Paschen emission-line profiles.
If this represents an excretion disk, the velocity of its rotation (plus
expansion) is
 $\sim 150$~km/s, in good agreement with absorption-line data.
As discussed above, gas flowing out in the plane of the accretion disk is
observed in the X-ray (via the absorption of the radiation of the receding
jet and the distortion of the orbital light curves), in optical photometric
data (distortion of the orbital light curves), and in VLBI images (the central
gap and equatorial disk); it has been predicted that this flow could also
be detected in the form of an extended H$\alpha$ disk around SS433
(Fabrika 1993).

The outflow velocity in the plane of the accretion disk in the immediate
vicinity of the system is $\sim 100$~km/s. If we consider the distribution of
the wind density along a fixed direction (the line of sight), the outflowing
gas should be distributed non-uniformly at small distances from the
system r~$\la 5\cdot 10^{13}$~cm, which corresponds to motion with velocities
100--150~km/s over several orbital periods. The regions of enhanced density
are modulated with the orbital period, and the distance between them is
$(1 - 1.5) \, \cdot 10^{13}$~cm. The distance between condensations and the
amplitude of the density variations should become smaller with increasing
distance from the system, since high-velocity gas ejected from the accretion
disk after slower gas (in the same direction) has had time to catch up with
this slower gas. At large distances from the system, the wind along the line
of sight is also modulated with the precessional period, and the distance
between gas condensations is $\approx 5 \cdot 10^{14}$~cm in radius.

The velocity of the wind at distances $\sim 10^{14}$~cm can be estimated from
Fig.~18. The radial velocity of the Fe\,II absorption (like the~H$\beta$ and He\,I
absorption) grows as the accretion disk turns toward the observer, but only
to --600~km/s ($\alpha \approx 75^{\circ}$). Higher above the plane of the
disk, the wind temperature increases enough that it is probably no longer
possible for the Fe\,II ion to exist. However, further, the wind velocity
measured from the Fe\,II lines begins to decrease, and is only $V_t \approx
340$~km/s at phase $\psi = 0.95$ (47~days after the edge-on epoch $T_2$).
This final wind velocity results from the averaging of the impulses of
fast and slow gas moving along the line of sight. When the high-velocity wind
catches up to the slow wind emitted earlier in the plane of the accretion
disk, the slower wind is compressed. It is at these precessional phases
that we observe Fe\,II absorption at the greatest distances from the source.
The mean wind speed $V_t$ is observed at distances $\ga 1.4 \cdot 10^{14}$~cm,
covered by the gas moving with this speed over 47~days, in places where
the conditions for enhanced Fe\,II absorption are again created.

In VLBI images of the SS433 jets (``The Radio Jets and W50''), a gap or
sharp weakening of the radio emission is observed at the centre (Paragi
{\it et al.\/} 1999). The binary system is located on the jet axis, but not exactly
at the centre of the gap. The radius of the gap is $\approx 1.8 \cdot
10^{14}$~cm, in projection onto the plane of the sky for a distance to SS433
of 5~kpc. Given that the gas flows out within a rather broad range of angles
in the equatorial plane (the angular range due purely to precession of the
disk is $\pm 20^{\circ}$), the material absorbing the radio emission in
the central gap is located in the equatorial plane a distance $\sim 3 \cdot
10^{14}$~cm from the source. It is likely that this absorbing material
is gas in dense regions of the equatorial wind, observed in optical spectra
via the Fe\,II absorption.

The equatorial VLBI disk (Paragi {\it et al.\/} 1999; Blundell {\it et al.\/} 2001) is
observed out to appreciably larger distances from SS433, to $\sim (3-4) \cdot
10^{15}$~cm. The mechanism for this equatorial radio emission is not
entirely clear (``The Radio Jets and W50''); its spectrum is thermal, but
the brightness temperature is very high. Additional observations and
theoretical studies are required, but it is reasonable to say that the
conditions in the extended disk are made suitable for radio emission by
the dissipation of energy from shock waves. Due to the precession of SS433,
the equatorial wind is modulated by slow ($\sim 100$~km/s) and fast
($\sim 1\,500$~km/s) portions of material. The amplitude of this modulation
depends on the angle above the orbital plane. In particular, the slow wind
should disappear at angles $\ga \pm 20^{\circ}$, so that only the fast wind
remains at large heights above the orbital plane, and it is this fast wind
that compresses the dense equatorial wind. The fragments of equatorial
wind ($\sim 1\,200$~km/s) detected by Paragi {\it et al.\/} (2002) are consistent
with the model of the wind that follows from spectroscopic observations.

\bigskip

\subsection{Gas Streams in SS433}

\renewcommand{\thefigure}{19}
\begin{figure}[p]
\centerline{\psfig{figure=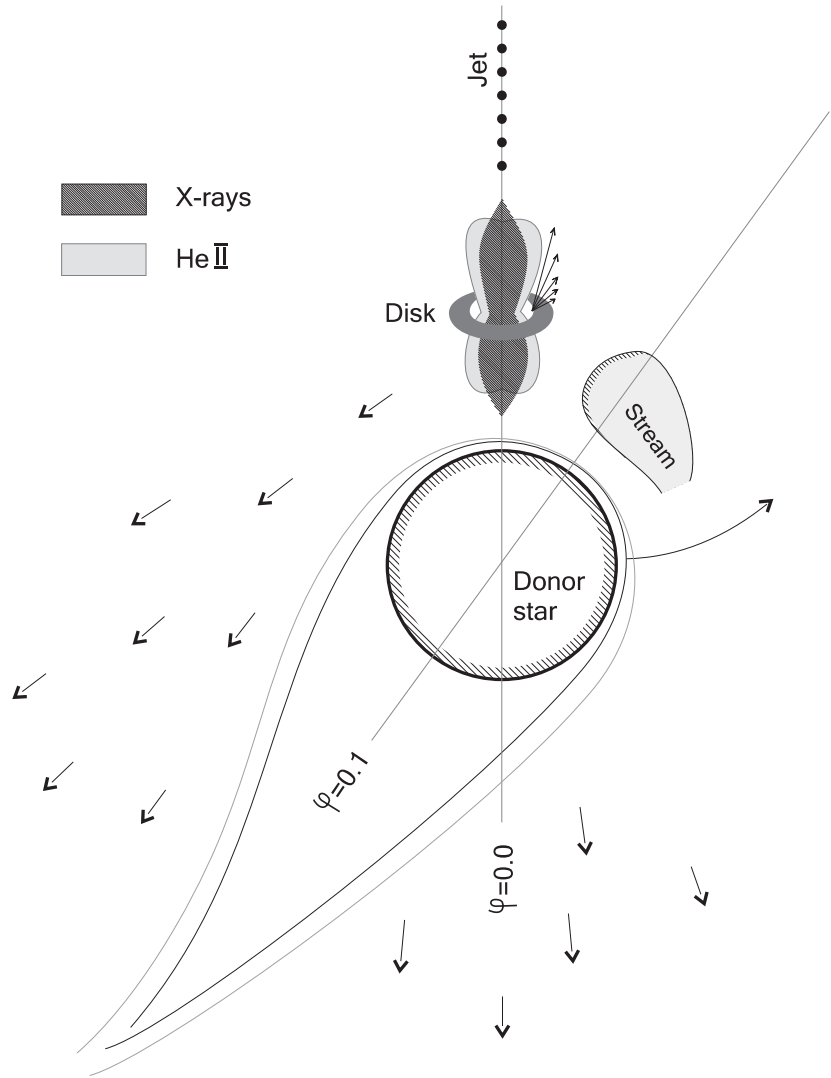,width=109mm}}
\caption{Schematic of the components and gaseous flows in the
SS433 system. The ``unobserved'' accretion disk is shown. The
straight arrows denote the wind from the disk, as well as the
wind blowing at the donor star.}
\end{figure}

Figure~19 presents a schematic of the SS433 system. With the exception of
the accretion disk itself and the donor star, for which no unambiguous
observational manifestations have been detected, all the remaining components
of the system have been studied observationally, and are presented roughly
to scale. The most recent observations of Gies {\it et al.\/} (2002b) indicate that
the donor star has an A-type spectrum. It follows from their estimates of the
component mass ratio $M_x/M_o = 0.57 \pm 0.11$ that the size of the donor
in units of the distance between the components is $R_o = 0.43 \pm 0.02$.
If some component of the system is not observed directly (for example,
a ``hot spot'' where a gas stream interacts with the
disk), it is not depicted in the schematic. The disk wind is shown by
the arrows directly above the disk and behind the optical star. We have not
shown the wind photosphere or the flow from the point L2 in order to avoid
cluttering the figure. The gas stream to the disk may be drawn too far
from the star; it may be somewhat closer, but in that case its size must
be proportionally decreased. In the forward part of the stream, which
is eclipsed by the star at orbital phase $\phi \approx 0.1$, He\,II, He\,I and
hydrogen emission lines are formed, while there is only He\,I and hydrogen
line emission further from the disk in the stream
 (not shown in the figure).

The base of the jet is shown in Fig.~19 as an extended region of X-ray
emission surrounded by hot gas radiating in He\,II lines. The relative size
of this region and of the He\,II-emission region follow from analyses of
eclipse data. At the same time, it is not precisely known what fractions
of the eclipsed X-ray flux form in the ``slow'' wind gas and in the hot
jet gas. The model for the cocoon surrounding the bases of the jets will
depend on this. The 6.4~keV flourescent line of weakly ionized
iron has a small width, FWHM\,$\la 1\,000$~km/s (Marshall {\it et al.\/} 2002), and
could form either within the He\,II cocoon or in the outer wind. Future
observations of X-ray eclipses with high spectral resolution will provide
information about the location of the ``stationary'' X-ray source.

\bigskip

\section{SS433 and Microquasars}

\subsection{Microquasars}

In this section, we will briefly describe the properties of microquasars
as the closest relatives of SS433, as well as ultra-luminous X-ray
sources in other galaxies, which are probably directly related
to  both SS433 and microquasars. The current literature on
microquasars is very extensive; reviews are given by Mirabel and
Rodriguez (1999), Greiner (2000), Mirabel (2001), and Fender
(2001a, 2002).

Microquasars are believed to be X-ray binaries with relativistic jets.
SS433 should probably be considered the prototype for microquasars, since
it was the first stellar object in which relativistic jets were discovered.
However, the name ``microquasar'' was first applied to the X-ray binary
Sco\,X-1, which has radio lobes. The radio images of some X-ray binaries
with relativistic jets and of radio-loud quasars and radio galaxies
are so similar that they are difficult to tell apart without additional
information. Therefore, the term ``microquasar'' was, in turn, chosen
to emphasize the morphological similarities of the radio structure of
these objects (Mirabel {\it et al.\/} 1992).

As a rule, the term microquasars usually refers to X-ray binaries containing
neutron stars or black holes that display jet-like radio activity. This
class currently includes slightly fewer than 20 objects, with an additional
ten or so candidates (Tsarevsky 2002). For example, the massive X-ray
binary Cyg\,X-1 is also a microquasar. It is believed that jets are always
associated with accretion processes, however the variety of the objects
considered somewhat smears out the class of microquasars.

Two ``classical'' microquasars are GRS\,1915+105 and GRO  J1655--40.
Both of these X-ray binaries contain black holes, with masses of
$\sim 14 \,M{\sun}$ (Greiner {\it et al.\/} 2001) and $\approx 7.0 \,M{\sun}$
(Orosz and Bailyn 1997), respectively. Classical microquasars
are superluminal synchrotron radio sources. The velocities of
their jets are 0.92\,c--0.98\,c (Mirabel and Rodriguez 1999).
They are transient objects, whose  jets are ejected during
certain periods of activity; the life times  of individual
radio blobs in the jets are from several days to several weeks.
In active states, the X-ray luminosity is appreciably increased.
It is very probable that the jets in these microquasars are leptonic (in
contrast to the SS433 jets); i.e., there is acceleration and collimation
of relativistic particles in the inner regions near the black
 hole. Infrared
synchrotron radiation is observed during the formation of the relativistic
jet (over a time interval of several minutes;
 Mirabel {\it et al.\/} 1998).
Together with the characteristic behaviour of
 the X-ray emission during
flares (see below), this indicates that,
 in contrast to the situation
for SS433, the region in which the jets
 are generated is fully accessible
to observation. It is possible that optical synchrotron radiation during
the formation of the jets will also be detected.

The speeds of the jets in microquasars cover a broad range  
($\sim 0.1-0.9$\,c). However, in nearly all cases, the jet speed cannot be
determined accurately, since the distance to the objects and the
orientation of the jets are not known. As has already been noted, the
observational manifestations of microquasars are extremely varied, and
their X-ray variability and quasiperiodic oscillations have been especially
well studied. We refer the reader to the reviews listed above, as well
as to the recent review of the X-ray continua of microquasars and models
for their formation by Poutanen and  Zdziarski (2002).

In spite of the large amount of data available, no substantial differences
in the jet activity of microquasars containing neutron stars and black
holes have been noted. This is striking, since the jet ejection involves
the innermost regions, where we would expect the differences between a
neutron star and black hole to become very important. However, the
ratio of the radio to the X-ray luminosity at the peak of a flare is
appreciably higher for black holes than for neutron stars (Fender and
Kuulkers 2001). This could be associated both with a higher efficiency
for the generation of jets or, for example, with higher absorption of
the X-ray emission in the case of black holes.

Fender and Hendry (2000) analysed the radio emission of a number of
persistent (non-transient) X-ray binaries containing both black holes
(or black-hole candidates) and neutron stars. They concluded that the
formation of radio jets requires that the relativistic star not has
a strong magnetic field ($< 10^{10}$~G) and that the accretion rate
be high ($> 0.1 L_{crit}$); in addition, they found that there is a
dramatic physical change in the accretion flow at the time when the
jets are launched.  In the case of X-ray binaries with neutron stars,
the magnetic field must not be strong, so that the accretion flow is
not channeled by the field, right to the innermost regions. X-ray pulsars
(neutron stars with strong magnetic fields) do not show jet-like radio
activity (Fender and Hendry 2000).

A very important circumstance is that the jets in microquasars are
ejected during the low/hard X-ray state (Fender 2001a, 2002), which is
characterized by a hard, power-law spectrum and strong variability in
the X-ray. The presence of radio emission is directly correlated with
the source being in the low/hard state. This implies that the intensity
of the jet activity can be anticorrelated with the accretion rate.

In the famous flare of the microquasar GRS\,1915+105 on September 9, 1997
(Mirabel {\it et al.\/} 1998; Mirabel and Rodriguez 1999), against the background
of powerful, short ($\sim 50$~s) oscillations of the X-ray flux, there
was first the appreciable dip in the X-ray flux over a time comparable to the
oscillation time scale. At the same time, the X-ray spectrum hardened
[(13--60~keV)/(2--13~keV)], and the infrared and radio emission also
weakened, but more smoothly.  During  about 7--8~min of the X-ray dip, a sharp,
isolated X-ray spike appeared, comprised primarily of soft emission.
This is believed to mark the moment when the jet was launched, since the
infrared and X-ray fluxes began to grow immediately after the spike (with
the oscillations again appearing), and there was a radio flare shortly
after the infrared flare. This entire sequence from beginning to end
developed over 30--40~min. The X-ray behavior is usually  interpreted as reflecting
the rapid disappearance (emptying) and subsequent replenishment of the
inner accretion disk.

In the massive X-ray binary Cyg\,X--3, which is believed to probably
contain a neutron star in a very close pair with a Wolf--Rayet star
(the orbital period is 4.8~h), there is appreciable quenching of the
radio emission prior to powerful radio flares (Fender {\it et al.\/} 1997). During
strong radio activity, the soft X-ray flux usually grows, interpreted as
a substantial growth in the rate at which matter arrives at the accretion
disk.

Very interesting behaviour was observed during flares accompanying the
ejection of radio jets in Cyg\,X--3, reminiscent of behaviour shown by
GRS\,1915+105, but occurring on a substantially longer time scale. The
hard X-ray flux at 20--100~keV indicated by BATSE data (McCollough {\it et al.\/}
1999) is anticorrelated with the radio flux in the quiescent state, but
becomes clearly correlated with the radio flux during periods of strong
activity.

The pattern shown by strong flares of Cyg\,X--3 (McCollough
 {\it et al.\/} 1999)
is such that there is first a very large weakening of the hard radiation.
The radio emission also drops, followed by a powerful radio and X-ray
flare or series of several flares, during which the radio and X-ray fluxes
are correlated. The ejection of the jet probably occurs at the moment of
the sudden dip in the hard X-ray and radio emission.

The existence of such strong correlations suggests (McCollough {\it et al.\/}
1999) that both the accretion disk and the jets contribute to the
generation of the hard X-ray radiation. It is also possible that inverse
Compton scattering on the radio (synchrotron) electrons plays a determining
role in the generation of the hard X-ray radiation.

In the microquasar model developed by Markoff {\it et al.\/} (2001), the synchrotron
radiation of the jets and inverse Compton radiation are ascribed a
determining role in the formation of virtually the entire spectrum from
radio to hard X-ray energies. In any case, it is clear we are dealing with
very powerful jets, whose contribution to the total energy released is
substantial, no lower than 5\,\% of the total accretion luminosity (Fender
2001b).

Thus, the behaviour of the X-ray emission during flares in microquasars
may be determined both by radiation associated with the emerging jet~-- the
synchrotron radiation of the relativistic electrons and inverse Compton
scattering of external photons on these electrons, and/or with the
emptying of the inner regions of the accretion disk (Greiner 2000). However,
it is also possible that the correlation of the X-ray and radio fluxes, the
sharp dip in the X-ray emission, and the growth of the hardness ratio of
the X-ray emission observed in the X-ray minima of classical microquasars
are associated with additional absorption of X-ray emission that arises
during the launching of the jets. At this point of the activity, there is
a sharp increase in the rate at which gas reaches the inner regions of the
accretion disk, or a sudden restructuring of the gaseous flows in these
regions. Of course, we must find quantitative and not only qualitative
answers to these questions.

The entire period of flare activity (the cycle of activity) in
 Cyg\,X-3
occupies 80--100~days, while the characteristic time for the radio flares
(and correlated X-ray flares), and also for dips in the radio or X-ray fluxes,
is about 10~days.

In classical microquasars (GRS\,1915+105), the characteristic
 times for
the development of flares are minutes. Is it likely that we are seeing
``naked'' relativistic objects in classical microquasars? More precisely,
we observe all processes there ``in real time''. Therefore,
observations of microquasars are considered direct testing
grounds for the physics of black holes. In the case of Cyg\,X-3
(and all the more so SS433), this is not possible, since the
absorption by the surrounding gas (accretion flows) is
appreciably stronger, so that the inner regions are hidden from
the observer. Of course, the difference in the time scales cannot be
explained purely as an effect of absorption in inner regions of the
source. It is likely that the same process are manifest on various time
scales: the increase in the accretion rate, restructuring of the accretion
disk and flows, and the appearance of jet activity.

\bigskip

\subsection{Supercritical Transients}

Grimm {\it et al.\/} (2002) studied the luminosity function of X-ray binaries
in our Galaxy using RXTE All Sky Monitor data. The X-ray luminosity
depends primarily on the rate of accretion of gas onto the relativistic
star, which, in turn, is determined by the rate of loss of gas by the
donor. Therefore, we expect a continuous, generally power-law, distribution
of X-ray binaries in luminosity, right to the critical luminosity corresponding
to the mass of a neutron star ($L_{e} \sim 2 \cdot 10^{38}$~erg/s)
or slightly more than this limiting luminosity. Since the accretion luminosity
cannot appreciably exceed the Eddington limit, objects may ``accumulate''
near this critical luminosity, and we therefore expect a break in the X-ray
luminosity function (XLF) at high luminosities. This is approximately what is
observed (Grimm {\it et al.\/} 2002; see also references therein to studies of XLFs
for other galaxies).

Of course, in the case of ``moderate'' transient excesses over the critical
accretion rate $\dot M \le 10-100 \, \dot M_{e}$, approximately Eddington
or slightly super-Eddington sources can appear. Grimm {\it et al.\/} (2002) found
that the luminosity function for low-mass X-ray binaries (as earlier, this
term concerns only the mass of the donor) has a break near the
Eddington luminosity for a neutron star ($\approx 1.4 \, M{\sun}$), and
that at least 12 sources showed episodes of supercritical luminosity over
the observing time of the RXTE All Sky Monitor program.

The mass-loss rate of the donor and the rate at which mass is captured
by the relativistic star depend on many factors. In particular, an
increased outflow rate could be the reaction of the donor to accretion
activity; however, in general, the donor mass-loss rate is not related to
the state of the other component (it is completely unrelated to the
presence of an Eddington limit for the luminosity). A sharp increase
in the mass transfer could be a consequence of internal processes in
the donor atmosphere, the properties of the mass loss, possible precessional
motions, the passage of the components through periastron, etc.

In the presence of appreciable transient increases in the mass-accretion
rate over short times, an ``SS433 syndrome'' arises~-- a sharp dip in
the X-ray flux due to absorption in the wind from the accretion disk.
Matter is ejected from the system by radiation pressure (Shakura and
Sunyaev 1973). In addition, powerful disk-like flows screening the central
object or even the entire system can arise. Reprocessing of the X-ray
radiation generated in the central regions in the powerful wind should
lead to the appearance of a peculiar object that is very weak in the X-ray
but bright in the UV and optical. The spectrum of such an object should
show broad emission lines that  are formed in a wind with a velocity of
several thousands
 of km/s. It is obvious that we would expect the formation
of jets, with a sharp enhancement of the radio emission. In addition, it
is quite possible that, at such supercritical times, the hard radiation
of the object becomes collimated perpendicular to the disk.

The famous giant September 1999 flare of the unusually rapid transient
V4641\,Sgr (a black hole) was interpreted by Revnivtsev
 {\it et al.\/} (2002ab)
as a super-Eddington outburst. The unusually rapid and strong flare
of the transient CI\,Cam (a neutron star or black hole in a pair with
a B[e]-supergiant) was also interpreted by Hynes {\it et al.\/} (2002) as a
supercritical accretion episode. In both cases, the short time of the
X-ray flare was associated with the appearance of a wind and absorption
of the X-ray radiation. In addition, the corresponding optical flares were
unusually bright. The maximum bolometric radiation of such flares should
occur in the optical or UV. In both cases, broad optical emission lines
were observed, suggesting the formation of a wind in the inner regions
of the accretion structure.

Did ``heavy and cool'' jets like those in SS433 appear during flares
of V4641\,Sgr and CI\,Cam? There were no reports of unusual lines during
the flares in these objects, but the spectra themselves were very complex
and evolved rapidly (see Revnivtsev {\it et al.\/} 2002ab and Hynes {\it et al.\/} 2002
for references).  The data for SS433 suggest that the appearance of cool
jets during a supercritical-accretion episode is unlikely, since a persistent
channel and a well established disk wind are required to collimate the
jets and confine the cool~gas~clouds~in~them.

It is even more difficult to answer the question of whether collimated
radiation formed during the flares of 4641\,Sgr and CI\,Cam. It is possible
that we may have the opportunity of observing a supercritical flare in
a ``face-on'' X-ray transient in the relatively near future. If the solid
angle subtended by the opening angle of the channel during the supercritical
flare is $\Omega_c$ (and, of course, if collimated radiation can emerge
in such a situation), then for every $2 \pi / \Omega_c$ super-Eddington
outbursts, there should be one for which we detect an extremely bright
X-ray transient ($L_x \sim 10^{40}$~erg/s).

It may be possible to determine the distances to X-ray transients in
our Galaxy if we know the Eddington luminosity for the object (the mass
of the relativistic star) and assume that the peak luminosity during a
short, supercritical X-ray flare should be very close to the Eddington
luminosity. It may also be possible to devise a method for determining
the distances to other galaxies based on knowledge of
 the X-ray luminosity
at the break in the XLF, $\sim 2 \cdot 10^{38}$~erg/s (Sarazin {\it et al.\/} 2001).
Future X-ray missions should enable the determination of the XLFs of
a multitude of galaxies.

\bigskip

\subsection{A Face-on SS433 and Ultraluminous X-ray Sources in
Other Galaxies}

The total 2--10~keV X-ray luminosity of the X-ray sources in our Galaxy
is $\sim (2-3) \cdot 10^{39}$~erg/s (Grimm {\it et al.\/} 2002), with the total
luminosity determined primarily by the few brightest objects. Roughly
the same picture is observed for M31 (Makishima {\it et al.\/} 1989),
whose total 2--20~keV luminosity is $\sim 5 \cdot 10^{39}$~erg/s.

The brightest X-ray sources in our Galaxy and in the Local Group have
X-ray luminosities of a few $\times 10^{38}$~erg/s, while some microquasars
reach luminosities of $\sim 3 \cdot 10^{39}$~erg/s at the peaks of flares.
However, substantially brighter objects that are not active nuclei
(supermassive black holes) are encountered in other galaxies.

One of the most statistically complete surveys of X-ray sources
in nearby galaxies was obtained by Roberts and Warwick (2000)
using archival ROSAT HRI (High Resolution Imager) data and the
list of bright Northern galaxies compiled by Ho {\it et al.\/}
(1997). Roberts and Warwick (2000) distinguished 142 non-nuclear
X-ray sources in galaxies that were included in HRI
observations. Maximum luminosities of their sources reach $L_x
\sim 10^{40}$~erg/s. After adding data for M31, Roberts and
Warwick (2000) obtained the X-ray luminosity distribution of
discrete X-ray sources in 49 spiral galaxies. This distribution
normalised to the optical blue luminosity $10^{10}\, L_{\sun}$
has the form $dN/dL_{38} = (1.0 \pm 0.2) \cdot L_{38}^{-1.8}$,
where $L_{38}$ is the X-ray luminosity in units of
$10^{38}$~erg/s. It also follows from this distribution [Fig.~7
of Roberts and Warwick (2000)] that one source with luminosity
$L_x \ge 10^{40}$~erg/s is encountered in a collection of spiral
galaxies with total blue luminosity $L_B \ge 10^{12} \,
L_{\sun}$. This is in good agreement with the fact that there
are no sources with this luminosity in the Local Group. The mass
of the Local Group is $(1.3 \pm 0.3) \cdot 10^{12}\,M_{\sun}$
(Karachentsev {\it et al.\/} 2002).

The most recent studies of the XLFs for sources in other
galaxies (Sarazin {\it et al.\/} 2001; Kilgard {\it et al.\/}
2002; Zezas and Fabbiano 2002;
Kim and Fabbiano 2003; Colbert {\it et al.\/} 2003;
Grimm {\it et al.\/} 2003 and
references therein) are based on CHANDRA observations.  The
spatial resolution and spectral range of CHANDRA enable studies
of distributions of point sources, and make it possible to
eliminate contamination from supernova remnants to a
considerable degree. It was found that these XLFs depend
strongly on the starburst activity in the corresponding
galaxies. The slopes of the differential XLFs ($dN/dL \propto
L^{-\alpha}$) vary from $\alpha \approx 1.5$ in galaxies with
strong starburst activity to $\alpha = 2.0 - 2.5$ in spiral and
elliptical galaxies. Grimm {\it et al.\/} (2003), 
Gilfanov {\it et al.\/} (2003) discuss a
universal XLF scaled with the star-formation rate. Its slope is
$\alpha \approx 1.6$, and can, in principle, be understood in
terms of the population of high-mass X-ray binaries (Postnov
2003). The universal XLF can be fit with a single power law over
a very wide luminosity range,  $L_x \sim 10^{36} -
10^{40}$~erg/s, and has a cutoff at $L_x \sim$ a few $ \times
10^{40}$~erg/s.  The absence of breaks in the universal XLF is
strange (Zezas and Fabbiano 2002; Grimm {\it et al.\/} 2003); at
least three different populations of objects radiate in this
luminosity interval ($\sim 10^{36} - 10^{40}$~erg/s).  Neutron
stars cannot be more luminous than $\sim 2 \cdot 10^{38}$~erg/s,
and stellar-mass black holes cannot radiate appreciably more
than $ \sim 10^{39}$~erg/s.  In this sense, the universal XLF
(Grimm {\it et al.\/} 2003) imposes certain restrictions on the
nature of the radiating objects.  It may be that we must more
accurately consider the anisotropy of the radiation of X-ray
binaries and variability of the X-ray sources.  Further studies
of XLFs in individual galaxies will help reveal more detailed
behaviour of the XLFs associated with the star-formation history
of the galaxies.

Beginning in 2000, it became clear that ultraluminous X-ray  
sources ($L_x > (1-5) \cdot 10^{39}$~erg/s) in galaxies form a
separate class of objects. However, can we raise the Eddington
luminosity limit in order to avoid considering fundamentally new
types of objects?

Grimm {\it et al.\/} (2002) present several reasons why it may
be possible to increase this limit somewhat. (i) In the standard
theory of Shakura and Sunyaev (1973) with a quasi-flat accretion
disk, the radiation flux emerging perpendicular to the plane of
the disk exceeds the average value by about a factor of three.
(ii) If the chemical composition of the accreting material is
poor in hydrogen (the donor is enriched in helium), this will
raise the Eddington limit, since the Eddington limit for a
helium plasma is twice that of a hydrogen plasma. These two
factors can change the classical limit by up to a factor of~six.
(iii) In theories of supercritical accretion disks, of course,
it is possible for the accretion luminosity to exceed the
Eddington luminosity.  The factor by which the luminosity
exceeds the critical value depends on the model (and on whether
or not the model is stationary), but even in the earliest
supercritical accretion-disk model (Shakura and Sunyaev 1973),
the luminosity of the disk was exceeded by the logarithmic
factor $\ln(\dot M/\dot M_{Edd})$, which can be appreciable in
appreciably supercritical regimes.  In the slim-disk models
(Paczynsky and Wiita 1980; Abramowicz {\it et al.\/} 1988), the
resulting luminosity is likewise higher than the Eddington
luminosity.  (iv) There are mechanisms for exceeding the
accretion luminosity in the case of accretion onto a neutron
star with either a strong or weak magnetic field, due to the
formation of specific geometries in the accretion structures
(such as a magnetised accretion column).

Nevertheless, the collected observational data on
super-Eddington X-ray sources in galaxies (see below) forces us
to search for ``drastic'' solutions to this problem. Either (a)
these objects are not super-Eddington, and are black holes with
masses of $10^2 - 10^4 \,M_{\sun}$, between those for stellar
and supermassive black holes~-- so-called intermediate-mass
black holes, or (b) these objects are face-on supercritical
accretion disks in binary systems (SS433, microquasars) whose
radiation can be collimated by the geometry of a funnel and
beamed by the motion of the source along a direction close to
the line of sight.

Below, we will consider this second hypothesis -- that the
radiation of these objects is formed in the funnels of
supercritical accretion disks -- in more detail. Such objects
were first predicted by Katz (1987). We will not be concerned
with the relativistic beaming factor, since its magnitude could
be determined quite reliably in the case of a face-on SS433 (see
the section ``Structure and Formation of the Jets''). The
beaming factor depends on the spectrum of the emerging
radiation, and is $\approx 2$. The beaming factors for the
relativistic jets of microquasars may be $\sim 10 - 100$,
depending on the jet velocities.

\enlargethispage{5mm}

It was shown in the previous section that the mean radius of the wind
photosphere in SS433 is $R_{ph} \sim 1 \div 2 \cdot 10^{12}$~cm, or
$(0.2-0.5)a$ in terms of the distance between the components. Inside the
funnel of the SS433 supercritical disk with its solid angle of $\Omega_c$,
the radius (height) of the photosphere is
$$
R_{ph,\,j} \sim \dot
M_j \sigma_T/\Omega_c m_p V_j \sim 4 \cdot 10^9\mbox{cm}
$$

\noindent
for a mass-loss rate
in the funnel $\dot M_j \sim 5 \cdot 10^{-7}\,M_{\sun}$/yr and a funnel
opening angle $\theta_c \sim 40^{\circ}$. Collimated radiation could form
along such a funnel. For the bolometric luminosity of SS433 $L_{bol} \sim
10^{40}$~erg/s, which is nearly all radiated in the inner regions of the
accretion disk, we expect the luminosity of the collimated radiation to
also be of the order of $L_c \sim 10^{39} - 10^{40}$~erg/s.

For an observer able to directly see the central parts of the funnel
in SS433, this object would resemble an ultraluminous X-ray source, with
a luminosity of $L_x = (2 \pi/\Omega_c) L_c \sim 10^{40} - 10^{42}$~erg/s,
up to $\sim 10^4$ times brighter than Cyg\,X-1. The X-ray flux
of an
SS433 oriented face-on would vary with a characteristic time scale of
$R_{ph,\,j}/V_j \div R_{ph}/V_j \sim 0.1 - 10^2$~s. The orientation of
SS433 relative to the Earth does not enable us to directly study the
funnel  
(although we cannot consider this orientation unfortunate, since
we can
investigate the binary system itself and the accretion disk by
analysing various eclipses in the system). Objects similar to SS433 in
other galaxies could be manifest as extremely bright X-ray sources
(Katz 1987).

Ultraluminous X-ray sources (ULXs) are indeed observed in other galaxies
(Fabbiano 1998). As a rule, they are located in spiral and irregular
galaxies, in spiral arms and nuclear regions; i.e., in regions of active
star formation. This is consistent with the possibility that ULXs belong
to a young stellar population. In our Galaxy, we know the only SS433, 
and computed evolutionary synthesis models (Lipunov {\it et al.\/} 1996) 
also predict the presence of only a few objects of this type in a spiral galaxy
such as our own. However, in young star-forming regions, the density of
the most massive stars (from which a system such as SS433 could form) is
enhanced by a factor of a hundred compared to the mean density in the
galaxy. Fabrika and Mescheryakov (2001), King {\it et al.\/} (2001) and Koerding
{\it et al.\/} (2001) have suggested that ULXs are objects such as SS433 or
microquasars oriented face-on. Recently, ULXs have been studied very
actively using space-borne instruments. Their main properties (luminosity,
spectrum, variability) are consistent with the hypothesis that ULXs are
supercritical accretion disks oriented so that our line of sight is close
to the disk axis.

Virtually all well studied ULXs display appreciable variability of their
X-ray fluxes. This is a strong argument supporting the idea that
these objects are supercritical accretion disks oriented face-on. However,
when we imagine even such a well studied star as SS433 oriented face-on,
it becomes a hypothetical object whose properties (such as its
spectrum) can be predicted only with considerable  uncertainty.

The frequency with which ULXs are encountered in other galaxies (Roberts
and Warwick 2000; Fabrika and Mescheryakov 2001) is reasonably close to
the expected value if they are face-on supercritical accretion disks.
Fabrika and Mescheryakov (2001) carried out a cross-correlation of sources
from the ROSAT All Sky Survey Bright and Faint Source Catalogs (Voges {\it et al.\/}
1999, 2000) and the RC3 galaxy catalog (de Vaucouleurs {\it et al.\/} 1991), which
contains 16\,741 bright spiral and irregular galaxies. They distinguished
142 sources that were not known to be active nuclei in these galaxies, with
80 of these being located in non-nuclear regions. The X-ray luminosities of
these objects are $L_x \sim 10^{39} - 3 \cdot 10^{41}$~erg/s. The estimated
frequency with which non-nuclear sources are encountered was $\sim 0.05$, or
one object roughly per every 20 galaxies. This frequency can be understood
quantitatively if there are $\sim 1$ SS433-like objects in each galaxy,
with the opening angle of the cone collimating the radiation being equal to
$\theta_c = 30^{\circ} - 40^{\circ}$ and the orientation of these objects
being random. The X-ray spectra (hardness index) of the selected objects
show that they are, on average, hard sources.

The source samples of Roberts and Warwick (2000) and Fabrika
 and Mescheryakov
(2001) differ significantly. The former sample
 was constructed using
pointed HRI observations of a relatively small number of bright galaxies
($B<12.5$), making it possible to study sources with luminosities
$\sim 10^{38}-10^{39}$~erg/s in detail. However,
 observations of a larger
number of galaxies are required to collect
 an appreciable number of
substantially brighter objects,
 $L_x \ge 10^{40}$~erg/s. Fabrika and
Mescheryakov (2001) used catalogs based on the All Sky Survey, and the
sample of galaxies ($V<15.0$) was significantly more representative.
However, sources weaker
 than $\approx 10^{40}$~erg/s could be completely
selected only in nearby galaxies ($< 11$~Mpc). Therefore, this second
sample included, on average, the brightest ULXs.

ULXs have also been identified in elliptical galaxies (Colbert and Prak
2002; Colbert {\it et al.\/} 2003). 
This does not contradict the interpretation of ULXs as microquasars,
since the population of microquasars includes many low-mass X-ray binaries,
which are present in elliptical galaxies. 
However the brightest ULXs are detected in interacting and 
starbursting galaxies, for example the ``hyper-ULXs'' 
($L_x > 10^{41}$~erg/s) in the Cartwheel galaxy (Gao {\it et al.\/} 
2003). Nevertheless, it is possible that
ULXs do not represent a uniform class of objects.

The spectra of ULXs are very similar to those of X-ray binaries, and are
sometimes well fit by a so-called multicolor disk blackbody model 
(kT~$\sim$~1--3\,keV),
although a more complex description of the spectrum is
often required (Okada {\it et al.\/}
1998; Makishima {\it et al.\/} 2000; Kotoku {\it et al.\/} 2000; 
Kubota {\it et al.\/} 2002; Ebisawa {\it et al.\/} 2003). 
In recent studies a cool multicolor disk model (kT~$\sim$~0.1--0.3\,keV)
plus a power--law component are required 
(Miller {\it et al.\/} 2003abc) for the best representation of the ULXs
spectra. Some ULXs show very soft or extremely steep X--ray spectra
(Fabbiano {\it et al.\/} 2003a; Cagnoni {\it et al.\/} 2003). 
Like X-ray binaries, ULXs can display transitions
between the soft/high
and hard/low states of the spectrum (Kubota
{\it et al.\/} 2001; La Parola {\it et al.\/} 2001). As has already been noted, the
variability of the X-ray flux is very substantial (Mizuno {\it et al.\/} 2001;
Mukai {\it et al.\/} 2003; Fabbiano {\it et al.\/} 2003b; Roberts and 
Colbert 2003), and can reach a factor of two over a time
of about one hour. It is likely that studies of variability
on shorter time scales are limited by the sensitivity of modern detectors.
There have even been reports of periodic variability on time scales of
hours to days in some ULXs (Bauer {\it et al.\/} 2001; 
Sugiho {\it et al.\/} 2001; Liu {\it et al.\/} 2002a)
and short ($\sim$~20\,sec) quasi--periodical oscillations 
(Strohmayer and Mushotzky 2003).     

Radio emission has been detected from a ULX in the galaxy NGC\,5408
(Kaaret {\it et al.\/} 2003), with reports that the X-ray, radio and optical
fluxes of this object are consistent with the expectations for beamed
emission from a relativistic jet. It is very important to obtain optical
identifications for ULXs, since they could help provide more certain
answers about the nature of these sources. When ULXs are identified,
it is with very weak objects with optical magnitudes of 20--25, often
located in nebulae (Miller 1995; Roberts {\it et al.\/} 2001; 
Liu {\it et al.\/} 2002b;
Wang 2002; Wu {\it et al.\/} 2002; Roberts {\it et al.\/} 2003;
Holt {\it et al.\/} 2003; Zampieri {\it et al.\/} 2003). 
They are usually blue objects, and some are young clusters 
(Goad {\it et al.\/} 2002; Zezas {\it et al.\/} 2002).

Bubble--like nebulae around ULXs are frequently detected 
(Pakull and Mirioni 2003; Roberts {\it et al.\/} 2003).     
In the region of the ULX in the galaxy Holmberg\,II, Pakull and Mirioni
(2001) detected a nebula radiating in the He\,II\,$\lambda 4886$ excited by
the X-ray source.  They concluded that there was no strong beaming along
the line of sight.  However, strong collimation of the radiation or beaming
is not required to understand these objects.  Thus, the available radio and optical
identifications support the hypothesis that ULXs are objects with supercritical
accretion disks or microquasars, or at least do not contradict this hypothesis.

An alternative model for ULXs is that they contain intermediate-mass black
holes (IMBHs) with masses $\sim 10^3\,M_{\sun}$ (Colbert and
 Mushotzky 1999; van der Marel 2003; Miller and Colbert 2003),    
which could have formed from the very first
(Population~III) generation of stars (Madau and Rees 2001) or in globular
clusters (Miller and Hamilton 2002;). Such black holes could accrete
interstellar gas and become bright X-ray sources with luminosities
$\sim 10^{40}$~erg/s if the surrounding gas is sufficiently dense
($n > 10^2 - 10^3\,\mbox{cm}^{-3}$) and the velocity of the
IMBHs relative to this gas is sufficiently low ($\Delta V <
10$~km/s).  These last two conditions substantially limit the
number of IMBHs that are accessible to observation.

What criteria can be proposed to distinguish between these two alternative
models for ULXs? Investigations of the nebulosities surrounding these
sources should include searches for evidence of dynamical interactions
between jets that may be emerging from the objects and the interstellar
gas. By analogy with SS433 (see the section ``The Radio Jets and W50''),
we may expect perturbations of the interstellar medium with amplitudes of
tens of km/s on scales of tens of parsec. Such features around ULXs could
easily be detected in galaxies at distances of up to $\sim 10$~Mpc, even
in ground-based observations. On the other hand, IMBHs can only ionize
the interstellar medium, not dynamically perturb it. The radius of the
region in which interstellar gas would be captured around a black hole
with a mass of $10^3\,M_{\sun}$ moving with a relative speed of
$\Delta V = 10$~km/s  is only 0.1~pc.

As is mentioned above, one possible test is to search for variability.
Black holes are not able to produce strong variability on time scales
appreciably shorter than a few $\times 0.01 \, (M_{BH}/M_{\sun})$~s (Sunyaev and
Revnivtsev 2000). If we suppose that the black holes in ULXs radiate at
the Eddington limit, it is very unlikely that there will be brightness
variability associated with these black holes on time scales $< 1 \,L_{40}$~s,
where the X-ray luminosity is expressed in units
 of $10^{40}$~erg/s. Detailed
studies of the variability of ULXs on such short time scales must await
the next generation of X-ray telescopes.


A critical experiment that could enable identification of ULXs with
SS433-like objects oriented face-on would be observations
 indicating the
existence of a funnel in the supercritical accretion
 disk. In this case,
the presence of very broad X-ray absorption lines with complex profiles is
predicted. These absorption bands should
 belong to hydrogen- and helium-like
ions of the most abundant
 heavy elements (Fe, S, Si, Mg and others), and
should extend from the Kc to the K$\alpha$ energies of the corresponding
ions and transitions. Thanks to the Doppler decrease in the optical depth
of the material accelerated in the funnel, it may be possible to study the
funnel down to the depth of the photosphere via observations of these
 absorption lines.

Variations of the gas parameters along the funnel~-- its velocity, density,
temperature and volume filling factor~-- could make the absorption-line
profiles appreciably more complex, necessitating the use of X-ray spectra
with high signal/noise ratios in searches for these lines. For example, if,
as is proposed in the section ``The Structure and Formation of the Jets'',
the gas in the inner parts of the funnel is first accelerated to velocities
$\sim 10^{10}$~cm/s and then decelerates due to the action of the wind from
the walls, such that its velocity acquires the value $0.26 c$, we expect
the presence of a very broad K$\alpha$ absorption line shifted toward higher
energies, with the blue wing of this line extending to energies corresponding
to the Kc threshold. Such shallow, broad absorption lines could distort
the continuum near the K$\alpha$--Kc energies.

In essence, the predicted complexity of the dependence of the absorption-line
profiles on the structure of the funnel and mechanisms for acceleration and
collimation of the gas in the funnel present excellent opportunities for
direct probing of these structures in supercritical accretion disks and
for studies of mechanisms for formation of the jets.

\newpage 

\section{Acknowledgements}

The author thanks T.R. Irsmambetova, A.A. Panferov and
 K.A.Postnov for discussions, and O.N. Sholukhova, N.S.
Fabrika, A.E. Surkov, R.A. Karimova and E.A. Barsukova for help with the
preparation of the manuscript. The author is especially grateful
to T. Kotani, Z. Paragi, V. P. Goranskii and H.L. Marshall
for 
presenting figures, to Z. Paragi for useful comments concerning the
radio data, to G. Tsarevsky for providing a list of microquasars, and 
to R. Sunyaev for very valuable comments. This work was supported by the
Russian Foundation for Basic Research (projects N\,03-02-16341) 
and a grant from the State Russian Program
``Astronomy''.

\section{References}
\begin{list}{}{\topsep  0pt \parsep  0pt \itemsep  0pt \partopsep 0pt%
\settowidth{\labelwidth}{999.}%
\setlength{\labelsep}{1em}%
\setlength{\leftmargin}{\labelwidth}%
\addtolength{\leftmargin}{\labelsep}} \small \raggedright
\baselineskip 10pt

\item[1.]
Abell, G.O. and Margon, B. 1979, {\it Nature} 279, 701.
\item[2.]
Abramowicz, M., Czerny, B., Lasota, J. and Szuszkiewicz, E. 1988,
{\it Astrophys. J.} 332, 646.
\item[3.]
Abramowicz, M.A., Igumenshchev, I.V., Quataert, E. and Narayan, R. 2002,
{\it Astrophys. J.} 565, 1101.
\item[4.]
Anderson, S.F., Grandi, S.A. and Margon, B. 1983, {\it
Astrophys. J.} 273, 697.
\item[5.]
Antokhina, E.A. and Cherepashchuk, A.M. 1987, {\it Sov. Astron.} 31, 295.
\item[6.]
Antokhina, E.A., Seifina, E.V. and Cherepashchuk, A.M. 1992,
      {\it Sov. Astron.} 36, 143.
\item[7.]
Arav, N. and Belelman, M. 1992, {\it Astrophys. J.} 401, 125.
\item[8.]
Arav, N. and Belelman, M. 1993, {\it Astrophys. J.} 413, 700.
\item[9.]
Asadullaev, S.S. and Cherepashchuk, A.M. 1986, {\it Sov. Astron.} 30, 57.
\item[10.]
Aslanov, A.A., Cherepashchuk, A.M., Goranskij, V.P., Rakhimov, V.Yu.
     and Vermeulen, R.C. 1993, {\it Astron. Astrophys.} 270, 200.
\item[11.]
Band, D.L. 1987, {\it Publ. Astr. Soc. Pac.} 99, 1269.
\item[12.]
Band, D.L. and Grindlay, J.E. 1984, {\it Astrophys. J.} 285, 702.
\item[13.]
Band, D.L. and Grindlay, J.E. 1986, {\it Astrophys. J.} 311, 595.
\item[14.]
Bauer, F.E., Brandt, W.N., Sambruna, R.M., Chartas, G., Garmire, G.P.,
Kaspi, S. and Netzer, H. 2001, {\it Astron. J.} 122, 182.
\item[15.]
Baykal, A., Anderson, S.F. and Margon, B. 1993, {\it Astron. J.} 106, 2359.
\item[16.]
Begelman, M.C., Hatchett, S.P., McKee, C.F.,
    Sarazin, C.L. and Arons, J. 1980, {\it Astrophys. J.} 238, 722.
\item[17.]
Begelman, M. and Rees, M.J. 1984, {\it Mon. Not. R. Astron. Soc.} 206, 209.
\item[18.]
Begelman, M.C., Blandford, R.D. and Rees, M.J. 1984, {\it Rev.
Mod.  Phys.} 56, N\,2, 1.
\item[19.]
Bisikalo, D.V., Boyarchuk, A.A., Kuznetsov, O.A. and Chechetkin, V.M. 1999,
     {\it Astron. Rep.} 43, 587.
\item[20.]
Blandford, R.D. and K\"onigl, A. 1979, {\it Astrophys. J.} 232, 34.
\item[21.]
Blundell, K.M., Mioduszewski, A.J., Podsiadlowski, P., Muxlow, T.W.B. and
    Rupen, M.P. 2001, {\it Astrophys. J.} 562, L79.
\item[22.]
Blundell, K.M., Rupen, M.P., Mioduszewski, A.J., Muxlow, T.W.B.
and Podsiadlowski, P. 2002, In {\it `New Views on
Microquasars''}, the Fourth Microquasars Workshop,
    Ph.\,Durouchoux, Y.\,Fuchs, J.\,Rodriguez (eds). Center for
    Space Physics, Kolkata (India), p. 249; astro-ph/0209365.
\item[23.]
Bodo, G., Ferrari, A., Massaglia, S. and Tsinganos, K. 1985,
{\it Astron.  Astrophys.} 149, 246.
\item[24.]
Bodo, G., Ferrari, A., Massaglia, S. and Brinkmann, W. 1988,
      {\it Astrophys. Lett. Commun.} 27, 5.
\item[25.]
Bohannan, B. and Crowther, P.A. 1999, {\it Astrophys. J.} 511, 374.
\item[26.]
Bonsignori-Facondi, S.R., Padrielli, L.,
    Montebugnoli, S. and Barbieri, R. 1986, {\it Astron. Astrophys.} 166, 157.
\item[27.]
Borisov, N.V. and Fabrika, S.N. 1987, {\it Sov. Astron. Lett.} 13, 200.
\item[28.]
Brinkmann, W., Fink, H.H, Massaglia, S., Bodo, G. and Ferrari, A. 1988,
               {\it Astron. Astrophys.} 196, 313.
\item[29.]
Brinkmann, W., Kawai, N. and Matsuoka, M. 1989, {\it Astron. Astrophys.} 218, L13.
\item[30.]
Brinkmann, W., Kawai, N., Matsuoka, M. and Fink, H.H. 1991,
{\it Astron. Astrophys.} 241, 112.
\item[31.]
Brinkmann, W., Aschenbach, B. and Kawai, N. 1996, {\it Astron. Astrophys.} 312, 306.
\item[32.]
Brinkmann, W. and Kawai, N. 2000, {\it Astron. Astrophys.} 363, 640.
\item[33.]
Brown, J.C., Cassinelli, J.P. and Collins, G.W. II. 1991,
        {\it Astrophys. J.} 378, 307.
\item[34.]
Brown, J.C. and Fletcher, L. 1992, {\it Astron. Astrophys.} 259, L43.
\item[35.]
Brown, J.C., Mundell, C.G., Petkaki, P. and Jenkins, G. 1995,
{\it Astron. Astrophys.} 296, L45.
\item[36.]
Bursov, N.N. and Trushkin, S.A. 1995, {\it Astron. Lett.} 21, 145.
\item[37.]
Cagnoni, I., Turolla, R., Treves, A., Huang, J.-S., Kim, D. W.,
Elvis, M. and Celotti, A. 2003, {it Astrophys. J}. 582, 654
\item[38.]
Calvani, M. and Nobili, L. 1981, {\it Astrophys. Space Sci.} 79,
387.
\item[39.]
Cant\'{o}, J., Tenorio-Tagle, G. and R\'{o}\v{z}yczka, M. 1988,
{\it Astron. Astrophys}. 192, 287.
\item[40.]
Ciatti, F., Mammano, A. and Vittone, A. 1978, {\it IAU Circ.}
N\,3305, 3.
\item[41.]
Ciatti, F., Mammano, A. and Vittone, A. 1981, {\it Astron. Astrophys.} 94, 251.
\item[42.]
Chakrabarti, S.K. and Matsuda, T. 1992, {\it Astrophys. J.} 390, 639.
\item[43.]
Chakrabarti, S.K. Goldoni, P., Wiita, P.J., Nandi, A. and Das, S.
    2002, {\it Astrophys. J.} 576, L45.
\item[44.]
Chattopadhyay, I. and Chakrabarti, S.K. 2002,
    {\it Mon. Not. R. Astron. Soc.} 333, 454.
\item[45.]
Cherepashchuk, A.M. 1981, {\it Mon. Not. R. Astron. Soc.} 194, 761.
\item[46.]
Cherepashchuk, A.M., Aslanov, A.A. and Kornilov, V.G. 1982,
{\it Sov. Astron.} 26, 697.
\item[47.]
Cherepashchuk, A.M. 1989, {\it Astrophys. Space Phys. Rev.} 7,
185.
\item[48.]
Cherepashchuk, A.M., Bychkov, K.V. and Seifina, E.V. 1995,
            {\it Astrophys. Space Sci.} 229, 33.
\item[49.]
Cherepaschuk, A. 2002, Space Sci. Rev. 102, 23. 
\item[50.]
Cherepashchuk, A.M., Sunyaev, R.A., Seifina, E.V., Panchenko, I.E.,
Molkov, S.V. and Postnov, K.A. 2003, {\it Astron. Astrophys.} 411, L441.
\item[51.]
Clark, D.H. and Murdin, P. 1978, {\it Nature} 276, 45.
\item[52.]
Clark, D.H. 1985, {\it The Quest for SS433}. Viking, New York.
\item[53.]
Colbert, E.J.M. and Mushotzky, R.F. 1999, {\it Astrophys. J.}
519, 89.
\item[54.]
Colbert, E.J.M. and Ptak, A.F. 2002, {\it Astrophys. J. Suppl.
Ser.} 143, 25.
\item[55.]
Colbert, E.J.M., Heckman, T.M., Ptak, A.F., Strickland D.K. and Weaver, 
K.A. 2003, astro--ph/0305476; 2004, {\it Astrophys. J.} 602, 231.
\item[56.]
Collins, G.W., II 1985. {\it Mon. Not. R. Astron. Soc.} 213, 279.
\item[57.]
Collins, G.W., II and Newsom, G.H. 1986, {\it Astrophys. J.} 308, 144.
\item[58.]
Collins, G.W., II and Newsom, G.H. 1988, {\it Astrophys. J.} 331, 486.
\item[59.]
Collins, G.W. II and Scher, R.W. 2002, {\it Mon. Not. Roy.
Astron.  Soc.} 336, 1011.
\item[60.]
Crampton, D., Cowley, A.P. and Hutchings, J.B. 1980, {\it Astrophys. J.} 235, L131.
\item[61.]
Crampton, D. and Hutchings, J.B. 1981a, {\it Astrophys. J.} 251, 604.
\item[62.]
Crampton, D. and Hutchings, J.B. 1981b, {\it Vistas Astron.} 25, 13.
\item[63.]
Crowther, P.A. and Smith, L.J. 1997, {\it Astron. Astrophys.} 320, 500.
\item[64.]
D'Odorico, S., Oosterloo, T., Zwitter, T. and Calvani, M.
       1991, {\it Nature} 353, 329.
\item[65.]
Davidson, K. and  McCray, R. 1980, {\it Astrophys. J.} 241, 1082.
\item[66.]
de Vaucouleurs, G., de Vaucouleurs, A., Corwin, H.G., Jr.,
Buta, R.J., Paturel G., Fouque P.
 1991, {\it Third Reference Cataloque of Bright
Galaxies}. Springer-Verlag, Berlin, Heidelberg, New York.
\item[67.]
Dolan, J.F., Boyd, P.T., Fabrika, S., Tapia, S.,
   Bychkov, V., Panferov, A.A., Nelson, M.J.,
   Percival, J.W., van Citters, G.W., Taylor, D.C. and
   Taylor, M.J. 1997, {\it Astron. Astrophys.} 327, 648.
\item[68.]
Dopita, M.A. and Cherepashchuk, A.M. 1981, {\it Vistas Astron.} 25, 51.
\item[69.]
Drake, S.A. and Ulrich, R.K. 1980, {\it Astrophys. J. Suppl. Ser.} 42, 351.
\item[70.]
Dubner, G.M., Holdaway, M., Goss, W.M. and Mirabel, I.F.
    1998, {\it Astron. J.} 116, 1842.
\item[71.]
Ebisawa, K., \.{Z}ycki, P., Kubota, A., Mizuno, T. and Watarai, K.
2003, {\it Astrophys. J.} 597, 780.
\item[72.]
Efimov, Yu.S., Shakhovskoi, N.M. and Piirola, V. 1984, {\it Astron. Astrophys.}
      138, 62.
\item[73.]
Eggum, G.E.,  Coroniti, F.V. and Katz, J.I. 1985, {\it Astrophys. J.} 298, L41.
\item[74.]
Eggum, G.E., Coroniti, F.V. and Katz, J.I. 1988, {\it Astrophys. J.} 330, 142.
\item[75.]
Eikenberry, S.S., Cameron, P.B., Fierce, B.W., Kull, D.M.,
 Dror, D.H., Houck, J.R. and  Margon, B. 2001, {\it Astrophys. J.} 561, 1027.
\item[76.]
Fabbiano, G. 1998, In {\it ``Hot Universe''}. Proceedings of
IAU Symp. N\,188, K.\,Koyama, S.\,Kitamoto, M.\,Itoh (eds).
    Kluwer Acad.  Press, Dordrecht, p.93.
\item[77.]
Fabbiano, G. and White N.E. 2003, astro--ph/0307077.
\item[78.]
Fabbiano, G., King, A.R., Zezas, A., Ponman, T.J., Rots, A. and
Schweizer, F. 2003a, {\it Astrophys. J}. 591, 843.
\item[79.]
Fabbiano, G., Zezas, A., King, A.R., Ponman, T.J., Rots, A. and Schweizer,
F. 2003b, {\it Astrophys. J.} 584, L5.
\item[80.]
Fabian, A.C. and Rees, M.J. 1979, {\it Mon. Not. R. Astron. Soc.} 187, 13P.
\item[81.]
Fabrika, S.N. 1984, {\it Sov. Astron. Lett.} 10, 16.
\item[82.]
Fabrika, S.N. and Borisov, N.V. 1987, {\it Sov. Astron. Lett.} 13, 279.
\item[83.]
Fabrika, S.N., Kopylov, I.M. and Shkhagosheva, Z.U. 1990,
     preprint N\,61 of Special Astrophysical Observatory.
\item[84.]
Fabrika, S.N. and Bychkova, L.V. 1990, {\it Astron. Astrophys.} 240, L5.
\item[85.]
Fabrika, S.N. 1993, {\it Mon. Not. R. Astron. Soc.} 261, 241.
\item[86.]
Fabrika, S.N. 1997, {\it Astrophys. Space Sci.} 252, 439.
\item[87.]
Fabrika, S.N., Bychkova, L.V. and Panferov, A.A. 1997a,
     {\it Bull. Spec. Astrophys. Obs.} 43, 75.
\item[88.]
Fabrika, S.N., Goranskij, V.P., Rakhimov, V.Y., Panferov, A.A.
    Bychkova, L.V., Irsmambetova, T.R., Shugarov, S.Y. and Borisov, G.V.
    1997b, {\it Bull. Spec. Astrophys. Obs.} 43, 109.
\item[89.]
Fabrika, S.N., Panferov, A.A., Bychkova, L.V. and Rakhimov, V.Yu.
       1997c, {\it Bull. Spec. Astrophys. Obs.} 43, 95.
\item[90.]
Fabrika, S.N. 1998, Doct. Diss. Special Astrophysical Observatory, RAS.
\item[91.]
Fabrika, S. and Mescheryakov, A. 2001, In {\it ``Galaxies and
their Constituents at the Highest Angular Resolution''}. IAU
Symp. N\,205, R.T.\,Schilizzi (ed.), Manchester, United
Kingdom, p.268; astro-ph/0103070.
\item[92.]
Fabrika, S.N. and Irsmambetova T.R. 2002; In {\it ``New Views on
Microquasars''}, the Fourth Microquasars Workshop,
    Ph.\,Durouchoux, Y.\,Fuchs, J.\,Rodriguez (eds). Center for
    Space Physics, Kolkata (India), p.268; astro-ph/0207254.
\item[93.]
Falomo, R., Boksenberg, A., Tanzi, E.G., Tarenghi, M. and Treves, A.
    1987, {\it Mon. Not. R. Astron. Soc.} 224, 323.
\item[94.]
Fejes, I., Schilizzi, R.T. and Vermeulen, R.C. 1988, {\it Astron.
Astrophys.} 189, 124.
\item[95.]
Feldman, P.A., Purton, C.R., Stiff, T. and Kwok, S. 1978,
{\it IAU Circ.} N\,3258, 1.
\item[96.]
Fender, R. 2001a, in {\it ``High Energy Gamma-Ray Astronomy''},
F.A.\,Aharonian and H.J.\,V\"olk (eds). American
Institute of Physics Proc., 558, p.221; astro-ph/0101233.
\item[97.]
Fender, R.P. 2001b, {\it Mon. Not. R. Astron. Soc.} 322, 31
\item[98.]
Fender, R. 2002, in {\it ``Relativistic Flows in
Astrophysics''}, A.W.\,Guthmann, M.\,Georganopoulos,
A.\,Marcowith and K.\,Manolakou (eds); {\it Lecture Notes in
Physics} 589, 101.
\item[99.]
Fender, R.P., Hendry, M.A. 2000, {\it Mon. Not. R. Astron. Soc.}
317, 1.
\item[100.]
Fender, R.P., Kuulkers, E. 2001, {\it Mon. Not. R. Astron. Soc.}
324, 923.
\item[101.]
Fender, R.P., Bell Burnell, S.J., Waltman, E.B., Pooley G.G., Ghiggo,
F.D. and Foster, R.S. 1997, {\it Mon. Not. R. Astron. Soc.} 288,
849.
\item[102.]
Fender, R., Rayner, D., Norris, R., Sault, R.J. and Pooley, G. 2000,
      {\it Astrophys. J.} 530, L29.
\item[103.]
Ferrari, A., Trussoni, E., Rosner, R. and Tsinganos, K. 1985, {\it Astrophys. J.}
       294, 397.
\item[104.]
Fiedler, R.L., Johnston, K.J., Spencer, J.H., Waltman, E.B.,
 Florkowski, S.R., Matsakis, D.N., Josties, F.J., Angerhofer, P.E.,
       Klepczynski, W.J. and McCarthy, D.D. 1987, {\it Astron. J.} 94, 1244.
\item[105.]
Filippenko, A.V., Romani, R.W., Sargent, W.L.W. and Blandford, R.D.
     1988, {\it Astron. J.} 96, 242.
\item[106.]
Frasca, S., Ciatti, F. and Mammano, A. 1984 {\it Astrophys. Space Sci.} 99, 329.
\item[107.]
Fuchs, Y., 2002, astro-ph/0207429.
\item[108.]
Fuchs, Y., Koch-Miramond, L. and \'Abrah\'am, P. 2002, in {\it
     ``Neutron Stars in Supernova Remnants''}, P.O.\,Slane and
     B.M.\,Gaensler (eds), ASP Conf. Ser. N\,271. ASP, San
     Francisco, p.369; astro-ph/0112339; in {\it Proceedings of
     the 4th Microquasar Workshop}, Ph.\,Durouchoux, Y.\,Fuchs
     and J.\,Rodrigueez (eds). Center for Space
     Physics, Kolkata, p.261; astro-ph/0208432.
\item[109.]
Fukue, J. 1987a, {\it Publ. Astron. Soc. Japan} 39, 679.
\item[110.]
Fukue, J. 1987b, {\it Publ. Astron. Soc. Japan} 39, 895.
\item[111.]
Fukue, J.,  Nakashima, R., Arimoto, J., Awano, Y., Honda, S., Ishikawa, K.,
    Kato, T., Kawai, N., Matsumoto, K., Okugami, M., Sakaguchi, T.,
    Tajima, Y., Tanabe, K., Tsuda, K., Watanabe, Y., Yamada, Y.  and
    Yokoo, T. 1997, {\it Publ. Astron. Soc. Japan} 49, 93.
\item[112.]
Fukue, J. 2000, {\it Publ. Astron. Soc. Japan} 52, 829.
\item[113.]
Gao, Yu., Wang, Q.D., Appleton, P.N. and Lucas, R.A. 2003,
{\it Astrophys. J}. 596, L171.
\item[114.]
Geldzahler, B.J., Share, G.H., Kinzer, R.L.,
     Magura, J., Chupp, E.L. and Rieger, E. 1989, {\it Astrophys. J.} 342, 1123.
\item[115.]
Gies, D.R., McSwain, M.V., Riddle, R.L., Wang, Z., Wiita, P.J. and
Wingert, D.W. 2002a, {\it Astrophys. J.} 566, 1069.
\item[116.]
Gies, D.R., Huang, W. and McSwain, M.V. 2002b, {\it Astrophys.
     J.} 578, L67.
\item[117.]
Giles, A.B., King, A.R., Jameson, R.F., Sherrington, M.R., Hough, J.H.,
     Bailey, J.A. and Cunningham, E.C., 1980, {\it Nature} 286, 689.
\item[118.]
Gilfanov, M., Grimm, H.-J. and Sunyaev, R. 2003, {\it Mon. Not. R. 
Astron. Soc.} 347, L57.
\item[119.]
Gladyshev, S.A., Goranskii, V.P. and Cherepashchuk, A.M. 1987,
     {\it Sov. Astron.} 31, 541.
\item[120.]
Goad, M.R., Roberts, T.P., Knigge, C. and Lira, P. 2002,
     {\it Mon. Not. R. Astron. Soc.} 335, L67.
\item[121.]
Goranskij, V.P., Kopylov, I.M., Rakhimov, V.Yu., Borisov, N.V.,
     Bychkova, L.V., Fabrika, S.N. and Chernova, G.P. 1987, {\it Commun.
     Spec. Astrophys. Obs.} 52, 5.
\item[122.]
Goranskii, V.P., Fabrika, S.N., Rakhimov, V.Yu., Panferov, A.A.,
    Belov, A.N. and Bychkova, L.V. 1997, {\it Astron. Rep.} 41, 656.
\item[123.]
Goranskii, V.P., Esipov, V.F. and Cherepashchuk, A.M. 1998a,
    {\it Astron. Rep.} 42, 336.
\item[124.]
Goranskii, V.P., Esipov, V.F. and Cherepashchuk, A.M. 1998b,
    {\it Astron. Rep.} 42, 209.    
\item[125.]
Goranskii, V.P. 2002, private communication.
\item[126.]
Grandi, S.A. and Stone, R.P.S. 1982, {\it Publ. Astr. Soc. Pac.} 94, 80.
\item[127.]
Greiner, J. 2000, in {\it ``Cosmic Explosions: Tenth
   Astrophysics Conference''},  S.S.\,Holt and W.W.\,Zhang (eds).
   American Institute of Physics Proc., 522, p.307.
\item[128.]
Greiner, J., Cuby, J.G., McCaughrean, M.J. 2001, {\it Nature}
   414, 522.
\item[129.]
Grimm, H.-J., Gilfanov, M. and Sunyaev, R. 2002, {\it Astron.
   Astrophys.} 391, 923.
\item[130.]
Grimm, H.-J., Gilfanov, M. and Sunyaev, R. 2003, {\it Mon. Not.
R.  Astron. Soc.} 339, 793.
\item[131.]
Grindlay, J.E., Band, D., Seward, F.,  Leahy, D.,
  Weisskopf, M.C. and Marshall, F.E. 1984, {\it Astrophys. J.} 277, 286.
\item[132.]
Henson, G., Kemp, J. and Kraus, D. 1982, {\it IAU Circ.} N\,3750.
\item[133.]
Hirai, Y. and Fukue, J. 2001, {\it Publ. Astron. Soc. Japan} 53, 679.
\item[134.]
Hjellming, R.M. and Johnston, K.J. 1981, {\it Astrophys. J.} 246, L141.
\item[135.]
Hjellming, R.M. and Johnston, K.J. 1988, {\it Astrophys. J.} 328, 600.
\item[136.]
Ho, L.C., Filippenko, A.V. and Sargent, W.L.W. 1997,
{\it Astrophys. J.} 487, 658.
\item[137.]
Holt, S.S., Schlegel, E.M., Hwang, U. and Petre, R. 2003,
{\it Astrophys. J}. 588, 792.
\item[138.]
Humphreys, R.M., and Davidson, K. 1994, {\it Publ. Astr. Soc. Pac.} 106, 1025.
\item[139.]
Hut, P. and van den Heuvel, E.P.J. 1981, {\it Astron. Astrophys.} 94, 327.
\item[140.]
Hynes, R.I., Clark, J.S., Barsukova, E.A., Callanan, P.J., Charles, P.A.,
Collier Cameron, A., Fabrika, S.N., Garcia, M.R., Haswell, C.A., Horne,
K., Miroshnichenko, A., Negueruela, I., Reig, P., Welsh, W.F. and Witherick,
D.K. 2002, Astron. Astrophys. 392, 991.
\item[141.]
Icke, V. 1989,  {\it Astron. Astrophys.} 216, 294.
\item[142.]
Inoue, H., Shibazaki, N. and Hoshi, R. 2001, {\it Publ. Astron. Soc. Japan} 53, 127.
\item[143.]
Irsmambetova, T.R. 1997, {\it Astron. Lett.} 23, 299.
\item[144.]
Irsmambetova, T.R. 2001, {\it Astrophysics} 44, 243.
\item[145.]
Jaroszynski, M., Abramowicz, M.A. and Paczynski, B. 1980, {\it Acta Astron.} 30, 1.
\item[146.]
Johnston, K.J., Santini, N.J., Spencer, J.H., Klepczynski, W.J.,
    Kaplan, G.H., Josties, F.J., Angerhofer, P.E., Florkowski, D.R. and
    Matsakis, D.N. 1981, {\it Astron. J.} 86, 1377.
\item[147.]
Johnston, K.J., Geldzahler, B.J., Spencer, J.H.,
      Waltman, E.B., Klepczynski, W.J., Josties, F.J.,
       Angerhofer, P.E., Florkowski, D.R., McCarthy, D.D.
      and Matsakis, D.N. 1984, {\it Astron. J.} 89, 509
\item[148.]
Jowett, F.H. and Spencer, R.E. 1995, in {\it ``Proc. 27th
YERAC''}, D.A.\,Green and W.\,Steffen (eds). Cambridge
		 University Press, Cambridge, p.12
\item[149.]
Kaaret, P., Corbel, S. Prestwich, A.H., Zezas, A. 2003,
{\it Science} 299, 365.
\item[150.]
Karachentsev, I.D., Sharina, M.E., Makarov, D.I., Dolphin, A.E.,
Grebel, E.K., Geisler, D., Guhathakurta, P., Hodge, P.W.,
Karachentseva, V.E., Sarajedini, A. and Seitzer, P. 2002,
{\it Astron. Astrophys.} 389, 812.
\item[151.]
Katz, J.I. 1980, {\it Astrophys. J.} 236, L127.
\item[152.]
Katz, J.I. 1986, {\it Comments Astrophys.} 11, 201.
\vspace*{0.3cm}
\item[153.]
Katz, J.J. 1987, {\it Astrophys. J.} 317, 264.
\item[154.]
Katz, J.I., Anderson, S.F., Grandi, S.A. and Margon, B.
     1982, {\it Astrophys. J.} 260, 780.
\item[155.]
Kawai, N., Matsuoka, M., Pan, H. and Stewart, G.C.
       1989, {\it Publ. Astron. Soc. Japan} 41, 491.
\item[156.]
Kemp, J.C., Henson, G.D., Kraus, D.J., Carroll, L.C., Beardsley, I.S.,
    Takagishi, K., Jugaku, J., Matsuoka, M., Leibowitz, E.M.,
     Mazeh, T. and  Mendelson, H. 1986, {\it Astrophys. J.} 305, 805.
\item[157.]
Kilgard, R.E., Kaaret, P., Krauss, M.I., Prestwich, A.H., Raley, M.T
and Zezas, A. 2002, {\it Astrophys. J.} 573, 138.
\item[158.]
Kim, D,-W. and Fabbiano, G. 2003, {\it Astrophys. J}. 586, 826.
\item[159.]
King, A.R., Davies, M.B., Ward, M.J., Fabbiano, G. and Elvis, M. 2001,
   {\it Astrophys. J.} 552, L109.
\item[160.]
Kirshner, R.P. and Chevalier, R.A. 1980, {\it Astrophys. J.} 242, L77.
\item[161.]
Kodaira, K., Nakada, Y. and Backman, D.E. 1985, {\it Astrophys. J.} 296, 232.
\item[162.]
Koerding, E., Falcke, H., Markoff, S. and Fender, R. 2001,
   {\it Astron.  Gesells. Meet. Abstr.} 18, 176.
\item[163.]
K\"onigl, A. 1983, {\it Mon. Not. R. Astron. Soc.} 205, 471.
\item[164.]
Kopylov, I.M., Kumaigorodskaya, R.N. and Somova, T.A. 1985, {\it Sov. Astron.}
   29, 186.
\item[165.]
Kopylov, I.M., Kumaigorodskaya, R.N.,  Somov, N.N., Somova, T.A. and
   Fabrika, S.N. 1986, {\it Sov. Astron.} 30, 408.
\item[166.]
Kopylov, I.M., Kumaigorodskaya, R.N., Somov, N.N., Somova, T.A. and
     Fabrika, S.N. 1987, {\it Sov. Astron.} 31, 410.
\item[167.]
Kopylov, I.M., Bychkova, L.V., Fabrika, S.N., Kumaigorodskaya, R.N.
     and Somova, T.A. 1989, {\it Sov. Astron. Lett.}  15, 474.
\item[168.]
Kotani, T., Kawai, N., Aoki, T., Doty, J., Matsuoka, M., Mitsuda, K.,
       Nagase, F., Ricker, G. and White, N.E. 1994,
       {\it Publ. Astron. Soc. Japan} 46, L147.
\item[169.]
Kotani, T., Kawai, N., Matsuoka, M. and Brinkmann, W.
        1996, {\it Publ. Astron. Soc. Japan} 48, 619.
\item[170.]
Kotani, T., Kawai, N., Matsuoka, M. and  Brinkmann, W. 1997a,
    in {\it ``X-ray Imaging and Spectroscopy of Cosmic Hot
		 Plasmas''}, F.\,Makino and K.\,Mitsuda (eds).
    Universal Academy Press, Tokyo, p.443.
\item[171.]
Kotani, T., Kawai, N., Matsuoka, M. and Brinkmann, W. 1997b,
     in {\it ``Accretion Phenomena and Related Outflows''}. IAU
     Coll. N\,163, D.T.\,Wickramasinghe, G.V.\,Bicknell and
     L.\,Ferrario (eds). Astron.  Soc. of the Pacific, San
     Francisco, p.370.
\item[172.]
Kotani, T. 1998, PhD. The Institute of Space and Astronautical Sciences.
    Japan.
\item[173.]
Kotani, T., Kawai, N., Matsuoka, M. and  Brinkmann, W.
    1998, in {\it ``The Hot Universe''}. Proc. of IAU Symp.
    N\,188.  K.\,Koyama, S.\,Kitamoto, M.\,Itoh (eds). Kluwer
    Acad. Press, Dordrecht, p.358.
\vspace*{0.3cm}
\item[174.]
Kotani, T., Trushkin, S. and Denissyuk, E.K. 2002, In {\it `New Views on
    Microquasars''}, the Fourth Microquasars Workshop,
    Ph.\,Durouchoux, Y.\,Fuchs, J.\,Rodriguez (eds). Center for
    Space Physics, Kolkata (India), p. 257 ; astro-ph/0208250.
\item[175.]
Kotoku, J., Mizuno, T., Kubota, A. and Makishima, K.
2000, {\it Publ. Astr. Soc. Japan} 52, 1081.
\item[176.]
Kubota, A., Mizuno, T., Makishima, K., Fukazawa, Y., Kotoku, J.,
Ohnishi, T. and Tashiro, M. 2001, {\it Astrophys. J.} 547, L119.
\item[177.]
Kubota, A., Done, C. and Makishima, K. 2002, {\it Mon. Not. R.
Astr.  Soc.} 337, L11.
\item[178.]
La Parola, V., Peres, G., Fabbiano, G., Kim, D. W. and Bocchino, F.
2001, {\it Astrophys. J.} 556, 47.
\item[179.]
Lebedev, S.V., Chittenden, J.P., Beg, F.N., Bland, S.N., Ciardi,
A., Ampleford, D., Hughes, S., Haines, M.G., Frank, A., Blackman, E.G.
and Gardiner, T. 2002, {\it Astrophys. J.} 564, 113.
\item[180.]
Leibowitz, E.M. and Mendelson, H. 1982, {\it Publ. Astr. Soc. Pac.} 94, 977.
\item[181.]
Leibowitz, E.M. 1984, {\it Mon. Not. R. Astron. Soc.} 210, 279.
\item[182.]
Leibowitz, E.M., Mazeh, T., Mendelson, H., Kemp, J.C., Barbour, M.S.,
     Takagishi, K., Jugaku, J. and Matsuoka, M. 1984,
     {\it Mon. Not. R. Astron. Soc.} 206, 751.
\item[183.]
Liebert, J., Angel, J.R.P., Hege, E.K.,
   Martin, P.G. and Blair, W.P. 1979, {\it Nature} 279, 384.
\item[184.]
Lind, K.R. and Blandford, R.D. 1985, {\it Astrophys. J.} 295, 358.
\item[185.]
Lipunov, V.M. and Shakura, N.I. 1982, {\it Sov. Astron.} 26, 386.
\item[186.]
Lipunov, V.M., Ozernoy, L.M., Popov, S.B., Postnov, K.A. and Prokhorov, M.E.
      1996, {\it Astrophys. J.} 466, 234.
\item[187.]
Lipunova, G.V. 1999, {\it Astron. Lett.} 205, 508.
\item[188.]
Liu, J.-F., Bregman, J.N., Irwin, J. and Seitzer, P. 2002a,
{\it Astrophys. J.} 581, L93.
\item[189.]
Liu, J.-F., Bregman, J.N. and Seitzer, P. 2002b, {\it Astrophys.
J.} 580, L31.
\item[190.]
Lubow, S.H. and Shu, F.H. 1975, {\it Astrophys. J.} 198, 383.
\item[191.]
Lyndel-Bell, D. 1978, {\it Phys. Scripta} 17, 185.
\item[192.]
Madau, P. 1988, {\it Astrophys. J.} 327, 116.
\item[193.]
Madau, P. and Rees, M.J. 2001, {\it Astrophys. J.} 551, L27.
\item[194.]
Makishima, K., Ohashi, T., Hayashida, K., Inoue, H., Koyama, K., Takano, S.,
Tanaka, Y., Yoshida, A., Turner, M.J.L., Thomas, H.D., Stewart, G.C.,
Williams, R.O., Awaki, H., Tawara, Y. 1989, {\it Pub. Astr. Soc.
Japan} 41, 697.
\item[195.]
Makishima, K., Kubota, A., Mizuno, T., Ohnishi, T., Tashiro, M.,
Aruga, Y., Asai, K., Dotani, T., Mitsuda, K., Ueda, Y., Uno, S.,
Yamaoka, K., Ebisawa, K., Kohmura, Y. and Okada, K. 2000,
{\it Astrophys. J.}  535, 632.
\item[196.]
Mammano, A. and Vittone, A. 1978, {\it IAU Circ.} N\,3308, 3.
\item[197.]
Margon, B. 1979, {\it IAU Circ.} N\,3345, 1.
\item[198.]
Margon, B., Grandi, S. and Ford, H. 1979a,
      {\it Bull. Am. Astron. Soc.} 11, 446.
\item[199.]
Margon, B., Stone, R.P.S., Klemola, A.,
     Ford, H.C., Katz, J.I., Kwitter, K.B. and  Ulrich, R.K.
     1979b, {\it Astrophys. J.} 230, L41.
\item[200.]
Margon, B., Grandi, S.A., Stone, R.P.S. and
   Ford, H.C. 1979c, {\it Astrophys. J.} 233, L63.
\item[201.]
Margon, B. and Anderson, S.F. 1989, {\it Astrophys. J.} 347, 448.
\item[202.]
Margon, B. 1984, {\it Ann. Rev. Astron. Astrophys.} 22, 507.
\item[203.]
Markoff, S., Falcke, H., and Fender, R. 2001, {\it Astron.
Astrophys.} 372, L25.
\item[204.]
Marshall, F.E., Mushotzky, R.F., Boldt, E.A.,
      Holt, S.S. and Serlemitsos, P.J. 1978, {\it IAU Circ.} N\,3314, 2.
\item[205.]
Marshall, H.L., Canizares, C.R. and Schulz, N.S. 2002,
{\it Astrophys. J.} 564, 941.
\item[206.]
Matese, J.J. and Whitmire, D.P. 1982, {\it Astron. Astrophys.} 106, L9.
\item[207.]
Matese, J.J. and Whitmire, D.P. 1983, {\it Astrophys. J.} 266,
776.
\item[208.]
Matese, J.J. and Whitmire, D.P. 1984, {\it Astrophys. J.} 282,
522.
\item[209.]
Mazeh, T., Leibowitz, E.M. and Lahav, O. 1981, {\it Astrophys. Lett.} 22, 55.
\item[210.]
Mazeh, T., Aguilar, L.A., Treffers, R.R., K\"onigl A. and
Sparke, L.S.  1983,  {\it Astrophys. J.} 265, 235.
\item[211.]
Mazeh, T., Kemp, J.C., Leibowitz, E.M., Meningher, H. and
      Mendelson, H. 1987, {\it Astrophys. J.} 317, 824.
\item[212.]
McAlary, C.W. and McLaren, R.A. 1980, {\it Astrophys. J.} 240, 853.
\item[213.]
McCollough, M.L., Robinson, C.R., Zhang, S.N., Harmor, V.A., Hjellming,
R.M., Waltman, E.B., Foster, R.S., Ghiggo, F.D., Briggs, M.S., Pendleton,
G.N. and Johndton K.J. 1999, {\it Astrophys. J.} 517, 951.
\item[214.]
McLean, I.S. and Tapia, S. 1980, {\it Nature} 287, 703.
\item[215.]
Migliari, S., Fender, R. and Mendez, M. 2002, {\it Science} 297,
167.
\item[216.]
Milgrom, M. 1979a, {\it Astron. Astrophys.}  76, L3.
\item[217.]
Milgrom, M. 1979b, {\it Astron. Astrophys.} 78, L9.
\item[218.]
Milgrom, M. 1981, {\it Vistas Astron.} 25, 141.
\item[219.]
Miller, B.W. 1995, {\it Astrophys. J.} 446, L75.
\item[220.]
Miller, M.C. and Hamilton, D.S. 2002, {\it Mon. Not. R. Astron.
Soc.} 330, 232.
\item[221.]
Miller, M.C. and Colbert, E.J.M. 2003, astro-ph/0308402.
\item[222.]
Miller, J.M., Fabbiano, G., Miller, M.C. and Fabian, A.C. 2003a,
{\it Astrophys. J.} 585, L37.
\item[223.]
Miller, J.M., Zezas, A., Fabbiano, G. and Schweizer, F. 2003b,
astro-ph/0302535.
\item[224.]
Miller, J.M., Fabian, A.C. and Miller, M.C. 2003c, astro-ph/0310617.
\item[225.]
Mirabel, I.F., Rodriguez, L.F., Cordier B., Paul, J. and Lebrun, F.
1992, {\it Nature} 358, 215.
\item[226.]
Mirabel, I.F., Dhavan, V., Chaty, S., Rodriguez, L.F., Marti, J.,
Robinson, C.R., Swank, J. and Geballe, T. 1998, {\it Astron.
Astrophys.} 330, L9
\item[227.]
Mirabel, I.F. and Rodriguez, L.F. 1999, {\it Ann. Rev. Astron. Astrophys.} 37, 409.
\item[228.]
Mirabel, I.F. 2001, {\it Astrophys. and Space Sci. Suppl.} 276,
319.
\item[229.]
Mizuno, T., Kubota, A. and Makishima, K. 2001, {\it Astrophys.
J.}  554, 1282.
\item[230.]
Molteni, D., Lanzafame, G. and Chakrabarti, S.K. 1994,
{\it Astrophys. J.} 425, 161.
\item[231.]
Mukai, K., Pence, W.D., Snowden, S.L. and Kuntz, K.D.
2003, {\it Astrophys. J.} 582, 184.
\item[232.]
Murdin, P., Clark, D.H. and Martin, P.G. 1980, {\it Mon. Not. R.
     Astron. Soc.} 193, 135.
\item[233.]
Namiki, M., Kawai, N., Kotani, T., Mamauchi, S. and
     Brinkmann, W. 2000, {\it Adv. Space Res.} 25, 709.
\item[234.]
Namiki, M., Kawai, N., Kotani, T. and Makishima, K. 2003,
{\it Publ. Astron. Soc. Japan} 55, 281.
\item[235.]
Narayan, R., Nityananda, R. and Wiita, P.J. 1983, {\it Mon. Not. R.
Astron. Soc.} 205, 1103.
\item[236.]
Niell, A.E., Preston, R.A. and Lockhart, T.G. 1981, {\it Astrophys. J.} 250, 248.
\item[237.]
Okada, K., Dotani, T., Makishima, K., Mitsuda, K. and Mihara, T.
1998, {\it Publ. Astr. Soc. Japan} 50, 25.
\item[238.]
Okuda, T. and Fujita, M. 2000, {\it Publ. Astron. Soc. Japan} 52, L5.
\item[239.]
Okuda, T. 2002, {\it Publ. Astron. Soc. Japan}, {\bf 54, 253  }.
\item[240.]
Orosz, J.A. and Bailyn, C.D. 1997, {\it Astroph. J.} 477, 876.
\item[241.]
Paczynsky, B. and Wiita, P. 1980, {\it Astron. Astrophys.}
88, 23.
\item[242.]
Pakull, M.W. and Mirioni, L. 2001, {\it Astron.
Gesells. Meet. Abstr.} 18, 12.
\item[243.]
Pakull, M.W. and Mirioni, L. 2003, {\it Rev. Mex. de Astron. y
Astrof. (Serie de Conferencias)}. 15, 197
\item[244.]
Panferov, A.A. and Fabrika, S.N. 1993, {\it Astron. Lett.} 19, 41.
\item[245.]
Panferov, A.A. and  Fabrika, S.N. 1997, {\it Astron. Rep.} 41, 506.
\item[246.]
Panferov, A.A., Fabrika, S.N. and Rakhimov, V.Yu. 1997, {\it Astron. Rep.} 41, 342.
\item[247.]
Panferov, A.A. 1999, {\it Astron. Astrophys.} 351, 156.
\item[248.]
Papaloizou, J.C.B. and Pringle, J.E. 1982, {\it Mon. Not. R. Astron. Soc.} 200, 49.
\item[249.]
Papaloizou, J.C. and Pringle, J.E. 1983, {\it Mon. Not. R. Astron. Soc.} 202, 1181.
\item[250.]
Paragi, Z., Vermeulen, R.C., Fejes, I., Schilizzi, R.T., Spencer, R.E.
  and  Stirling, A.M. 1999, {\it Astron. Astrophys.} 348, 910.
\item[251.]
Paragi, Z., Fejes, I., Vermeulen, R.C., Schilizzi, R.T., Spencer, R.E.
     and Stirling, A.M. 2000, In {\it ``Galaxies and their
      Constituents at the Highest Angular Resolution''}. IAU
     Symp. N\,205, R.T.\,Schilizzi (ed.), Manchester, United
     Kingdom, p.266.
\item[252.]
Paragi, Z., Fejes, I., Vermeulen, R.C., Schilizzi, R.T., Spencer, R.E.
     and Stirling, A.M. 2002, in {\it ``6th VLBI Network''
     Symposium}, E.\,Ros, R.W.\,Porcas, A.P.\,Lobanov and
     J.A.\,Zensus (eds), p.263; astro-ph/0207061.
\item[253.]
Pekarevich, M., Piran, T. and Shaham, J. 1984, {\it Astrophys. J.} 283, 295.
\item[254.]
Peter, W. and Eichler, D. 1993, {\it Astrophys. J.} 417, 170.
\item[255.]
Peter, W. and Eichler, D. 1996, {\it Astrophys. J.} 466, 840.
\item[256.]
Petterson, J.A. 1981, {\it Adv. Space Res.} 1, 49.
\item[257.]
Postnov, K.A. 2003, {\it Astron. Lett.} 29, 1.
\item[258.]
Poutanen, J. and Zdziarski, A.A. 2002, in {\it ``New Views on
Microquasars''}, the Fourth Microquasars Workshop,
Ph.\,Durouchoux, Y.\,Fuchs, and J.\,Rodriguez (eds). Center for
Space Physics, Kolkata (India), p.268; astro-ph/0209186.
\item[259.]
Rees, M.J., Phinney, E.S., Begelman, M.C. and Blandford, R.D. 1982,
     {\it Nature} 295, 17.
\item[260.]
Revnivtsev, M., Sunyaev, R., Gilfanov, M. and Churazov, E.
2002a, {\it Astron. Astrophys.} 385, 904.
\item[261.]
Revnivtsev, M., Gilfanov, M., and Churazov, E. and Sunyaev, R. 2002b,
{\it Astron. Astrophys.} 391, 1013.
\item[262.]
Roberts, W.J. 1974, {\it Astrophys. J.} 187, 575.
\item[263.]
Roberts, T.R. and Warwick, R.S. 2000, {\it Mon. Not. R. Astron. Soc.} 315, 98.
\item[264.]
Roberts, T.P. and Colbert, E.J.M. 2003, {\it Mon. Not. R. Astron. Soc}.
341, L49.
\item[265.]
Roberts, T.P., Goad, M.R., Ward, M.J., Warwick, R.S., O'Brien, P.T.,
Lira, P. and Hands, A.D.P. 2001, {\it Mon. Not. R. Astron. Soc.}
325, L7.
\item[266.]
Roberts, T.P., Goad, M.R., Ward, M.J. and Warwick, R.S. 2003
{\it Mon. Not. R.  Astron. Soc.} {\bf 342, 709.  }
\item[267.]
Romney, J.D., Schilizzi, R.T., Fejes, I. and Spencer, R.E.
         1987, {\it Astrophys. J.} 321, 822.
\item[268.]
Rowell, G.P. 2001, astro-ph/0104288. 
\item[269.]
Safi-Harb, S. and Oegelman, H.  1997, {\it Astrophys. J.} 483, 868.
\item[270.]
Safi-Harb, S. and Petre, R. 1999, {\it Astrophys. J.} 512, 784.
\item[271.]
Safi-Harb, S. and Kotani, T. 2002, in {\it ``New Views on
Microquasars''}, the Fourth Microquasars Workshop,
Ph.\,Durouchoux, Y.\,Fuchs, and J.\,Rodriguez (eds). Center for
Space Physics, Kolkata (India), 271; astro-ph/0210396.
\item[272.]
Sarazin, C.L., Irwin, J.A. and Bregman, J.N. 2001,
{\it Astrophys. J.}  556, 533.
\item[273.]
Sawada,K., Matsuda, T. and Hachisu, I. 1986. {\it Mon. Not. R. Astron. Soc.} 221, 679.
\item[274.]
Shapiro, P.R.,  Milgrom, M. and Rees, M.J. 1986, {\it Astrophys. J. Suppl.
     Ser.} 60, 393.
\item[275.]
Seaquist, E.R., Gregory, P.C. and Crane, P.C. 1978, {\it IAU Circ.} N\,3256, 2.
\item[276.]
Seaquist, E.R., Gilmore, W., Nelson, G.J., Payten, W.J. and Slee, O.B.
    1980, {\it Astrophys. J.} 241, L77.
\item[277.]
Seaquist, E.R. 1981, {\it Vistas Astron.} 25, 79.
\item[278.]
Seaquist, E.R., Gilmore, W.S., Johnston, K.J. and  Grindlay, J.E.
    1982, {\it Astrophys. J.} 260, 220.
\item[279.]
Seifina, E.V., Shakura, N.I., Postnov, K.A. and Prokhorov, M.E. 1991.
     {\it Lect. Notes Phys.} 385, 151.
\item[280.]
Seward, F., Grindlay, J., Seaquist, E. and Gilmore, W. 1980, {\it Nature}
287, 806.
\item[281.]
Shakura, N.I. 1972, {\it Sov. Astron.} 16, 756.
\item[282.]
Shakura, N.I. and Sunyaev, R.A. 1973, {\it Astron. Astrophys.} 24, 337.
\item[283.]
Shklovskii, I.S. 1960, {\it Sov. Astron.} 4, 243.
\item[284.]
Shklovskii, I.S. 1981, {\it Sov. Astron.} 25, 315.
\item[285.]
Sikora, M. 1981, {\it Mon. Not. R. Astron. Soc.} 196, 257.
\item[286.]
Spencer, R. E. and Waggett, P. 1984, in {\it ``VLBI and Compact
Radio Sources''}. Proc.IAU Symp. N\,110, R.\,Fanti (ed.), p.297.
\item[287.]
Stephenson, C.B. and Sanduleak, N. 1977, {\it Astrophys. J. Suppl. Ser.} 33, 459.
\item[288.]
Stewart, G.C., Watson, M.G., Matsuoka, M., Brinkmann, W., Jugaku, J.,
     Takagishi, K., Omodaka, T., Kemp, J.C., Kenson, G.D.,
     Kraus, D.J., Mazeh, T. and Leibowitz, E.M. 1987,
     {\it Mon. Not. R. Astron. Soc.} 228, 293.
\item[289.]
Stone, J.M., Pringle, J.E. and Begelman, M.C. 1999, {\it Mon.
Not.  R. Astron. Soc.}  310, 1002.
\item[290.]
Strohmayer, T.E. and Mushotzky, R.F. 2003, {\it Astrophys. J}. 586, L61.
\item[291.]
Sugiho, M., Kotoku, J., Makishima, K., Kubota, A., Mizuno, T.,
Fukazawa, Y. and Tashiro, M. 2001, {\it Astrophys. J.} 561, L73.
\item[292.]
Sunyaev, R. \& Revnivtsev, M. 2000, {\it Astron. Astrophys.}
358, 617.
\item[293.]
Trushkin, S.A., Bursov, N.N. and Smirnova, Yu.V. 2001, {\it Astron. Rep.}
 45, 804.
\item[294.]
Tsarevsky, G. 2002, private communication.
\item[295.]
van den Heuvel, E.P.J., Ostriker, J.P. and Petterson, J.A.
    1980, {\it Astron. Astrophys.} 81, L7.
\item[296.]
van den Heuvel, E.P.J. 1981, {\it Vistas Astron.} 25, 95.
\item[297.]
van der Laan, H. 1966, {\it Nature} 211, 1131.
\item[298.]
van der Marel, R.P. 2003, astro-ph/0302101.
\item[299.]
Velazquez, P.F. and Raga, A.C. 2000, {\it Astron. Astrophys.} 362, 780.
\item[300.]
Vermeulen, R. C. 1996, in: {\it ``Jets from stars and Galactic
Nuclei''}, W.\,Kundt (ed.). Springer-Verlag, Berlin, Heidelberg,
New York; also {\it Lect. Notes Phys.} 471, 122.
\item[301.]
Vermeulen, R.C., Schilizzi, R.T., Icke, V.,  Fejes, I. and
       Spencer, R.E. 1987, {\it Nature} 328, 309.
\item[302.]
Vermeulen, R.C., Murdin, P.G., van den Heuvel, E.P.J., Fabrika,
      S.N., Wagner, B., Margon, B., Hutchings, J.B., Schilizzi, R.T.,
       van Kerkwijk, M.H., van den Hoek, L.B., Ott, E., Angebault, L.P.,
     Miley, G.K., D'Odorico, S. and Borisov, N. 1993a,
    {\it Astron. Astrophys.} 270, 204.
\item[303.]
Vermeulen, R.C., Schilizzi, R.T., Spencer, R.E.,
    Romney, J.D. and  Fejes, I.  1993b, {\it Astron. Astrophys.} 270, 177.
\item[304.]
Vermeulen, R.C., McAdam, W.B., Trushkin, S.A., Facondi, S.R.,
 Fiedler, R.L., Hjellming, R.M., Johnston, K.J and Corbin, J.
 1993c, {\it Astron. Astrophys.} 270, 189.
\item[305.]
Voges, W., Aschenbach, B., Boller, Th., Bra\"uninger, H., Briel,
U., Burkert, W., Dennerl, K., Englhauser, J., Gruber, R.,
Haberl, F., Hartner, G., Hasinger, G., K\"urster, M.,
Pfeffermann, E., Pietsch, W., Predehl, P., Rosso,
C., Schmitt, J.H.M.M., Tr\"umper, J., Zimmermann, H.U.
1999, Astron. Astrophys. 349, 389.
\item[306.]
Voges, W., Aschenbach, B., Boller, Th., Bra\"uninger, H., Briel,
U., Burkert, W., Dennerl, K., Englhauser, J., Gruber, R.,
Haberl, F., Hartner, G., Hasinger, G.,
Pfeffermann, E., Pietsch, W., Predehl, P.,
Schmitt, J.H.M.M., Tr\"umper, J., Zimmermann, H.U.
2000, http://wave.xray.mpe.mpg.de/rosat/catalogues/rass-fsc.
\item[307.]
Wagner, R.M. 1986, {\it Astrophys. J.} 308, 152.
\item[308.]
Wang, Q.D. 2002, {\it Mon. Not. R. Astron. Soc.} 332, 764.
\item[309.]
Watson, M.G., Willingale, R., Grindlay, J.E. and  Seward, F.D. 1983,
{\it Astrophys. J.} 273, 688.
\item[310.]
Watson, M.G., Stewart, G.C., King, A.R. and Brinkmann, W. 1986,
  {\it Mon. Not. R. Astron. Soc.} 222, 261.
\item[311.]
Whitmire, D.P. and Matese, J.J. 1980, {\it Mon. Not. R. Astron. Soc.} 193, 707.
\item[312.]
Wu, H., Xue, S.J., Xia, X.Y., Deng, Z.G. and Mao, S. 2002,
{\it Astrophys. J.} 576, 738.
\item[313.]
Yamauchi, S., Kawai, N. and Aoki, T. 1994. {\it Pub. Astr. Soc. Japan} 46, L109.
\item[314.]
Yuan, W., Kawai, N., Brinkmann, W. and Matsuoka, M. 1995,
{\it Astron. Astrophys.} 297, 451.
\item[315.]
Zampieri, L., Mucciarelli, P., Falomo, R., Kaaret, P., Di Stefano, R.,
Turolla, R., Chieregato, M. and Treves, A. 2003, astro-ph/0309687.
\item[316.]
Zealey, W.J., Dopita, M.A. and Malin, D.F. 1980, {\it Mon.
                Not. R. Astron. Soc.} 192, 731.
\item[317.]
Zezas, A. and Fabbiano, G. 2002, {\it Astrophys. J.} 577, 726.
\item[318.]
Zezas, A., Fabbiano, G., Rots, A.H. and Murray, S.S. 2002,
{\it Astrophys. J.} 577, 710.
\item[319.]
Zwitter, T., Calvani, M., Bodo, G. and Massaglia, S. 1989,
      {\it Fundam. Cosmic Phys.} 13, 309.
\item[320.]
 Zwitter, T., Calvani, M. and D'Odorico, S. 1991,
{\it Astron. Astrophys.} 251, 92.

\end{list}

\end{document}